\documentclass[fleqn,11pt]{article}

\usepackage{arxiv}
\usepackage{{jabbrv}}
\usepackage{cite}
\usepackage{filecontents}
\usepackage{titletoc}

\usepackage[utf8]{inputenc}
\usepackage[T1]{fontenc}

\usepackage{gensymb}

\usepackage{hyperref}

\usepackage{color}

\usepackage{amsmath,amsfonts,amssymb}
\usepackage{graphicx}
\usepackage{booktabs}

\usepackage{csquotes}

\usepackage[per-mode=symbol-or-fraction,separate-uncertainty=true]{siunitx}
\DeclareSIUnit{\molar}{M}

\newcommand{\etal}{\emph{et al}}

\newcommand{\Rtub}{r_\text{t}}
\newcommand{\Ltub}{\ell_\text{t}}
\newcommand{\Rmt}{R_\text{MT}}
\newcommand{\lmax}{\ell_\text{max}}
\newcommand{\Lmt}{\ell_\text{MT}}
\newcommand{\Glong}{\Delta G_\text{long}^{0*}}
\newcommand{\GlongVal}[1]{\Delta G_\text{long}^{0*} = \SI{#1}{\kBT}}
\newcommand{\Glat}{\Delta G_\text{lat}^{0}}
\newcommand{\Elat}{E_\text{lat}}
\newcommand{\klat}{k_\text{lat}}
\newcommand{\Erep}{E_\text{rep}}
\newcommand{\krep}{k_\text{rep}}
\newcommand{\Ecurl}{E_\text{bend}}
\newcommand{\kcurl}{\kappa}
\newcommand{\Emt}{E_\text{MT}}
\newcommand{\kon}{k_\text{on}}
\newcommand{\konc}{k_+}
\newcommand{\koncVal}[1]{\konc = \SI[per-mode=reciprocal]{#1}{\per\micro\molar \per\second}}
\newcommand{\ctub}{c_\text{tub}}
\newcommand{\ctubc}{c_\text{tub,c}}
\newcommand{\ctubVal}[1]{c_\text{tub} = \SI{#1}{\micro\molar}}
\newcommand{\cdil}{c_\text{dil}}
\newcommand{\tdil}{t_\text{dil}}

\newcommand{\koff}{k_\text{off}}
\newcommand{\koffOneD}{\tilde{k}_\text{off}}
\newcommand{\cstd}{c_0}
\newcommand{\kform}{k_\text{form}}
\newcommand{\katt}{k_\text{att}}
\newcommand{\krup}{k_\text{rup}}
\newcommand{\Gmech}{\Delta G_\text{mech}}

\newcommand{\Flat}{F_\text{lat}}
\newcommand{\lrup}{\ell_\text{rup}}
\newcommand{\khydr}{k_\text{hydr}}
\newcommand{\khydrVal}[1]{\khydr = \SI[per-mode=reciprocal]{#1}{\per\second}}
\newcommand{\khydrN}{k_\text{hydr}^0}
\newcommand{\khydrNVal}[1]{\khydrN = \SI[per-mode=reciprocal]{#1}{\per\second}}
\newcommand{\Ehydr}{E_\text{hydr}}
\newcommand{\DEhydr}{\Delta E_\text{hydr}}
\newcommand{\Fhydr}{F_\text{hydr}}
\newcommand{\NGTP}{N_\text{GTP}}
\newcommand{\NGDP}{N_\text{GDP}}
\newcommand{\vgro}{v_\text{gro}}
\newcommand{\vshr}{v_\text{shr}}
\newcommand{\kBT}{\ensuremath{\mathit{k}_\text{B} \mathit{T}}}
\newcommand{\tsim}{t_\text{sim}}
\newcommand{\agro}{a_\text{gro}}
\newcommand{\bgro}{b_\text{gro}}

\newcommand{\Fcurl}{F_\text{bend}}
\newcommand{\Frup}{F_\text{rup}}
\newcommand{\avgFrup}{\langle \Frup \rangle}

\newcommand{\dDepoly}{d_\text{depoly}}

\newcommand{\bendAngle}{\Delta \theta}
\newcommand{\restBendAngle}{\Delta \theta_0}
\newcommand{\restBendAngleBarrier}{\Delta \theta_0^\text{barrier}}
\newcommand{\FhydrBarrier}{\Delta F_\text{hydr}^\text{barrier}}

\newcommand{\FhydrRelease}{\Delta F_\text{hydr}^\text{release}}
\newcommand{\dcutoff}{d_\text{cutoff}}
\newcommand{\Ncap}{N_\text{cap}}
\newcommand{\porousCapLength}{N_\text{pcap}}
\newcommand{\helicalShift}{\Delta z_\text{h}}
\newcommand{\persistenceL}{L_\text{p}}
\newcommand{\probGTP}{p_\text{GTP}}
\newcommand{\avgCapLength}{\bar{\ell}_\text{cap}}
\newcommand{\Dtdelay}{\Delta t_\text{delay}}

\usepackage{tikz}
\usepackage{pgfplots,ifthen}
\pgfplotsset{compat=1.15}
\usetikzlibrary{angles,arrows,patterns}

\pgfplotsset{%
       legend image code/.code={%
               \draw[mark repeat=2,mark phase=2] plot coordinates {
                       (0cm,0cm)
                       (0.125cm,0cm)   
                       (0.25cm,0cm)    
               };%
       },%
       legend style={%
               /tikz/every even column/.append style={%
                       column sep=0.15cm%
               }%
       }%
}

\pgfplotsset{every axis legend/.append style={cells={anchor=west}}}

\tikzset{font={\small}}

\usepgfplotslibrary{external,groupplots}
\usepackage[left=3cm,right=3cm,top=3.25cm,bottom=3.25cm,headheight=12pt,letterpaper]{geometry}

\title{Chemomechanical simulation of microtubule dynamics with
explicit lateral bond dynamics} 

\author{
  Matthias Schmidt \\
  Physics Department \\
  TU Dortmund University \\
   \And
  Jan Kierfeld \\
  Physics Department \\
  TU Dortmund University \\
  \texttt{jan.kierfeld@tu-dortmund.de} 
}

\usepackage{url,hyperref,lineno,microtype,subcaption}
\usepackage{textcomp}
\usepackage[onehalfspacing]{setspace}
\usepackage{bm}

\begin{document}

\flushbottom
\maketitle

\begin{abstract}
We introduce and parameterize a chemomechanical model  of 
microtubule dynamics on the dimer level, which is based on the
allosteric tubulin model and  includes attachment, 
detachment and hydrolysis of tubulin dimers as well as
stretching of lateral bonds, bending at longitudinal junctions,
and the possibility of lateral bond rupture and formation.
The model is computationally efficient such that  we reach
sufficiently long simulation times to observe repeated
catastrophe and rescue events at realistic tubulin concentrations
and hydrolysis rates, which allows us 
to deduce catastrophe and rescue rates. 
The chemomechanical model also allows us to  gain insight into
microscopic features of the GTP-tubulin cap structure and 
microscopic structural features triggering microtubule catastrophes
and rescues. 
Dilution simulations  show qualitative agreement with experiments. 
We also explore the consequences of a possible
feedback of mechanical forces onto the
hydrolysis process and the GTP-tubulin cap structure.

\end{abstract}

\newpage


\section{Introduction}

Microtubule (MT) dynamics is essential for many cellular processes, such as the
positioning and separation of chromosomes in mitosis \cite{McIntosh2002}, or
maintenance of cell polarity and cell shape \cite{Siegrist2007}.
An important feature, which enables MTs to exert pulling and pushing forces in
these cellular processes, is their dynamic instability, which is the stochastic
switching of MTs between states of growth by polymerization and states
of fast shrinkage by depolymerization \cite{Mitchison1984}.

Switching from growth into shrinkage happens in catastrophe events, whose
mechanism and triggers are not completely understood on the molecular level,
but they  are associated with a loss of the GTP-cap by hydrolysis within
the MT \cite{Carlier1984,Walker1991} (see Refs.\ \cite{Howard2009,VanHaren2019}
for reviews). 
Hydrolysis is strongly
coupled to  mechanics of the MT, as 
 is clearly seen in the curling of MT protofilaments
into a \enquote{ram's horn} conformation after the catastrophe and during the
shrinking phase \cite{Mandelkow1991}. 
The loss of the stabilizing GTP-cap triggers a release of binding energy and
stored mechanical energy in the tubular MT structure.
Therefore, shrinkage following a catastrophe is more than simple
depolymerization of the MT;
it is rather a rupture  or crack propagation process
between protofilaments, which releases chemical
and mechanical energy while it propagates towards the minus end.
The energy released during shrinking has biological functions and 
can be employed to exert
pulling forces onto kinetochores during separation of MTs in mitosis
\cite{McIntosh2010}. 

The curling of hydrolyzed protofilaments into a ram's horn structure
shows that
  GDP-tubulin dimers have a bent conformation
  \cite{Mandelkow1991,Mueller-Reichert1998,Downing1998,Nogales2006}.
  Tubulin dimers assembled within the MT body are
  in a straight conformation, on the other hand \cite{Nogales1999}.
Hydrolysis of tubulin dimers embedded in a straight MT causes mechanical
strains in the tubular structure because the surrounding MT lattice
prevents these GDP-tubulin dimers from assuming their preferred bent
conformation.
This mechanical strain 
   is released in a catastrophe via the rupture of lateral bonds.

There are different models explaining
     how the mechanical strain is increased by hydrolysis or
  how lateral bonds are weakened by hydrolysis such that the strained MT
  becomes more prone for catastrophes.
  The first  cryo-electron microscopy (EM) studies showed
    blunt tips for growing MTs but curved tips for shrinking MTs
    \cite{Mandelkow1991}
    suggesting that GTP-protofilaments are straight while
    GDP-protofilaments are curved. 
    Later evidence from cryo-EM  showed 
    that GTP-protofilaments are also curved, but significantly
    less
    than GDP-protofilaments \cite{Mueller-Reichert1998}.
  The \emph{allosteric model} is based on the assumption that 
  hydrolysis of a tubulin dimer changes the  dimer
  conformation from a rather
  straight GTP-conformation to a bent GDP-conformation.
 Hydrolysis of tubulin dimers embedded in a straight MT causes mechanical
  strain in the tubular structure because the surrounding MT lattice
  prevents these GDP-tubulin dimers from assuming their preferred bent
  conformation.
  This model was employed in almost all previous MT simulation models
 that consider MT mechanics
  \cite{Molodtsov2005,VanBuren2005,Coombes2013,Mueller2014,Zakharov2015,Jain2015}
  The \emph{lattice model}, on the other hand, 
  is based on evidence from X-ray and cryo-EM structures
  \cite{Buey2006,Rice2008,Alushin2014,Manka2018}
   and simulations \cite{Ayoub2015,Fedorov2019} 
    that
   also  GTP-tubulin dimers assume a bent conformation and
   that
  hydrolysis rather affects the  lateral and longitudinal dimer
  interaction energies.
   It is supported by  recent experimental observations  that both
   growing and shrinking MTs have
   bent protofilament ends \cite{McIntosh2018}.
   Ref.\ \cite{McIntosh2018} also 
     presents  first simulation results
     with a lattice model.
     But there is also recent evidence from
         molecular dynamics (MD) simulation pointing in a different
        direction and  supporting  an intermediate  model, where hydrolysis
   affects interactions but also 
   lowers GDP-tubulin flexibility \cite{Igaev2018}.
  If hydrolysis weakens lateral interaction energies,
  hydrolysis makes the structure more prone for a catastrophe.
  While in the allosteric model, the mechanical strain in the structure
  is increased by  hydrolysis, in the  lattice model, the mechanical
  strain that the MT  structure can tolerate is reduced by hydrolysis.
  In both models, the result
  is an increased propensity for lateral bonds to rupture. Therefore,
  chemomechanical MT models with explicit bond rupture are
  a necessity to reproduce catastrophes.
  We build on existing modelling approaches based on the allosteric model
 \cite{Molodtsov2005,VanBuren2005,Coombes2013,Mueller2014,Zakharov2015,Jain2015}
 and include lateral bond rupture as explicit stochastic events with
 force-dependent rates, which
 can give important clues about how catastrophes
 are triggered in the MT structure.

The influence of tubulin dimer hydrolysis onto the mechanics of the MT lattice
suggests that, vice versa, mechanical forces and torques acting on tubulin
dimers via strains in the tubular structure could also affect hydrolysis rates,
an effect which has been explored only in Ref.\ \cite{Mueller2014} previously.
Although this interplay is  plausible from a mechanochemistry
point of view, experimental
  verification on the dimer level is extremely difficult and
  not possible yet, but we can employ chemomechanical
  MT models to explore and suggest possible implications for
  the dynamic instability.

  The coupling between chemical events -- namely polymerization events,
 dimer hydrolysis, bond rupture -- and
mechanical forces because of
conformational changes due to these
chemical events, is a characteristic of MTs and
requires chemomechanical MT models on the dimer level in order to develop a
microscopic understanding of their dynamic instability including catastrophe and
rescue events \cite{Zakharov2016}.
In this respect, chemomechanical models go beyond a phenomenological description
of MT dynamics in a four-parameter model based on growth and shrinking
velocities and phenomenological catastrophe and rescue
rates \cite{Dogterom1993}.
The challenge for microscopic chemomechanical models is to include all chemical
events as stochastic processes, to perform conformational relaxation governed by
MT mechanics following each chemical event, and, eventually, to also include the
feedback of mechanical forces within the MT onto reaction rates of the chemical
events.

We present a stochastic chemomechanical MT model on the dimer level. 
Our model includes
(i) a mechanical model of the MT containing lateral elastic bonds between
tubulin monomers in neighboring protofilaments and a harmonic bending energy
between tubulin monomers with a nonzero equilibrium angle after hydrolysis
(allosteric model),
(ii) stochastic addition and removal of tubulin dimers,
(iii) explicit stochastic lateral bond rupture and bond formation; the bond
rupture rate is coupled to the mechanical stress state of the bond and thus via
elastic interactions within the MT lattice also to the other bonds,
(iv) stochastic hydrolysis of dimers with a rate that can also couple to
the mechanical bending stress in the dimer. 
The stochastic kinetics (ii)-(iv) is handled by a Gillespie algorithm and after
each stochastic event, a  mechanical energy minimization mimicking the
relaxational dynamics of the structure is applied to the MT.

In order to parameterize our model, we will focus on the simplified scenarios of
a growing MT consisting of GTP-tubulin only and a shrinking MT consisting of
GDP-tubulin only. 
In both cases, we can neglect hydrolysis (iv);
in the growing GTP-MT, we can also neglect mechanics, which is generated
by hydrolysis.
In the presence of mechanics and hydrolysis, repeated catastrophe and rescue
events are obtained and will be described and analyzed.
One problem in chemomechanical MT models is the computational effort associated
with the mechanical relaxation.
We investigate in detail, which level of computational effort is necessary in
our model to obtain a sufficient mechanical relaxation following each chemical
event, on the one hand, and which simplifications can be taken to assure a
finite simulation time for growing MTs, on the other hand. 
This will allow us to
simulate arbitrarily long growing MTs at fixed computational speed.

Our chemomechanical model has to be compared to previous modelling approaches,
which include the mechanics of the MT
\cite{Molodtsov2005,VanBuren2005,Coombes2013,Mueller2014,Zakharov2015,Jain2015}:

\begin{itemize}
	\item	Refs.\ \cite{VanBuren2005,Coombes2013} employ the allosteric
              model for dimer bending
              and include stochastic addition and removal of dimers.
	     Hydrolysis is random. 
	      Mechanical energy minimization is  performed only
	      locally on randomly selected dimers.
              Lateral bond rupture is not implemented as explicit
                stochastic process but only included using a threshold
               energy  criterion.

	\item	The models in Refs.\ \cite{Molodtsov2005,Jain2015}
           focus on mechanics and do not include dimer addition and  removal.
           They are 
           also based on the allosteric model but  consider
           fixed hydrolysis states.
            In Ref.\ \cite{Molodtsov2005},
		the lateral bond energy landscape is harmonic around a minimum
		but includes an energy barrier and a dissociated, i.e., ruptured
		state.
		Global energy minimization gives the final state of the static
		structure.

	\item	In Ref.\ \cite{Zakharov2015}, the stochastic kinetics is added
		to a mechanical model similar to \cite{Molodtsov2005}.
		Here, the mechanical relaxation and lateral bond rupture is
		performed using Brownian dynamics (which include thermal
		fluctuations) with small time steps (equivalent to
		$2\times 10^7$ minimization steps), which is only applied to 300
		tubulin dimers at the plus end.
		Stochastic addition of dimers and removal by rupture of lateral
		and longitudinal bonds is included.
		The rupture of lateral bonds happens by 
		activation over the bond energy barrier, the longitudinal
		rupture by a threshold criterion.
		Hydrolysis is random and stochastic with a rate that is
		independent of mechanics.

	\item	Ref.\ \cite{Mueller2014} is also based on the allosteric model.
		Lateral bond rupture is possible using a threshold criterion.
                Mechanical energy minimization was performed globally. 
  		There is no addition
  		or removal of dimers, but hydrolysis is included.
		In a first attempt to include a coupling of the hydrolysis rate
		to mechanical forces, the hydrolysis kinetics remained
		deterministic, however, with the most probable hydrolysis event
		determined by mechanical forces. 
     In the present paper, we will add addition and removal of dimers 
		and a fully stochastic hydrolysis kinetics.
\end{itemize}

Our chemomechanical model has also to be compared to previous purely chemical
modelling approaches on the dimer level but without explicit mechanical model
\cite{VanBuren2002,Piette2009,Margolin2011,Margolin2012,Li2014}.
These models include attachment and detachment of tubulin dimers; some of these
models \cite{Margolin2011,Margolin2012,Li2014} also include lateral bond rupture
and are thus able to produce crack-like catastrophe events.
Crack-like catastrophe events are, however, triggered by adjusting chemical
rupture rates rather than including MT mechanics.
The model by Margolin \etal~\cite{Margolin2012} has successfully reproduced
features of the experimentally observed MT dynamic instability
\cite{Mahserejian2019} but relies on a heuristic tuning of simulation
parameters.

\section{Materials and Methods}

\subsection{Microtubule structure and energy}
\label{sec:3d_model}

Our MT model is formulated on the dimer level. 
The base units of the model are alpha- and beta-tubulin monomers.
In our model, we represent each monomer as cylinder with radius
$\Rtub = \SI{2}{\nano\meter}$ and height $\Ltub = \SI{4}{\nano\meter}$
(see \autoref{tab:geometric_parameters}).
Alpha- and beta-tubulin monomers form unbreakable tubulin dimers, which are
arranged head-to-tail into protofilaments.
13 protofilaments  form a 13$\_$3 MT, i.e., a MT with
a helical shift of 3 tubulin monomer lengths per turn.

\begin{table}[h!]
	\centering
	\caption{
		Geometric parameters of our MT model.
	}
	\label{tab:geometric_parameters}
	\begin{tabular}{@{}lll}
		\hline
		Parameter & Symbol & Value \\
		\hline
		mean MT radius & $\Rmt$ & \SI{10.5}{\nano\meter} \\
		\hline
		tubulin monomer radius & $\Rtub$ & \SI{2}{\nano\meter} \\
		\hline
		tubulin monomer length & $\Ltub$ & \SI{4}{\nano\meter} \\
		\hline
		helical shift between protofilaments & $\helicalShift$ & \SI{0.92}{\nano\meter} \\
		\hline
		rest length of lateral springs & $s_0$ & \SI{1.47}{\nano\meter} \\
		\hline
		straight equilibrium bending angle & $\restBendAngle$ & \SI{0}{\degree} \\
		\hline
		curved equilibrium bending angle & $\restBendAngle$ & \SI{11}{\degree} \\
		\hline
	\end{tabular}
\end{table}

For the remainder of this paper, we will use triples $(p,d,t)$ to address
specific tubulin monomers within the MT with
$p \in \{ 1, 2, \dots, 13 \}$ as protofilament
number,  $d \in \{ 1, 2, \dots, d(p) \}$ as  tubulin layer (with
$d = 1$ denoting the minus end and $d = d(p)$ denoting the plus end of
the protofilament $p$), and 
$t \in \{ 1, 2 \}$ denoting the tubulin monomer within the
dimer with $t = 1$ for the alpha-   and $t = 2$ for the beta-tubulin
monomers.
For simplicity, we assume periodicity in $p$ (i.e., $p = 0 \equiv 13$ and
$p = 14 \equiv 1$) and combined periodicity in $d$ and $t$ (i.e.,
$(p,d,3) \equiv (p,d+1,1)$ and $(p,d,0) \equiv (p,d-1,2)$).
We will also generally refer to the lateral neighbors of tubulin monomer
$(p,d,t)$ using $(p \pm 1,d,t)$ even though at the seam, lateral neighbors
differ in all three indices.

\begin{figure}[!ht] 
	\centering
	\includegraphics{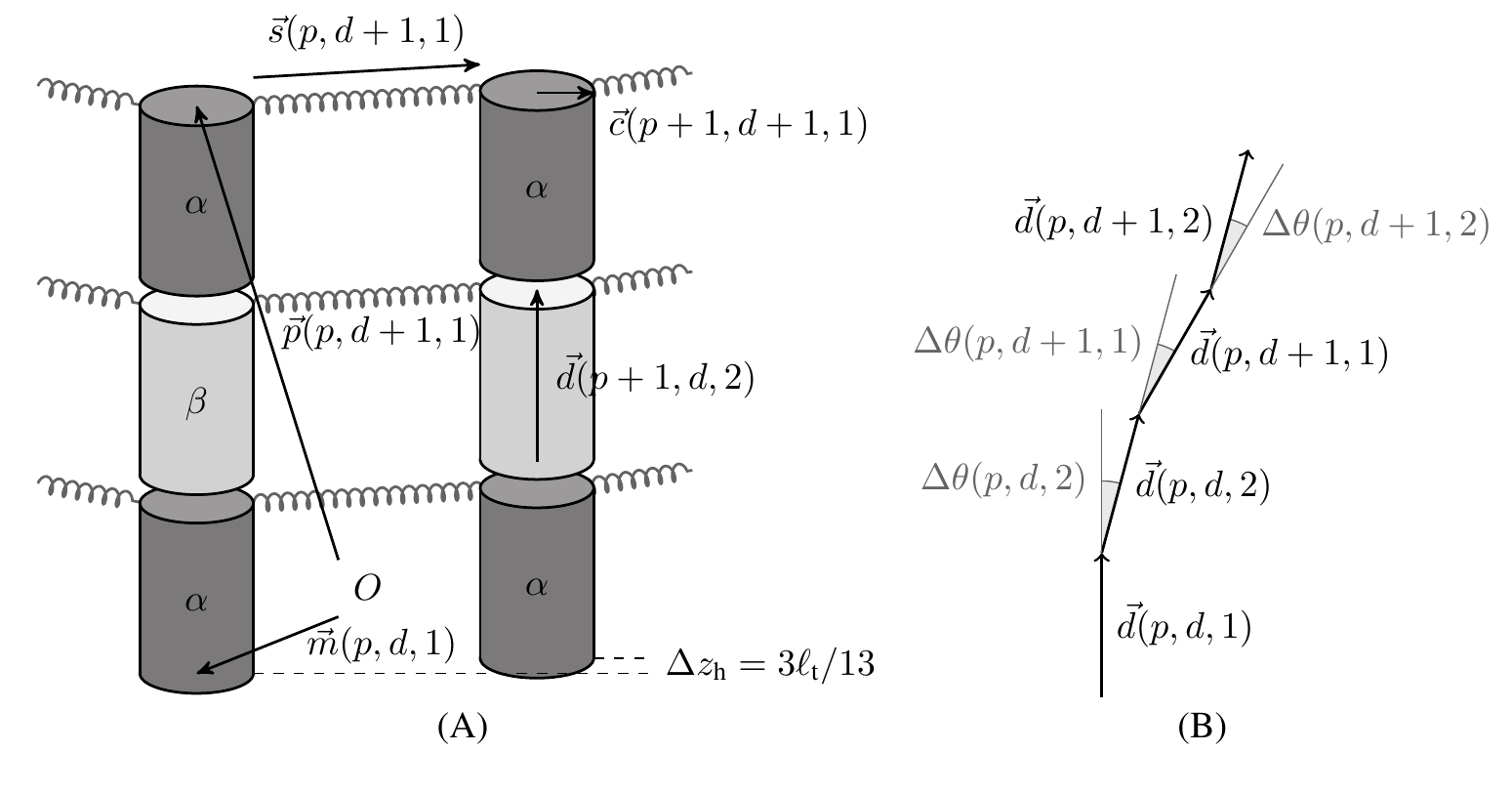}
	\caption{
		(A) Schematic illustration of the different vectors with the
		origin $O$. (The vertical gaps between tubulin cylinders are for
		illustration purposes only.) (B) Bending angles between the
		tubulin monomer direction vectors.
	}
	\label{fig:model_geometry}
\end{figure}

The MT is straight and oriented along the $z$-axis with the positive
$z$-direction pointing to the plus end.
Vectors $\vec{m}(p,d,t)$ and $\vec{p}(p,d,t)$
point to the  to the lower (minus end) and upper (plus end)
circular base of the  tubulin monomer $(p,d,t)$.
The direction vector
\begin{equation}
	\vec{d}(p,d,t)
	= \vec{p}(p,d,t) - \vec{m}(p,d,t)
	= \Ltub \begin{pmatrix}
		\cos \phi(p) \sin \theta(p,d,t) \\
		- \sin \phi(p) \sin \theta(p,d,t) \\
		\cos \theta(p,d,t)
	\end{pmatrix}
\end{equation}
with length $\Ltub= \SI{4}{\nano\meter}$ points from $\vec{m}(p,d,t)$
to $\vec{p}(p,d,t)$ and is
specified using spherical coordinates, i.e., azimuthal and
polar angles, see \autoref{fig:model_geometry}(A).
 The polar angle $\theta(p,d,t)$
is the only degree of freedom of each 
monomer, because we assume that monomers can only be displaced in radial
direction, i.e., all azimuthal angles are fixed to
$\phi(p) = 2 \pi (p - 1) / 13$.
As both alpha- and beta-tubulin have their polar angles as a degree of freedom,
the model supports intra- and inter-dimer curling \cite{Wang2005}. 

At the minus end of the MT each  protofilament $p$ starts with
an alpha-tubulin arranged in a circle with
mean MT radius $\Rmt = \SI{10.5}{\nano\meter}$ and with an
offset $z(p,1,1)=3 \Ltub(p - 1)  / 13$ in $z$-direction, such that
the seam is 
between the 13th and the 1st protofilament. 
The protofilament length that will be used to
calculate the growth and shrinkage
velocities is the maximum $z$-coordinate $\lmax(p)$ of all tubulin
monomers within the protofilament (see Supplementary Material
for more details).
The MT length is given by the average
\begin{equation}
	\Lmt
	= \frac{1}{13} \sum_{p = 1}^{13} \lmax(p) .
		\label{eq:average_microtubule_length}
\end{equation}

Every tubulin monomer has four interaction points: two in longitudinal direction
and two in lateral direction.
The longitudinal bond between alpha- and beta-tubulin monomers of the same dimer
is considered unbreakable but the orientation of this junction can change via
the beta-tubulin's polar angle $\theta(p,d,2)$.
In contrast, the longitudinal bond between adjacent tubulin monomers of
different dimers can break and is modeled via the bond energy $\Glong$ (where
the \enquote{0} refers to it being a standard energy \cite{VanBuren2004} and the
asterisk to the fact that it also includes the entropic cost of
\enquote{immobilization} \cite{VanBuren2002}).
The lateral interaction points are located at the edge of the upper base (see
\autoref{fig:model_geometry}(A)).
If there is a lateral bond between tubulin monomer $(p,d,t)$ and its neighbor in
the $(p+1)$-th protofilament, the bond is modeled as a harmonic spring with base
energy $\Glat$:
\begin{equation}
	\Elat(p,d,t)
	= \Glat + \frac{1}{2} \klat \left( | \vec{s}(p,d,t) | - s_0 \right)^2 
\end{equation}
with the spring constant $\klat$  of the bond and
the vector $\vec{s}(p,d,t)$
connecting the lateral interaction points;
 $s_0 \simeq \SI{1.47}{\nano\meter}$ is the
rest length of the spring (see \cite{Mueller2014} and also consider the helical
shift between two neighboring tubulin monomers of $3 \Ltub / 13$).
Lateral  bonds at the seam are assumed to have identical mechanical
properties as other lateral bonds based on evidence
that they do not constitute a  weaker bond \cite{Alushin2014,Harris2018}.
Additionally, there is a lateral repulsion term between neighboring tubulin
monomers (regardless of whether they are bonded or not) to ensure a cylindrical
form \cite{Mueller2014}:
\begin{equation}
	\Erep(p,d,t)
	= \krep \left( | \vec{p}(p,d,t) - \vec{p}(p+1,d,t) |
		- 2 \Rtub \right)^{-12}.
\end{equation}
The bending of monomer junctions is described by a harmonic
potential with bending constant $\kcurl$:
\begin{equation}
	\Ecurl(p,d,t)
	= \frac{1}{2} \kcurl \left( \bendAngle(p,d,t)
          - \restBendAngle(p,d,t) \right)^2 .
        \label{eq:Ecurl}
\end{equation}
The bending angle $\bendAngle(p,d,t) = \theta(p,d,t) - \theta(p,d,t-1)$
(see \autoref{fig:model_geometry}(B)) is calculated with
the neighboring monomer in
the minus direction (using the periodicity convention in $d$ and $t$,
$(p,d,0) \equiv (p,d-1,2)$), and $\restBendAngle(p,d,t)$ is its equilibrium
value.
For hydrolyzed beta-tubulin monomers and for alpha-tubulin monomers on top of a
hydrolyzed beta-tubulin (and for the first alpha-tubulin monomers of each
protofilament if the beta-tubulin in the same dimer is hydrolyzed), we use a
rest angle $\restBendAngle(p,d,t) = \SI{11}{\degree}$ in order to reproduce the
experimentally measured radius of curvature of \SI{21}{\nano\meter}
corresponding to an angle of \SI{22}{\degree} per dimer for a GDP-protofilament
curling into the ram's horn configuration
\cite{Mueller-Reichert1998,ElieCaille2007}.
Otherwise (for an unhydrolyzed beta-tubulin monomer or an alpha-tubulin monomer
on top of an unhydrolyzed beta-tubulin monomer), we assume a straight
equilibrium configuration with $\restBendAngle(p,d,t) = \SI{0}{\degree}$.
This choice of rest angles implements the allosteric model, where
GTP-hydrolysis leads to bending of tubulin dimers.

Our mechanical MT model is defined by the total energy
\begin{equation}
	\Emt
	= \sum_{p = 1}^{13} \sum_{d = 1}^{d(p)} \left( \Glong
		+ \sum_{t = 1}^2 \Bigl[ \Elat(p,d,t) + \Erep(p,d,t)
		+ \Ecurl(p,d,t) \Bigr] \right),
		\label{eq:total_microtubule_energy}
\end{equation}
where $\Elat(p,d,t)$ only contributes if there is a lateral bond between
tubulin monomers $(p,d,t)$ and $(p+1,d,t)$ and $\Erep(p,d,t)$ only contributes
if tubulin monomer $(p,d,t)$ has a lateral partner $(p+1,d,t)$.

There are four free parameters in our mechanical MT model (see
\autoref{tab:parameters}):
the longitudinal bond energy $\Glong$, the lateral bond energy $\Glat$, the
lateral spring constant $\klat$, and the bending constant $\kcurl$.
For the repulsion constant $\krep$, we use the same value
$\krep = \SI{e-6}{\square\radian \nano\meter\tothe{12}} \kappa$
that has been found previously to ensure the overall cylindrical shape of the MT
and only contributes a small portion to the MT energy \cite{Mueller2014}.

\begin{table}[h!]
	\centering
	\caption{
		Free parameters of our MT model and the \enquote{standard set}
		of their values that we will focus on in the rest of the paper.
	}
	\label{tab:parameters}
	\begin{tabular}{@{}lll}
		\hline
		Parameter & Symbol & standard set of values \\
		\hline
		longitudinal bond energy & $\Glong$ & \SI{-9.3}{\kBT} \\
		\hline
		lateral bond energy & $\Glat$ & \SI{-1.58}{\kBT} \\
		\hline
		lateral spring constant & $\klat$ &
			\SI{100}{\kBT \per \nano\meter\squared} \\
		\hline
		bending constant & $\kcurl$ &
			\SI{149}{\kBT \per \radian\squared} \\
		\hline
		pseudo-first-order polymerization rate & $\konc$ &
			\SI[per-mode=reciprocal]{4}{\per\micro\molar \per\second} \\
		\hline
		lateral bond formation attempt rate & $\katt$ &
			\SI[per-mode=reciprocal]{258}{\per\second} \\
		\hline
		constant hydrolysis rate & $\khydr$ &
			\SIrange[per-mode=reciprocal]{0.1}{0.5}{\per\second} \\
		\hline
		base hydrolysis rate & $\khydrN$ &
			\SIrange[per-mode=reciprocal]{1}{5}{\per\second} \\
		\hline
	\end{tabular}
\end{table}

In the simulation model, we do not use this mechanical energy to calculate
forces for a microscopic dynamics such as Brownian dynamics on the dimer level
(as opposed to \cite{Zakharov2015}).
We rather assume that mechanical relaxation dynamics is fast compared to
chemical changes in the MT due to tubulin attachment and detachment, bond
rupture and formation, or hydrolysis.
  The slowest mechanical process is relaxation of bending modes
  of protofilaments governed by small restoring bending moments.
  The basic  time scale for this process can be estimated as
  $\tau \sim \eta \Ltub^3/\kappa$ \cite{Kroy1997}, where $\eta \sim
  \SI{e-3}{\pascal\second}$ is the
  viscosity of water. This  gives
  $\tau \sim \SI{e-10}{\second}$, which is orders of magnitude smaller
  than typical time scales of seconds for chemical events. Therefore,
  even longer protofilaments  relax fast compared to chemical
  changes. 
  There is additional evidence from Brownian dynamics
  that bending mode relaxation is
  also  much faster than immobilization in cryo-EM \cite{Ulyanov2021}.
Therefore, we perform a  quasi-instantaneous energy minimization
of \eqref{eq:total_microtubule_energy} between these chemical simulation steps.
This is the computationally more efficient strategy to achieve mechanical
relaxation.
The rates of all chemical simulation events themselves determine the dynamics of
the MT and are handled by a Gillespie algorithm as explained in more detail
below.

\subsection{Chemical simulation events}
\label{sec:simulation_events}

\begin{figure}[!ht] 
	\centering
        \includegraphics{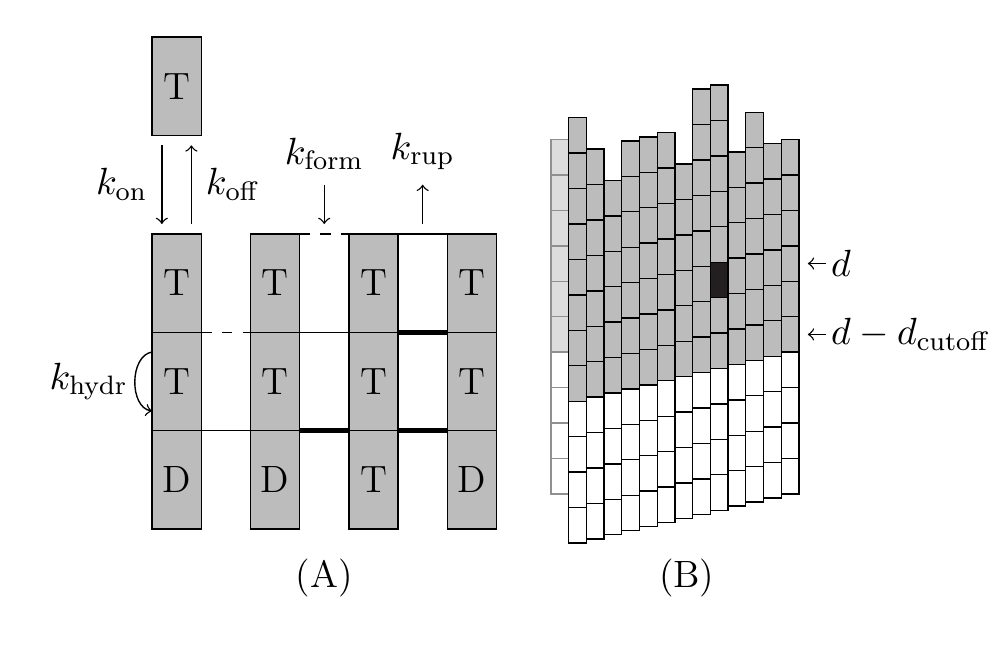}
	\caption{
		(A) Schematic illustration of the different simulation events
		with their rates.
		Dashed lateral bonds can be formed with rate $\kform$, thin
		solid lateral bonds can rupture with rate $\krup$, and thick
		bond cannot rupture.
		\enquote{T} and \enquote{D} correspond to the hydrolysis state
		of beta-tubulin of the dimers.
		(B) If the black tubulin dimer in layer $d$ is affected by an
		event and $\dcutoff = 2$ was used, all of the gray (and the
		black) tubulin dimers are used for energy minimization.
	}
	\label{fig:simulation_events}
\end{figure}

To simulate the dynamics of MTs, we include attachment of individual GTP-tubulin
dimers and detachment of (laterally unbonded) tubulin dimers or whole (laterally
unbonded) protofilament segments at
the plus end, as well as lateral bond rupture and formation, and hydrolysis of
tubulin dimers as stochastic chemical events into the simulation;
\autoref{fig:simulation_events}(A) summarizes the different possible
events and the associated rates.

\subsubsection{Attachment and detachment}

At the plus end of each protofilament, a GTP-tubulin dimer can attach with an
on-rate
\begin{equation}
	\kon
	= \konc \ctub
		\label{eq:kon}
\end{equation}
where $\konc$ is the pseudo-first-order polymerization rate and $\ctub$ is the
concentration of free GTP-tubulin dimers.
The on-rate is assumed to be
independent of the hydrolysis state of the protofilament end.

For depolymerization, we assume that a tubulin dimer at the plus end can only
detach if it has no lateral bonds.
We also allow for detachment of whole protofilament segments starting from an
interior dimer ($d < d(p)$) if the whole segment has no lateral bonds.
Laterally unbounded dimers or segments can detach with a rate
\begin{equation}
	\koff
	= \konc \cstd \exp \left( \Glong \right)
		\label{eq:depolymerization_rate}
\end{equation}
as given by Kramers theory with the longitudinal standard bond energy $\Glong$
(including the entropic cost of \enquote{immobilization}) and the standard
concentration $\cstd = \SI{1}{\molar}$ \cite{VanBuren2004}.

This approach differs from other models
\cite{VanBuren2002,VanBuren2005,Gardner2011}, where tubulin dimers can detach
regardless of whether they have lateral bonds or not.
In such models, if a tubulin dimer still has lateral bonds, its detachment rate
decreases exponentially.
In our model, we rather include lateral bond rupture and formation as separate
stochastic events into the simulation (similarly to the purely chemical models
in \cite{Margolin2011,Margolin2012,Li2014}); bond rupture can then be followed
by detachment of laterally unbounded dimers or protofilament segments.
Bond rupture enables dimer detachment and is necessary prior to a catastrophe;
vice versa, bond reformation is necessary for a rescue event.
Therefore, it is essential to also include the process of bond formation
into the model.
Moreover, it has been observed in MD simulations in \cite{Kononova2014}
that lateral tubulin bonds can easily reform.
The restriction that only laterally unbonded dimers can detach also causes an
indirect increase of the effective off-rate if the last dimers of a
protofilament are hydrolyzed because this tends to create stretched bonds, which
rupture more easily.

\subsubsection{Zipper-like lateral bond rupture and bond formation}

We assume that bond rupture between protofilaments starts from the plus end and
proceeds by a rupture front monomer by monomer towards the minus end;
likewise, bonds can be reformed only monomer by monomer towards the
plus end in a zipper-like fashion.
As a result, we always have a rupture front between two neighboring
protofilaments such that all monomers on top of the front toward the plus end
are ruptured and all monomers below toward the minus end are intact.
If tubulin monomer $(p,d,t-1)$ has a lateral bond with its neighbor in
protofilament $p+1$ but the tubulin monomer on top of it, $(p,d,t)$, has no
lateral bond with this neighbor, the rupture front can recede towards the plus
end, and tubulin monomer $(p,d,t)$ can form a bond with rate
\begin{equation}
	\kform
	= \katt
\end{equation}
with the attempt rate $\katt$.
Vice versa, if the bond at $(p,d,t)$ is intact and bond $(p,d,t+1)$ is broken,
the rupture front can advance towards the minus end by rupturing this bond with
a rate
\begin{equation}
	\krup
	= \katt \exp \left( \Glat + \Gmech \right)
		\label{eq:krup}
\end{equation}
which contains a chemical bond energy $\Glat$ and a mechanical energy
$\Gmech$, which accounts for the weakening of the lateral bond due to mechanical
strain in the bond and enters according to Bell theory
\cite{Bell1978,Evans1997}.
In our model, $\Gmech$ is due to the stretching of the Hookean springs
representing the lateral bonds so that $\Gmech = \Flat \lrup$, where
$\Flat = - \partial \Elat / \partial |\vec{s}(p,d,t)|$ is the force currently
acting on the lateral bond and $\lrup$ is the characteristic bond rupture
length.
We define $\lrup$  as the length increase of the lateral bond from
its rest length $s_0$ at which the stretching energy of the spring cancels the
bond energy:
\begin{equation}
	\lrup
	= \sqrt{\frac{- 2 \Glat}{\klat}} .
		\label{eq:lrup}
\end{equation}

\subsubsection{Hydrolysis without and with mechanical feedback}

Lastly, GTP in beta-tubulin monomers can hydrolyze into GDP via a random (or
scalar) hydrolysis rule meaning that almost every GTP-tubulin dimer in the MT
can hydrolyze  with a fixed rate $\khydr$
regardless of the hydrolysis state of its 
longitudinal neighbor (which would be a vectorial hydrolysis
rule).
The \enquote{almost} in the previous sentence refers to the finding that the
polymerization of the tubulin dimer $(p,d)$ and thus the formation of a
longitudinal bond between beta-tubulin $(p,d-1,2)$ and alpha-tubulin $(p,d,1)$
catalyzes the hydrolysis reaction in beta-tubulin $(p,d-1,2)$
\cite{Nogales1999}.
As a consequence, only GTP-tubulin dimers that ever had another tubulin dimer on
top of them can be hydrolyzed in our model.

We also consider the possibility
that hydrolysis is mechanochemically coupled to the bending
strain \cite{Mueller2014}.
Then, the hydrolysis rate is modulated
\begin{equation}
	\khydr(p,d)
	= \khydrN \exp \left( - \DEhydr(p,d) \right)
		\label{eq:mechanics_hydrolysis_rate}
\end{equation}
with a dimer-specific change $\DEhydr(p,d)$ in the energy barrier height
of the hydrolysis reaction, which
depends on the bending state of dimer $(p,d)$.
Because this bending state also depends via lateral bonds on the bending states
in all neighboring dimers, and because the bending state of all neighboring
dimers strongly depends on their hydrolysis state, the hydrolysis dynamics
becomes effectively non-random but depends on the hydrolysis state of the
neighbors.

\begin{figure}[ht!]
	\centering
	\includegraphics{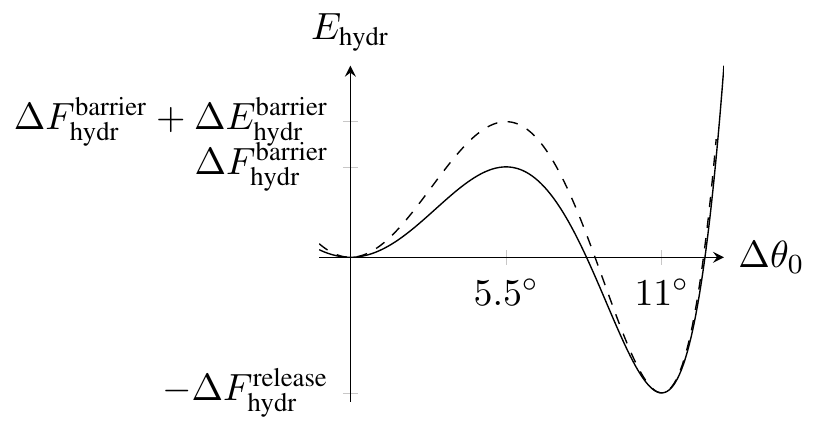}
	\caption{
		Schematic hydrolysis energy
		landscape with two local minima corresponding to the straight
		conformation ($\restBendAngle = \SI{0}{\degree}$) and the bent
		conformation ($\restBendAngle = \SI{11}{\degree}$) and an energy
		barrier $\FhydrBarrier$ at $\restBendAngle = \SI{5.5}{\degree}$
		between them.
		$\FhydrRelease$ is the energy released by hydrolysis.
		The dashed line represents the modified energy landscape due to
		the dimer-dependent contribution $\Ehydr(\restBendAngle)$.
	}
	\label{fig:mechanical_hydrolysis}
\end{figure}

The basis for our assumption of a tubulin dimer-specific mechanochemical
hydrolysis rate is to
view the equilibrium bending angle $\restBendAngle$ of a dimer
as the reaction coordinate
for hydrolysis which can be described by an energy profile
$\Fhydr(\restBendAngle)$.
$\Fhydr(\restBendAngle)$ has two local minima corresponding to the straight
conformation with $\restBendAngle = \SI{0}{\degree}$ and the curved
conformation with $\restBendAngle = \SI{11}{\degree}$ and a 
rate-limiting energy barrier of unknown height $\FhydrBarrier$ in between.
We propose that hydrolysis of a tubulin dimer is eased if its 
 actual bending angle $\bendAngle$  is closer to
the equilibrium angle $\restBendAngle= \SI{11}{\degree}$ in the hydrolyzed
state.
We model this dependency by adding a dimer-specific bending
energy contribution
$\Ehydr(\restBendAngle)$ to $\Fhydr(\restBendAngle)$,
which
changes the energy barrier height from $\FhydrBarrier$ to
$\FhydrBarrier + \DEhydr(p,d)$, see \autoref{fig:mechanical_hydrolysis}. 
$\FhydrBarrier$  can be
absorbed  into the constant rate $\khydrN$
so that only $\DEhydr(p,d)$ remains in the Arrhenius factor in
\eqref{eq:mechanics_hydrolysis_rate}.

To calculate the change in the energy barrier height $\DEhydr(p,d)$, we
now consider
the total MT energy in \eqref{eq:total_microtubule_energy}
as a function of the hydrolysis
reaction coordinate $\restBendAngle$ while keeping all polar angles
$\{ \theta(p,d,t) \}$ fixed.
We
simply assume that the energy barrier  is centered between the minima at
$\restBendAngleBarrier = \SI{5.5}{\degree}$ resulting in
\begin{equation}
	\DEhydr
	= \Emt(\restBendAngle = \SI{5.5}{\degree})
		- \Emt(\restBendAngle = \SI{0}{\degree}) .
		\label{eq:DEhydr_abstract}
\end{equation}
Because hydrolysis of tubulin dimer $(p,d)$ affects the rest bending angles of
beta-tubulin monomer $(p,d,2)$ and alpha-tubulin monomer $(p,d+1,1)$
and the rest bending angles only affect the bending energies \eqref{eq:Ecurl},
we finally obtain 
\begin{align}
	\DEhydr(p,d)
	&= \frac{1}{2} \kcurl \Bigl[ (\bendAngle(p,d,2) - \SI{5.5}{\degree})^2
     		- \bendAngle^2(p,d,2) + \nonumber\\
	&\qquad\qquad
   (\bendAngle(p,d+1,1) - \SI{5.5}{\degree})^2
		- \bendAngle^2(p,d+1,1) \Bigr]\nonumber\\
	&= \frac{1}{2} \kcurl \Bigl[ -(\bendAngle(p,d,2)+\bendAngle(p,d+1,1))
		\cdot \SI{11}{\degree} +2\cdot(\SI{5.5}{\degree})^2 \Bigr]
	\label{eq:mechanics_hydrolysis_rate2}
\end{align}
so that only a local bending energy change has to be calculated.
As a result,  tubulin monomers in the MT lattice with larger bending angles
$\bendAngle(p,d,t)$ tend to hydrolyze preferentially.
For the terminal tubulin dimer of a protofilament $(p,d(p))$,
the $d+1$-term in 
\eqref{eq:mechanics_hydrolysis_rate2} is missing because tubulin monomer
$(p,d(p)+1,1)$ does not exist.
This results in an overall smaller energy barrier and, thus, a higher
hydrolysis rate of the terminal tubulin dimer.

We also see  that the base hydrolysis rate $\khydrN$ in
\eqref{eq:mechanics_hydrolysis_rate} is not the hydrolysis rate for a perfectly
straight MT ($\bendAngle(p,d,t) = \SI{0}{\degree}$ for all tubulin monomers)
because  there is
still the constant contribution $\kcurl(\SI{5.5}{\degree})^2$ to the energy
barrier in \eqref{eq:mechanics_hydrolysis_rate2}
 that reduces the hydrolysis
rate.
As these terms are proportional to the bending constant $\kappa$, we cannot
simply absorb them into the constant factor $\khydrN$.

We note that for almost all GTP-tubulin dimers in the
GDP-body of the MT, we will typically find \emph{negative} bending angles; these
dimers bend inward in order to allow the longitudinal GDP-dimer neighbors to
further bend outwards.
For such negative bending angles the hydrolysis rate is reduced according to
\eqref{eq:mechanics_hydrolysis_rate2}.

In addition to the previous four free parameters
from the MT energy,
the simulation events add three additional
free parameters: the pseudo-first-order polymerization rate $\konc$, the attempt
rate $\katt$, and the hydrolysis rate $\khydr$ (or $\khydrN$).
In total, there are now seven free parameters, which are listed in
\autoref{tab:parameters}.

\subsection{Simulation and parameter determination}
\label{sec:simulation}

The actual MT simulation (implemented in C++) works as follows:
\begin{enumerate}
\item 	Initially, a MT with $\NGDP$ GDP-tubulin dimers
          followed by $\NGTP$ GTP
       tubulin dimers per protofilament is constructed with 
		$\theta(p,d,t) = \SI{0}{\degree}$ for all $(p,d,t)$.
	\item	Using the tubulin monomers' polar angles $\{ \theta(p,d,t) \}$,
		the MT's actual initial configuration is determined by
		minimizing its mechanical energy.
		Details on the minimization procedure will be discussed in the
		next section.
	\item	For all of the events described in the previous section, a list
		of possible events is determined and based on their rates $k_i$,
		a \enquote{tentative} event time $t_i$ is calculated using
		Gillespie's first reaction method \cite{Gillespie1976}:
		\begin{equation}
			t_i= \frac{1}{k_i} \ln \frac{1}{r}
		\end{equation}
		where $r$ is a uniformly distributed random number from $0$ to
		$1$.
		The event $i$ with the shortest event time $t_i$ is executed and
		the simulation time is increased by $t_i$.
	\item	Assuming fast mechanical relaxation the MT's energy is
		minimized after any event.
	\item	The simulation terminates if a protofilament is shorter than
          two tubulin dimers.
		\footnote{To calculate shrinkage velocities of shrinkage simulations via a
 		simple linear fit, it has proven to be easier to
		stop simulations if a protofilament still contains one tubulin
		dimer instead of zero tubulin dimers as the last tubulin dimer
		requires more time to depolymerize creating a \enquote{tail} in
		the length-versus-time plot.
		This time increase is due to lateral springs being stretched
		less because there is no additional tubulin dimer below the
		terminal tubulin dimer that would exert an additional bending
		moment.
		In practice, for determining parameters and when running full
		simulations, this first layer at the minus end is irrelevant and
		could be regarded as a \enquote{seed} on which the MT grows.
              }
		Otherwise we go back to the third step to determine
		the next event.
\end{enumerate}

There is a general agreement between different experiments
\cite{Mitchison1984,Walker1988,OBrien1990,Drechsel1992,Trinczek1993,Chretien1995,Pedigo2002}
that the MT growth velocity $\vgro$ increases linearly with the tubulin dimer
concentration $\ctub$ and that the shrinkage velocity $\vshr$ is independent of
$\ctub$.
We will use the results by Walker \etal~\cite{Walker1988}, which were measured
for $\ctub \in [ \SI{7.7}{\micro\molar}, \SI{15.5}{\micro\molar} ]$,
\begin{align}
	\vgro(\ctub)
	&= \SI{0.33 +- 0.01}{\micro\meter \per\minute \per\micro\molar} \ctub
		- \SI{1.59 +- 0.50}{\micro\meter \per\minute} ,
		\label{eq:walker_growth} \\
	\vshr
	&= \SI{-27 +-1}{\micro\meter \per\minute} ,
		\label{eq:walker_shrinkage}
\end{align}
and  lead to an individual critical concentration
$\ctubc \simeq \SI{5}{\micro\molar}$
(below which $\vgro < 0$).

To determine the values of the model parameters, we use a \enquote{divide and
conquer} approach \cite{VanBuren2002,VanBuren2005}.
First, we consider MT growth, where mechanics are assumed not to play a
significant role as protofilaments are not curling outward so that $\Gmech = 0$.
Thus, we use a GTP-only MT ($\NGDP = 0$) and set $\klat = 0$ and $\kcurl = 0$ so
that the  only free parameters left are $\konc$, $\Glong$,
$\Glat$, and $\katt$.
The goal of these simulations is to reproduce the measured growth velocity in
\eqref{eq:walker_growth} as function of the free tubulin dimer concentration
$\ctub$.
Secondly, we consider MT shrinkage, where mechanics are now assumed to play a
significant role, i.e., $\klat > 0$ and $\kcurl > 0$.
For a shrinking MT, we use $\NGTP = 0$, $\NGDP > 0$, and the parameter values
already determined by the growth simulations to reproduce the
shrinkage velocity in \eqref{eq:walker_shrinkage}.
In both cases, hydrolysis is ignored.
A schematic overview of the entire
parameter determination procedure can be found in
Figure S7 in the Supplementary Material. 

Comparing the number of free parameters and the amount of experimental data, we
can already predict that we will not be able to determine one set of fixed
parameter values but only restrict some parameter values to specific values if
other parameter values are set to (arbitrarily but reasonably)
chosen values.
We will discuss this issue in more detail in the conclusion.

\subsection{Energy minimization}
\label{sec:energy_minimization}

In previous three-dimensional models, different energy minimization approaches
have been used.
VanBuren \etal~\cite{VanBuren2005} used a local minimization approach in which
they randomly selected individual tubulin dimers and then only locally minimized
with respect to the parameters of this dimer.
On average, each tubulin dimer was visited three times for minimization.
Zakharov \etal~\cite{Zakharov2015} employed a completely different approach by
explicitly modelling the stochastic motion of tubulin monomers in space using
Brownian dynamics (applied to the first 300 tubulin dimers at the plus end).
They solve Langevin equations every \SI{2e-10}{\second} while using
\SI{e-3}{\second} as the time step for the events in their simulation resulting
in $\mathcal{O}(\num{e7})$ dynamics steps between actual events.
Using a parallel implementation run on a supercomputer, their simulation took
more than a day to simulate \SI{1}{\second} of MT dynamics.
There are drawbacks for both approaches: a local energy minimization scheme
might not come close enough to a mechanically relaxed configuration, whereas a
full Brownian dynamics simulation is computationally very costly.
In this paper, we employ a systematic mechanical energy
minimization between each stochastic chemical simulation event.
We try to achieve a better mechanical energy relaxation than
VanBuren \etal~\cite{VanBuren2005} with significantly less computational steps
than Zakharov \etal~\cite{Zakharov2015}.

In our simulation, we use the Broyden--Fletcher--Goldfarb--Shanno (BFGS)
algorithm, a quasi-Newton method, provided by the GNU Scientific Library (GSL)
\cite{GSL} to minimize the total mechanical MT energy in
\eqref{eq:total_microtubule_energy} as a function of the polar angles
$\{ \theta(p,d,t) \}$.
If each protofilament in the simulated MT contains $\NGDP + \NGTP$ tubulin
dimers, there are a total of $26 (\NGDP + \NGTP)$ polar angles and thus the same
number of minimization parameters.
In realistic simulations,
MTs can stay in the growing phase for a very long time
resulting in an unbounded increase in the number of minimization parameters
drastically slowing down the simulation.
In essence, the average time for one minimization step increases with the MT
length in this scenario making long-running simulations impossible.

To overcome this limitation, we will explore two possibilities to avoid having
a MT length-dependent number of minimization parameters:
\begin{enumerate}
	\item	restricting the number of minimization steps per energy
		minimization to a small value but still
		considering all minimization parameters (this approach is
		similar to the strategy in \cite{VanBuren2005}),
	\item	restricting the number of minimization parameters by only
		considering the tip of the MT but not restricting the number of
		minimization steps.
\end{enumerate}
While the first strategy is easy to understand and implement, the second
 needs further specifications in terms of how we define the tip of
the MT here.
If a certain event is executed that affects tubulin dimer $(p,d)$,
we include all layers starting from
$\min(0, d - \dcutoff)$  into mechanical energy minimization
because  mechanical interactions within the MT have a certain range,
where  $\dcutoff$ is a  cutoff \emph{layer} distance.

Below, we will compare these approaches of restricted minimization with respect
to accuracy and speed 
and find that
  we obtain accurate energy minimization at a high simulation
  speed by using the second approach and restricting the number of
  minimization parameters with $\dcutoff=10$.
  We can compare with the approaches of
  Zakharov \etal~\cite{Zakharov2015} and 
  VanBuren \etal~\cite{VanBuren2005} in terms of the average
  number of minimization steps between chemical events.

  Zakharov \etal~\cite{Zakharov2015} use $\mathcal{O}(\num{e7})$
  Brownian dynamics
steps between events and restrict 
the number of simulation parameters to 300 tubulin dimers at the plus end.
With  $\dcutoff=10$
we minimize on average with respect to a comparable number of
150 tubulin dimers at the plus end.
To compare the efficiency, we consider a single
quasi-Newton minimization step in our simulation to be equivalent to one time
step of their Brownian dynamics
(if we ignore the random thermal fluctuations in their Langevin equations,
they are basically using a gradient descent method).
We compare the event time $t_i$  divided by the number of
minimization steps after the execution of that event
to their Brownian dynamics time step of \SI{2e-10}{\second}.
For shrinking MTs, one minimization step takes
$\mathcal{O}(\SI{e-5}{\second})$ after polymerization events,
$\mathcal{O}(\SI{e-4}{\second})$ after depolymerization events, and
$\mathcal{O}(\SI{e-7}{\second})$ after lateral bond
events;
all of these time steps are orders of magnitude larger
  than \SI{2e-10}{\second} and, thus, the simulation proceeds orders of
  magnitude faster, while we still achieve an accurate energy minimization. 
As a comparison with the \SI{1}{\second} of MT dynamics simulated in
more than a day in a parallel computation in Ref.\ \cite{Zakharov2015}, we
generally do not require more than a few hours for \SI{1}{\minute} of MT
dynamics (for a constant hydrolysis rate) using just a single CPU core.

  VanBuren \etal~\cite{VanBuren2005} apply a local minimization procedure
  and restrict minimization to, on average,
  three minimizations 
  with respect to the parameters of each dimer. 
  Because  one step of their algorithm minimizes with respect
to the parameters of a single tubulin dimer, a comparison to
our quasi-Newton minimization steps which minimize the MT energy
with respect to the parameters of, on average, 
$\mathcal{O}(150)$ tubulin dimers
is not straightforward. 
In addition, VanBuren \etal's model also contains longitudinal springs so that
outward bending of single tubulin dimers as a consequence of local minimization
can be compensated by stretching the next longitudinal spring.
As our model does not contain such longitudinal springs, bending one tubulin
dimer causes the whole protofilament part above the tubulin dimer to also bend
outwards creating an effectively non-local, far-reaching interaction.
Consequently, we are not able to also implement a local minimization procedure
for comparison.
To make a qualitative comparison between the two approaches,
we assume that one minimization step  of our BFGS algorithm, which
acts on average on $300$ parameters, i.e., 
$150$ tubulin dimers,
corresponds to $100$ single tubulin dimer minimizations
 in the model of Ref.\ \cite{VanBuren2005} as they consider
three parameters per tubulin dimer.
Between chemical events, we perform on average $150$
BFGS minimization steps, which corresponds to $1.5 \times 10^4$
single tubulin dimer minimizations  in Ref.\ \cite{VanBuren2005}.
Therefore, we apply the equivalent of $15000/150 =100$
single tubulin minimizations to each of the 150 tubulin dimers
close to the plus tip on average
as compared to three single tubulin dimer minimizations
 in the simulation model of Ref.\ \cite{VanBuren2005}.
Accordingly, we should  achieve a more accurate mechanical energy relaxation.

  We also
  compared our chosen minimization method, the BFGS algorithm, against the
other multidimensional minimization algorithms using derivatives provided by GSL
\cite{GSL}, including the conjugate gradient method,
and found the BFGS
algorithm to perform better.
In particular, to fully minimize the initial configuration of a MT with
$\NGDP = 20$ and $\NGTP = 0$, BFGS only required about a third of the time
compared to the next best algorithm, a conjugate gradient method.

\section{Results}

\subsection{GTP-microtubule growth and model parameterization}
\label{sec:MTgrowth}

MT growth mainly depends on the four parameters $\konc$, $\Glong$, $\Glat$ and
$\katt$, because the growing MT tip mainly consists of straight GTP-tubulin
dimers.
Therefore, we consider growth of a GTP-only MT ($\NGDP = 0$) in the absence of
hydrolysis and set $\klat = 0$ and $\kcurl = 0$ so that the only free parameters
left are $\konc$, $\Glong$, $\Glat$, and $\katt$.
For $\koncVal{2}$ and $\koncVal{4}$, we scanned the parameter space
$(\Glong, \Glat, \katt)$  in steps  $\Delta \Glong = \SI{0.2}{\kBT}$.
to find parameter values that reproduce the
experimental growth velocity data of Walker \etal~in \eqref{eq:walker_growth}.
The growth velocity $\vgro$
for each simulation was determined by fitting $\Lmt(\tsim)$
with a linear function.
Experiments on MT growth show a linear dependence
$\vgro(\ctub) = \agro \ctub + \bgro $ characterized by two parameters $\agro$
and $\bgro$ from \eqref{eq:walker_growth}.
If simulations reproduce a linear dependence of $\vgro$ as a function of
$\ctub$, we can determine two of the three model parameters
$(\Glong, \Glat, \katt)$ by fitting to the experimental data
\eqref{eq:walker_growth} for  $\agro$ and $\bgro$, i.e.,  two
experimental constraints fix two
model parameters as a function of the third parameter.
This will allow us to parameterize a one-dimensional sub-manifold (a line)
within the three-dimensional parameter space $(\Glong, \Glat, \katt)$ where our
model agrees with experimental growth data.
This procedure is conceptually analogous to the approach of VanBuren \etal
~\cite{VanBuren2002}, but we work in a higher-dimensional (three-dimensional)
space of model parameters.

\begin{figure}[ht!]
	\centering
        \includegraphics[width=0.99\linewidth]{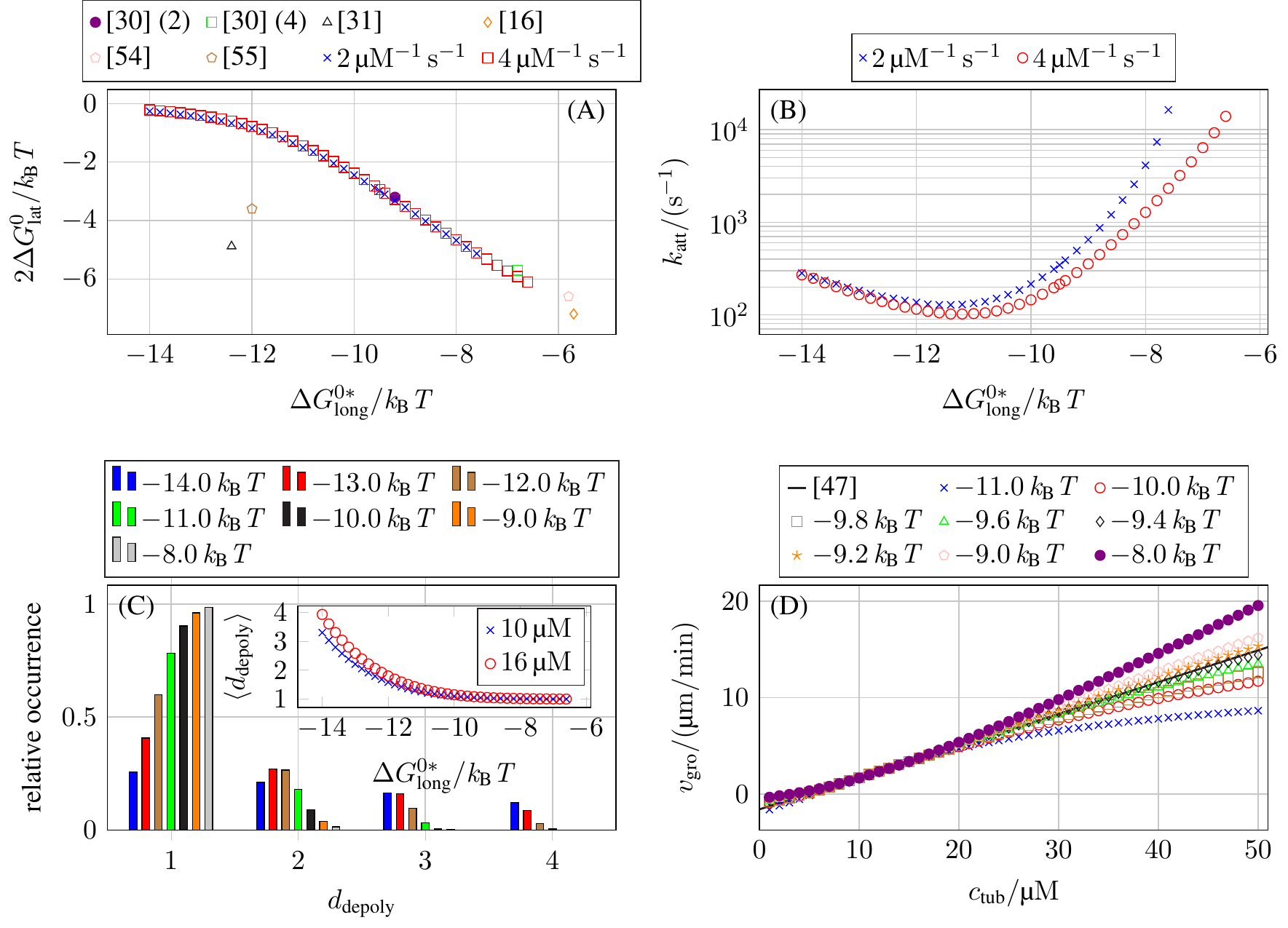}
	\caption{
		(A) Lateral bond energy $\Glat$ as a function of the
		longitudinal bond energy $\Glong$ from matching the
		concentration-dependent growth velocity data from
                Walker \etal~\cite{Walker1988}, see \eqref{eq:walker_growth}.
		To compare our lateral bond energies (per tubulin monomer) to
		other publications (lateral bond energy per tubulin
		\emph{dimer}), the $y$-axis shows $2 \Glat$.
		(The numbers behind Ref.\ \cite{VanBuren2002} refer
		to their value of $\konc$.)
		(B) Lateral bond attempt rate $\katt$ as a function of the
		longitudinal bond energy $\Glong$ for our two values of $\konc$
		from matching the concentration-dependent growth velocity data
		from Walker \etal~\cite{Walker1988},
                see \eqref{eq:walker_growth}.
		(C) Relative occurrence of
		different $\dDepoly$ values for MT growth with
		$\koncVal{4}$ and $\ctub = \SI{10}{\micro\molar}$.
		The inset shows the average $\dDepoly$ as a function of $\Glong$
		for $\koncVal{4}$ and $\ctub = \SI{10}{\micro\molar}$ and also
		for $\ctub = \SI{16}{\micro\molar}$.
		(D) MT growth velocity $\vgro$ as a function of a larger
		interval of free tubulin dimer concentration values $\ctub$ for
		$\koncVal{4}$ and different longitudinal bond energies
                $\Glong$.                
                 We also plot \eqref{eq:walker_growth} from 
                  the growth velocity data from Walker
                  \etal~\cite{Walker1988} over the larger concentration
                    interval.
	}
	\label{fig:growth_plots}
\end{figure}

As a result, we obtain a line in the three-dimensional parameter space,
which we parameterize by $\Glong$, i.e., for a given value of $\Glong$,
a value of $\Glat$
(see \autoref{fig:growth_plots}(A)) and a value of $\katt$ (see
\autoref{fig:growth_plots}(B)) is determined by the experimental growth data.

Afterwards, we will fix a particular value of $\Glong$ by the additional
requirement that the simulation should exhibit an as linear as possible
concentration dependence of the growth velocity $\vgro$ over a certain range of
tubulin concentrations $\ctub$ (see \autoref{fig:growth_plots}(D)) such that we
arrive at parameter sets $(\Glong, \Glat, \katt)$ for $\koncVal{2}$ and
$\koncVal{4}$, see \autoref{tab:most_linear_parameter_values}.

\begin{table}[ht!]
	\caption{
		Growth parameter values that generate the most linear
		dependence $v_\mathrm{gro}(c_\mathrm{tub})$.
	}
	\label{tab:most_linear_parameter_values}
	\begin{tabular}{@{}lll}
		\hline
		$\konc$ (\si[per-mode=reciprocal]{\per\micro\molar \per\second}) & $2$ & $4$ \\
		\hline
		$\Glong$ (\si{\kBT}) & $-9.7$ & $-9.3$ \\
		\hline
		$\Glat$ (\si{\kBT}) & $-1.38$ & $-1.58$ \\
		\hline
		$\katt$ (\si[per-mode=reciprocal]{\per\second}) & $281$ & $258$ \\
		\hline
	\end{tabular}
\end{table}

The results in \autoref{fig:growth_plots}(A) show that the values
of $\Glong$ and
$\Glat$ depend only weakly on our chosen $\konc$ values.
\autoref{fig:growth_plots}(A) also shows that our data matches
results obtained in
\cite{VanBuren2002} (this data was later re-used in
\cite{VanBuren2005,Coombes2013,Ayaz2014}) but also differs from other results
\cite{Piette2009,Mickolajczyk2019}, which were all obtained by the
same approach of fitting 
growth velocity data from Walker \etal~\cite{Walker1988} (or their own growth
data in \cite{Mickolajczyk2019}).
Kononova \etal~\cite{Kononova2014} obtained bond energies from
MD simulations of nano-indentation experiments;
the values from Kononova \etal~\cite{Kononova2014} are much larger for both
types of bonds ($\Glong \sim 2\Glat \sim 25 k_BT$) and, thus, not shown in
\autoref{fig:growth_plots}(A).

Qualitatively, the measured dependencies of $\Glat$ and $\katt$ on $\Glong$ can
be understood as follows:
the weaker longitudinal bonds are, the more likely it is that a tubulin dimer
will depolymerize.
To get the same growth velocity, this decrease in
\enquote{longitudinal stability} has to be compensated by an increase in
\enquote{lateral stability} by stronger lateral bonds (making it less likely
that lateral bonds break and, thus, enabling depolymerization)
or faster formation
of lateral bonds (to stabilize newly polymerized tubulin dimers).
\autoref{fig:growth_plots}(C) shows the number of tubulin dimers $\dDepoly$ that
detach at once during depolymerization events.
For increasingly stronger longitudinal bonds and, thus,
weaker lateral bonds, multi-dimer
depolymerization becomes more relevant.
The data in the inset in \autoref{fig:growth_plots}(C)  is also compatible
with
results in Ref.\ \cite{Margolin2012} obtained with a purely chemical model.

Until now, we only considered free tubulin dimer concentrations
$\ctub \in [ \SI{7}{\micro\molar}, \SI{16}{\micro\molar} ]$ to use similar
values as Walker \etal~\cite{Walker1988}, but there have also been
other measurements with a larger range of $\ctub$ values
\cite{Mitchison1984,OBrien1990,Chretien1995,Pedigo2002}.
In general, it is assumed that the growth velocity $\vgro$ increases linearly
with $\ctub$ for the whole MT just as the polymerization rate in \eqref{eq:kon}
increases linearly with $\ctub$ for individual protofilaments.
  Theoretically, it has been shown that,  for multistranded polymers,
lateral interactions
give rise to a non-linear relation between growth velocity on
monomer concentration \cite{Stukalin2004}. For MT growth, a non-linear
dependence on tubulin concentration was  found in Ref.\ \cite{Piette2009}
using a two-dimensional model based on Ref.\ \cite{VanBuren2002}.
Over a larger range of $\ctub$ values, our simulations also exhibit a 
non-linear relation between $\vgro$ and $\ctub$ depending
on the value of  $\Glong$, as shown in
\autoref{fig:growth_plots}(D).
  Data for different values of $\Glong$ (and correspondingly adjusted values of
$\Glat$ and $\katt$, see \autoref{fig:growth_plots}(A) and (B))
and the same value of $\konc$,
that was previously overlapping in the interval $\ctub \in [
\SI{7}{\micro\molar}, \SI{16}{\micro\molar} ]$ start to differentiate in
a larger concentration interval. 
While possible non-linear relations have been predicted  theoretically,
the available experimental data show a linear
  $\vgro(\ctub)$ dependence over a large range of $\ctub$ values
  \cite{Mitchison1984,OBrien1990,Chretien1995,Pedigo2002}.
Therefore, we determined the remaining free parameter value of $\Glong$ for the
two $\konc$ values from the condition that the concentration dependence of
$\vgro$ is as linear as possible up to \SI{50}{\micro\molar}.
To determine these values of $\Glong$, we ignored concentrations $\ctub$ below
the individual critical concentration (for which $\vgro < 0$) which violate our
fundamental assumption of a growing MT.

In summary, we find a triple $(\Glong, \Glat, \katt)$
that fits the growth
velocity data from Walker \etal~\cite{Walker1988} and that gives a linear
concentration dependence over a wide tubulin concentration range for two
representative values of $\konc$.
\autoref{tab:most_linear_parameter_values} lists these parameter triples for
$\koncVal{2}$ and $\koncVal{4}$.
For a given $\konc$, these results fix four of the seven model parameters in
\autoref{tab:parameters} using experimental data on MT growth.
To address the parameters $\kcurl$ and $\klat$, we now turn to MT shrinkage.

\subsection{GDP-microtubule shrinkage and model parameterization}

As opposed to MT growth, MT shrinkage
also depends on the bending constant $\kcurl$ and spring constant $\klat$ as
protofilament curling and bond rupture become relevant processes for a
shrinking MT.
We consider a shrinking MT that initially only consists of GDP-tubulin dimers
($\NGTP = 0$, $\NGDP > 0$) with parameter values $\konc$, $\Glong$, $\Glat$,
and $\katt$ as already determined by the growth simulations and in the absence
of hydrolysis
(a shrinking, initially GDP-only MT acquires some GTP-dimers
by attachment but remains GDP-dominated).
To investigate shrinkage, MTs with $\NGDP = 20$ and $\NGTP = 0$ were used.
For each parameter set, 20 simulations were run to get an average shrinkage
velocity $\vshr$.
Experimental data on shrinking MTs show a shrinkage speed $\vshr$ that is
independent of the tubulin dimer concentration.
For each value of $\konc$, we should be able to determine one of the two
parameters $(\kcurl, \klat)$ as a function of the other parameter by fitting
such that the experimental value of the shrinkage velocity is reproduced in
simulations (for parameters $\Glong$, $\Glat$, and $\katt$ fixed by the growth
velocity data).
We use the experimental shrinkage velocity of Walker \etal, see
\eqref{eq:walker_shrinkage}, for this fitting procedure.

\begin{figure}[ht!]
	\centering
        \includegraphics[width=0.99\linewidth]{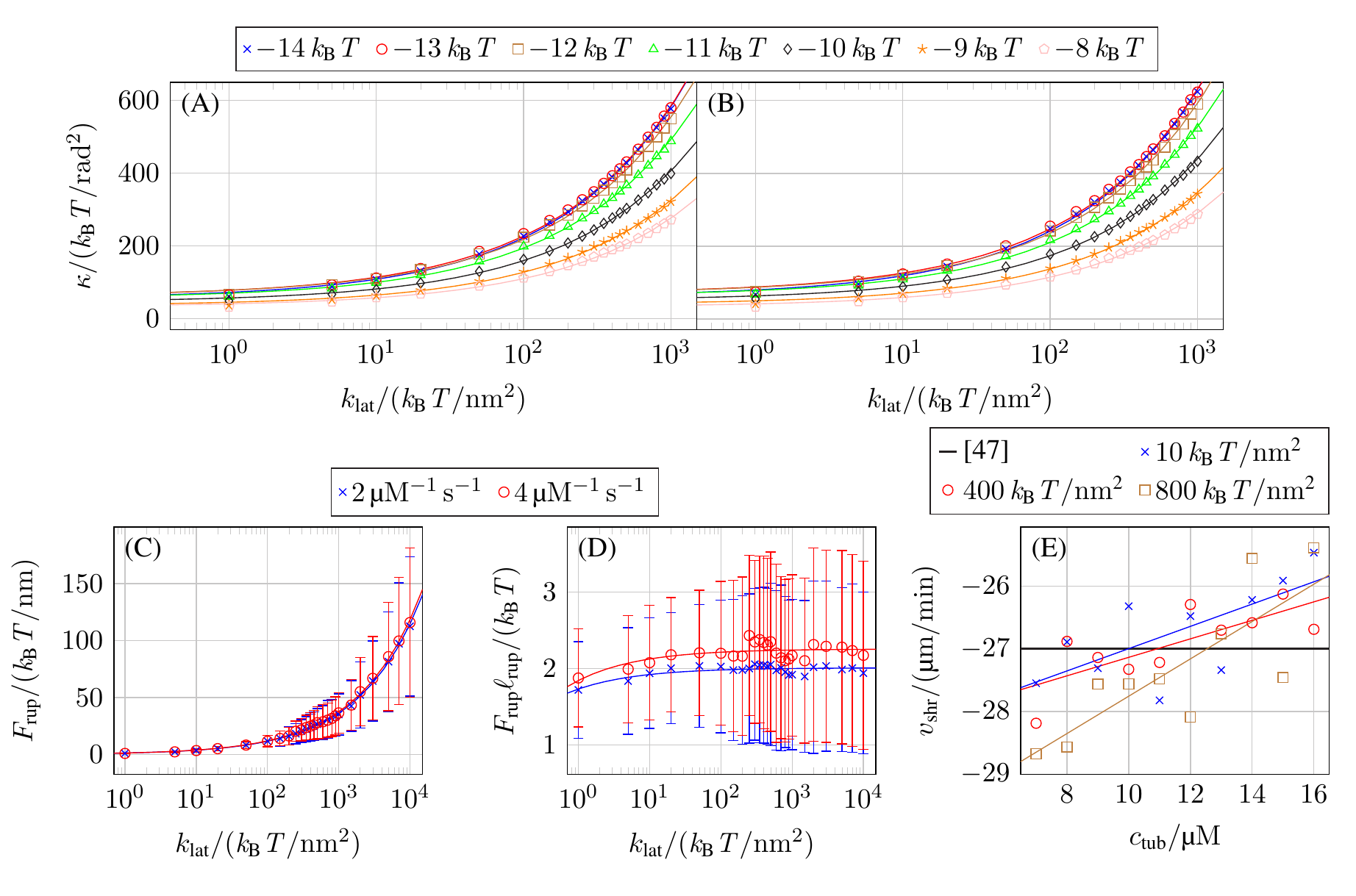}
	\caption{
		Mechanical parameter values reproducing the experimentally
		measured shrinkage velocity in \eqref{eq:walker_shrinkage} for
		(A) $\koncVal{2}$ and (B) $\koncVal{4}$ and different values of
		$\Glong$.
		(C) Force on lateral bonds at rupture $\Frup$ as a function of
		$\klat$ for $\koncVal{2}$ with $\GlongVal{-9.5}$ and
		$\koncVal{4}$ with $\GlongVal{-9.0}$, both at
		$\ctub = \SI{10}{\micro\molar}$.
		(D) Rupture energy $\Frup \lrup$ of lateral bonds as a function
		of $\klat$ for the same parameters as in (C).
		(E) Shrinkage velocity $\vshr$ as a function of the free tubulin
		dimer concentration $\ctub$ for $\koncVal{4}$,
		$\GlongVal{-9.3}$, and different values of $\klat$ and linear
		fits $\vshr(\ctub)$.
	}
	\label{fig:shrinkage_parameters}
\end{figure}

\autoref{fig:shrinkage_parameters}(A) and (B) show the values of $\klat$ and
$\kcurl$ for $\koncVal{2}$ and $\koncVal{4}$ and different values of $\Glong$
that reproduce the experimentally measured shrinkage velocity in
\eqref{eq:walker_shrinkage}.
All data points for each $\Glong$ fall on square root functions
\begin{equation}
	\kcurl(\klat)
	= a_\text{shr} \sqrt{\klat} + b_\text{shr} .
\end{equation}
This functional dependence can be understood qualitatively by considering the
mechanical contribution to the bond rupture rate \eqref{eq:krup},
$\exp(\Flat \lrup)$, which, on average, should have the same value for all
mechanical parameter combinations to produce the same shrinkage velocity.
As the characteristic bond rupture length
in \eqref{eq:lrup} depends on $\klat$ as
$\lrup \sim \sqrt{\klat}^{-1}$, the average lateral bond force at rupture
should depend on
$\klat$ like $\Frup \sim \klat \lrup \sim \sqrt{\klat}$.
The lateral bond force $\Flat$ is a consequence of the lateral bonds stretching
as the tubulin monomers curl outward to decrease the bending force
$\Fcurl = \kcurl \left( \bendAngle(p,d,t) - \restBendAngle(p,d,t) \right)$,
which leads to $\Frup \sim \Fcurl \sim \kcurl$ resulting in
$\kcurl \sim \sqrt{\klat}$ in accordance with
\autoref{fig:shrinkage_parameters}(A) and (B).

\autoref{fig:shrinkage_parameters}(C)  confirms that
the average force on lateral bonds
at rupture $\avgFrup$ has the functional dependence
$\avgFrup \sim \sqrt{\klat}$ predicted by our above qualitative argument
($\avgFrup$ and error bars
$\sigma_{\Frup}$ were determined by fitting normal distributions
to the histogram of the lateral bond rupture forces collected for 20 shrinkage
simulations per parameter set with $\NGDP = 20$).
Also, 
the resulting mechanical contribution $\Frup \lrup$ for the exponential function
of the lateral bond rupture rate in \autoref{fig:shrinkage_parameters}(D)
is approximately constant as  expected from our above argument.

As the experimentally measured shrinkage velocity $\vshr$ does not depend on the
free tubulin dimer concentration $\ctub$, we used constants to fit our
$\vshr(\ctub)$ data.
In reality, however, our data shows a linear dependence between $\vshr$ and
$\ctub$ as shown in \autoref{fig:shrinkage_parameters}(E)
 corresponding to a slowing down of depolymerization. 
This is caused by an increased probability for
intermediate addition of tubulin dimers and lateral bond formation between
them; these lateral bonds require additional time to rupture. 
While this dependency of $\vshr$ on $\ctub$ will have a small influence on the
concrete value of the shrinkage velocity, we expect it to not have any
qualitative effect on the overall MT dynamics. At higher
  tubulin concentrations, where the decrease of
  $|\vshr(\ctub)|$ would become significant, the catastrophe rates
  decrease dramatically so that shrinking will rarely occur.

Comparing our results from figures \ref{fig:shrinkage_parameters}(A) and (B) to
other results is not always directly possible due to different modelling
approaches but most find that
$\klat \ll \SI{1000}{\kBT \per \nano\meter\squared}$ and
$\kcurl \ll \SI{1000}{\kBT \per \radian\squared}$
\cite{VanBuren2005,Sim2013,Driver2017}, with some exceptions
\cite{Deriu2007,Kononova2014}.
Previously, we used MD simulation data from
Grafm{\"u}ller \etal~\cite{Grafmueller2011}
to calculate the bending constant $\kappa$ \cite{Mueller2014}. Compared to
Ref.\ \cite{Mueller2014}, we have to adjust the calculation to consider both
inter-dimer and intra-dimer bending resulting in
$\kappa \simeq \SI{50}{\kBT \per \radian\squared}$.
MD simulation in \cite{Kononova2014}, on the other hand, give a persistence
length of individual protofilaments of
$\persistenceL \simeq \SI{6}{\micro\meter}$, which corresponds to a
significantly larger value of
$\kcurl \simeq \SI{1500}{\kBT \per \radian\squared}$ for the bending constant.
This discrepancy cannot be resolved at present.
We use $\kcurl = \SI{149}{\kBT \per \radian\squared}$ in the following together
with the corresponding value of
$\klat = \SI{100}{\kBT \per \nano\meter\squared}$
according to \autoref{fig:shrinkage_parameters}(B) which are values close
to the ones used by \cite{VanBuren2005}.

\subsection{Restricted energy minimization for efficient simulation}
\label{sec:restricted_energy_minimization}

Until now, energy minimization was not restricted by either a maximum number of
minimization steps or by only considering a subset of tubulin dimers at
the MT tip so that we will consider this unrestricted minimization as the
\enquote{gold standard} to which we will compare the two restricted energy
minimization approaches described in Section \ref{sec:energy_minimization}.
We use the shrinkage velocity $\vshr$ as the observable by which we judge the
relevant cutoff values in the two approaches.

\begin{figure}[ht!]
	\centering
        \includegraphics[width=0.99\linewidth]{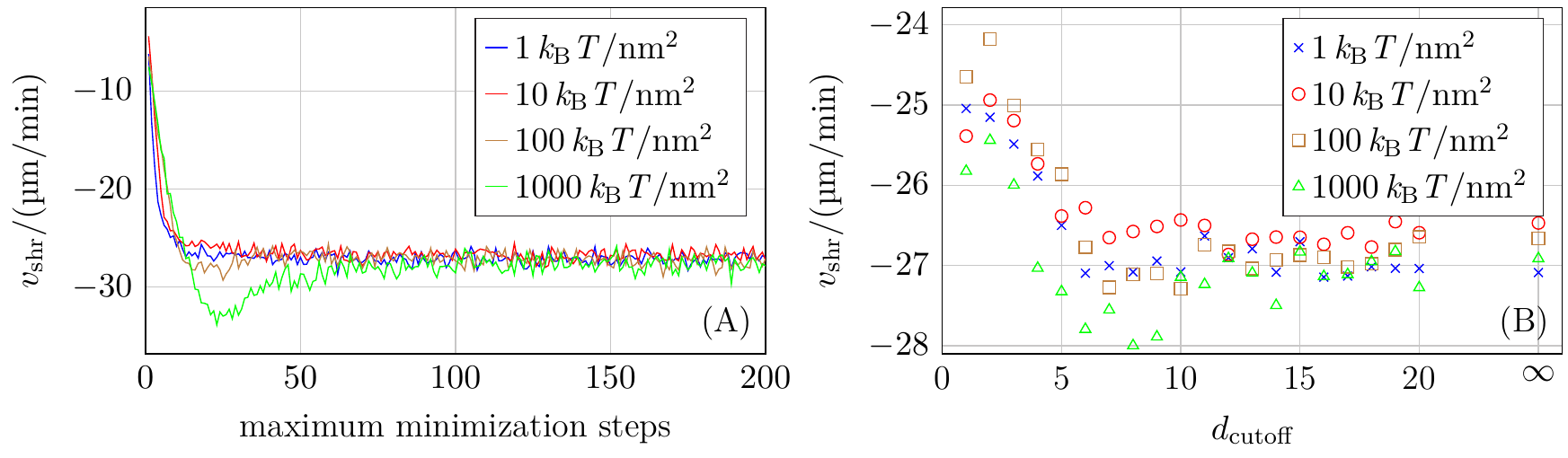}
        \caption{
		(A) Shrinkage velocity $\vshr$ as a function of the maximum
		number of minimization steps.
		(B) Shrinkage velocity $\vshr$ as a function of the layer cutoff
		distance $\dcutoff$ (where $\dcutoff = \infty$ means that no
		cutoff was used).
		20 simulations for each parameter set were run for both plots
		and both used $\koncVal{4}$, $\GlongVal{-9.3}$,
		$\ctub = \SI{10}{\micro\molar}$, $\NGDP = 20$, and different
		values of $\klat$.
	}
	\label{fig:restricted_energy_minimization}
\end{figure}

For restricting the number of quasi-Newton minimization steps,
\autoref{fig:restricted_energy_minimization}(A) shows that an acceptable maximum
number of minimization steps reproducing
$\vshr = \SI{-27}{\micro\meter\per\minute}$ depends on the chosen mechanical
parameters as  the higher their values are, the greater the energy and its
gradient.
A maximum number of minimization steps of around 100 should be an appropriate
value according to the results shown in
\autoref{fig:restricted_energy_minimization}(A).
The results in \autoref{fig:restricted_energy_minimization}(A) also
  show that reducing the number of minimization steps by a factor
  of 10 can lead to deviating growth velocities. Therefore,
  the improved  energy relaxation that we obtain in comparison
  to  Ref.\ \cite{VanBuren2005} by applying the equivalent of one order
  of magnitude more minimization steps should be relevant.

If minimization is restricted to a subset of minimization parameters at the tip
of the simulated MT, this subset is defined by the cutoff distance $\dcutoff$.
To have a maximum improvement in simulation speed, $\dcutoff$ should be as small
as possible.
It is evident from the data shown in
\autoref{fig:restricted_energy_minimization}(B)
that values $\dcutoff < 5$ have a
detectable influence on the shrinkage velocity.
We also ran some simulations with $\NGDP = 50$ and also for $\koncVal{2}$
(see Figure S8 in the Supplementary Material)
and based on all data, we choose $\dcutoff = 10$ as a
conservative value for the cutoff distance.

In summary, we are more confident in the second approach to only minimize the
MT tip where actual conformational changes happen, because for this subset, the
restricted energy is fully minimized.
Additionally, the first approach still has the issue of slowing down with an
increasing number of minimization parameters as all minimization parameters are
considered.
The second approach ensures that the number of minimization parameters does not
scale with the MT length but remains bounded, which assures that we can simulate
arbitrarily long growing MTs at a fixed minimal computational speed.
In the first approach, the quality of the minimization will probably also 
decline because the number of minimization parameters increases
while the number of minimization steps is kept constant.
Lastly, the first approach, in contrast to the second approach, does not
guarantee that the upper, i.e., the dynamic part of the MT is properly
minimized.

We also note that in the presence of
mechanical feedback onto hydrolysis, simulations take longer because
minimizations after hydrolysis events need to consider more tubulin dimers if
the hydrolyzed tubulin dimer is relatively deep in the MT lattice
(see Supplementary Material for more details).

\subsection{Full simulations exhibit repeated catastrophe and rescue events}

Based on the previous section on energy minimization, we use
$\dcutoff = 10$ for full simulations in which the initial MTs have both
a GDP body and a GTP cap, thus $\NGDP > 0$ and $\NGTP > 0$.
We now aim for realistic MT dynamics with
repeated phases of growth and shrinkage
in the same simulation and catastrophe and rescue events in between.
First, we only consider strictly random hydrolysis with a hydrolysis
rate $\khydr$ that is independent of tubulin dimers' position or
mechanical forces and which is another unknown free parameter in our model.
Hydrolysis coupled to mechanics via \eqref{eq:mechanics_hydrolysis_rate} will be
considered later.

It poses a computational challenge for chemomechanical MT models to reach time
scales of MT dynamics where  repeated catastrophe events occur at
realistic hydrolysis rates $\khydr$ and tubulin dimer concentrations $\ctub$.
In Ref.\ \cite{Zakharov2015}, where mechanics was implemented via full Brownian
dynamics, only short times scales could be reached (although the Brownian
dynamics was applied to only 300 tubulin dimers at the plus end).
Therefore, they increased the hydrolysis rate from their \enquote{normal} value
of \SI[per-mode=reciprocal]{0.5}{\per\second} (based on the \SI{2}{\second}
delay between polymerization and phosphate release measured by \cite{Melki1996},
which is also used by \cite{Aparna2017}) into a range of
\SIrange[per-mode=reciprocal]{3}{11}{\per\second}
in order to trigger catstrophe
events within computationally accessible time scales.
They found a linear scaling of catastrophe rate with $\khydr$ and employed a
linear extrapolation to obtain catastrophe rates for realistic hydrolysis rates
(see their Figure 3A).
In our simulations, we observe that increasing $\khydr$ beyond a certain
($\ctub$-dependent) value leads to immediate MT shrinkage because the initial
cap quickly hydrolyzes; this can be interpreted as an
instantaneous catastrophe. 
In such cases (like in \autoref{fig:full_simulation_klat100} for
$\ctub = \SI{7}{\micro\molar}$ and $\khydrVal{0.5}$), there is no real growth
phase based on which a catastrophe frequency could be determined.
For these hydrolysis rates, the individual critical concentration $\ctubc$
(where $\vgro = 0$ is reached) has apparently increased above the given tubulin
concentration.

\begin{figure}[ht!]
	\centering
	\includegraphics[width=0.99\linewidth]{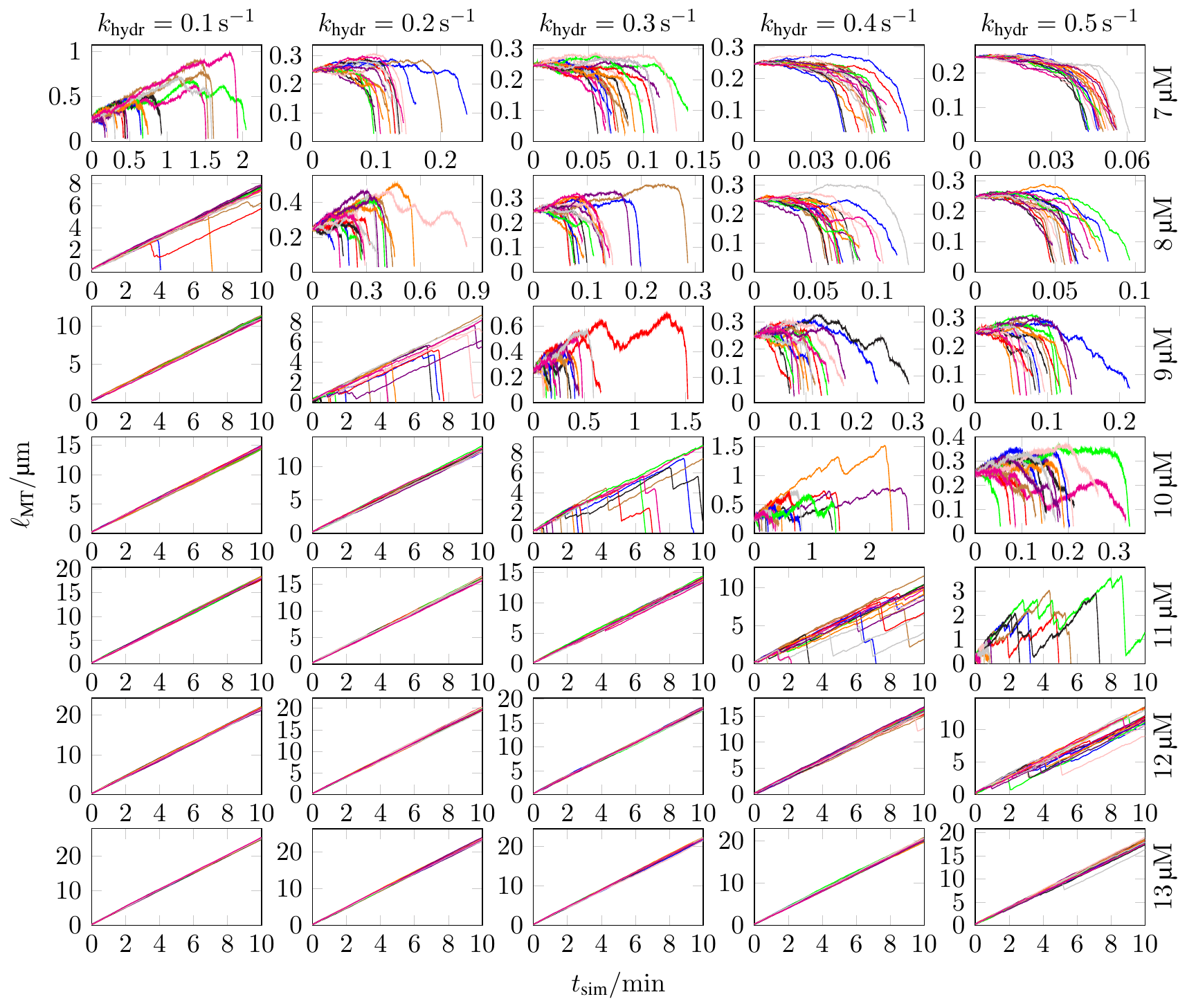}
	\caption{
		The MT length $\Lmt$ was measured as a function of the
		simulation time $\tsim$ for 20
		different simulations with $\koncVal{4}$, $\GlongVal{-9.3}$,
		$\klat = \SI{100}{\kBT \per\nano\meter\squared}$, seven
		different values of $\ctub$, and five different values of
		$\khydr$.
		MT growth trajectories for three additional $\ctub$ values can
		be found in Figure S9 in the Supplementary Material.
	}
	\label{fig:full_simulation_klat100}
\end{figure}

The experimental data on the hydrolysis rate 
is limited, so that many publications determine the hydrolysis rate themselves
by matching simulation results with experimental data
\cite{VanBuren2002,Piette2009,Margolin2011,Margolin2012,Padinhateeri2012,Bowne-Anderson2013,Coombes2013,Piedra2016}.
There are, however, more direct measurements in \cite{Melki1996}.
In most models and also in measurements from \cite{Melki1996}, the (random)
hydrolysis rate is in the range of
\SIrange[per-mode=reciprocal]{0.1}{0.5}{\per\second} (Ref.\
\cite{VanBuren2002} use a
relatively high value of \SI[per-mode=reciprocal]{0.95}{\per\second}).
We explore exactly this range of hydrolysis rates, see
\autoref{fig:full_simulation_klat100}.

\autoref{fig:full_simulation_klat100} shows MT growth curves (length vs.\ time)
over simulation times up to $\tsim = \SI{10}{\min}$ for several representative
tubulin concentrations and realistic hydrolysis rates.
MT growth trajectories as in \autoref{fig:full_simulation_klat100} for other
$\klat$ values can be found in Figures S10, S11, S12, and S13 in the
Supplementary Material.
Simulations in \autoref{fig:full_simulation_klat100}
were started with $\NGTP=10$ and $\NGDP=20$, but results
  are largely independent of the initial ratio $\NGTP/\NGDP$ (see, for example,
  Figure S11 in the Supplementary Material).

\begin{figure}[ht!]
	\centering
	\includegraphics[width=0.99\linewidth]{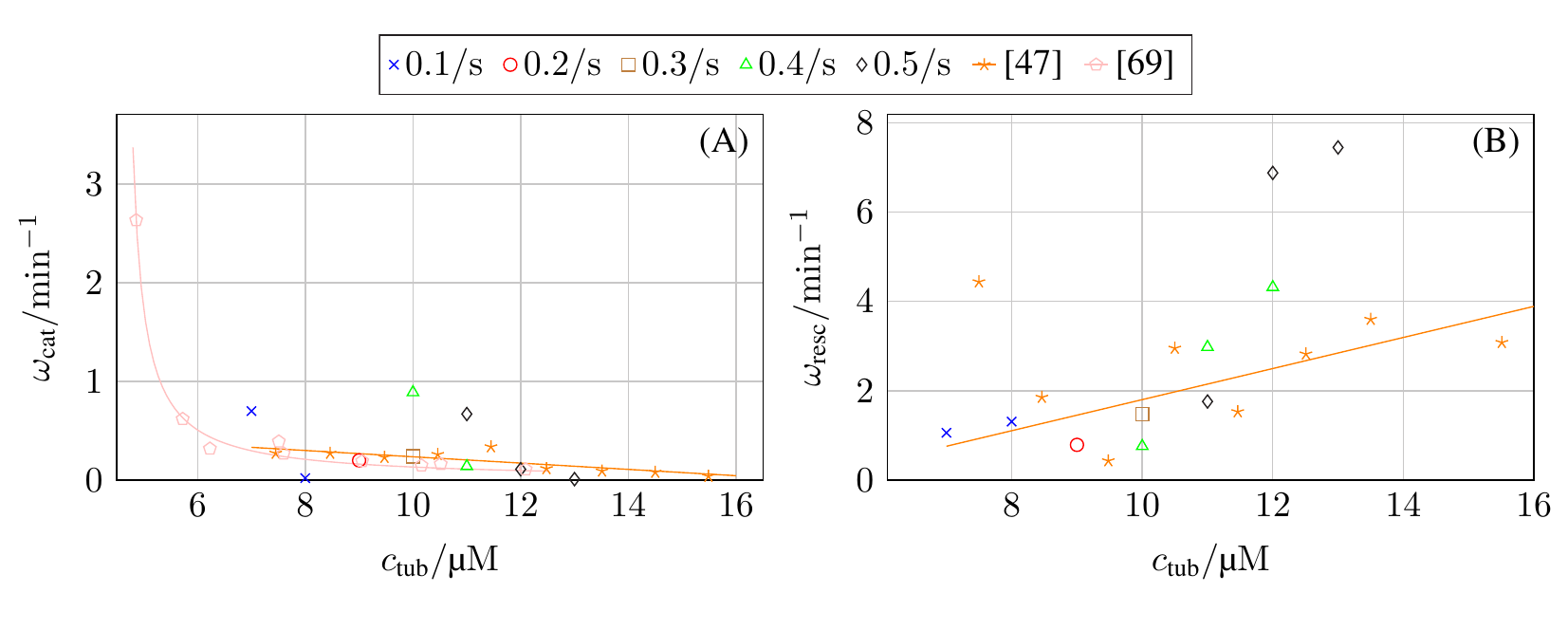}
	\caption{
          (A) Catastrophe rate $\omega_\text{cat}$ and (B)  rescue rate
          $\omega_\text{res}$
          as a function of
          GTP-tubulin concentration  $\ctub$ and in comparison
           with experimental data from Walker
         \etal~\cite{Walker1988} and Janson 
       \etal~\cite{Janson2003}.
	}
	\label{fig:catastropheRescueRates}
\end{figure}

Our chemomechanical MT model is computationally efficient such that
we can determine catastrophe and rescue rates as inverse
average growth and shrinking times between repeated
catastrophe and rescue events.
In the Supplementary Material, we explain the algorithm that
we used to identify catastrophe and rescue events and, thus,
growth and shrinking times from  MT 
simulation trajectories in detail. 
The results are shown in
\autoref{fig:catastropheRescueRates}.
In comparison to typical experimental data
\cite{Walker1988,Gardner2011_kinesins},
this decrease of the catastrophe rate
with tubulin concentration seems too steep.
Current phenomenological models for the MT catastrophe rate as a function of
tubulin concentration can be found in
\cite{Flyvbjerg1996,Zelinski2013}, experimental data in
\cite{Walker1988,Janson2003};
the decrease of the catastrophe rate with GTP-tubulin concentration
$\ctub$ appears steeper in the simulation for all hydrolysis
rates $\khydr=$ \SIrange[per-mode=reciprocal]{0.1}{0.5}{\per\second}.

In the following, we will discuss two aspects of MT growth and catastrophes in
more detail, namely the dependence of growth velocity on hydrolysis rate and the
detailed dynamics within single catastrophe events, which become accessible
within a computational model and are impossible to address experimentally.

\subsection{Growth velocity reduces linearly with hydrolysis rate because
of  cap structure}
\label{sec:vgro_of_ctub_with_hydrolysis}

So far, we parameterized the model by fitting the growth
velocity of GTP-only MTs, i.e., in the absence of hydrolysis to the
experimentally measured velocity in \eqref{eq:walker_growth}.
Hydrolysis reduces this growth velocity by increasing the probability of
GDP-dimers dimers at the plus end.
This increases the rate of bond rupture because hydrolyzed
dimers tend to create stretched bonds which rupture more easily (there
is no direct increase of the off-rate for hydrolyzed GDP-dimers
in our model).
As only laterally unbounded dimers can detach, hydrolyzed
GDP-dimers at the plus end have an
effectively higher detachment rate.

\begin{figure}[ht!]
	\centering
	\includegraphics[width=0.99\linewidth]{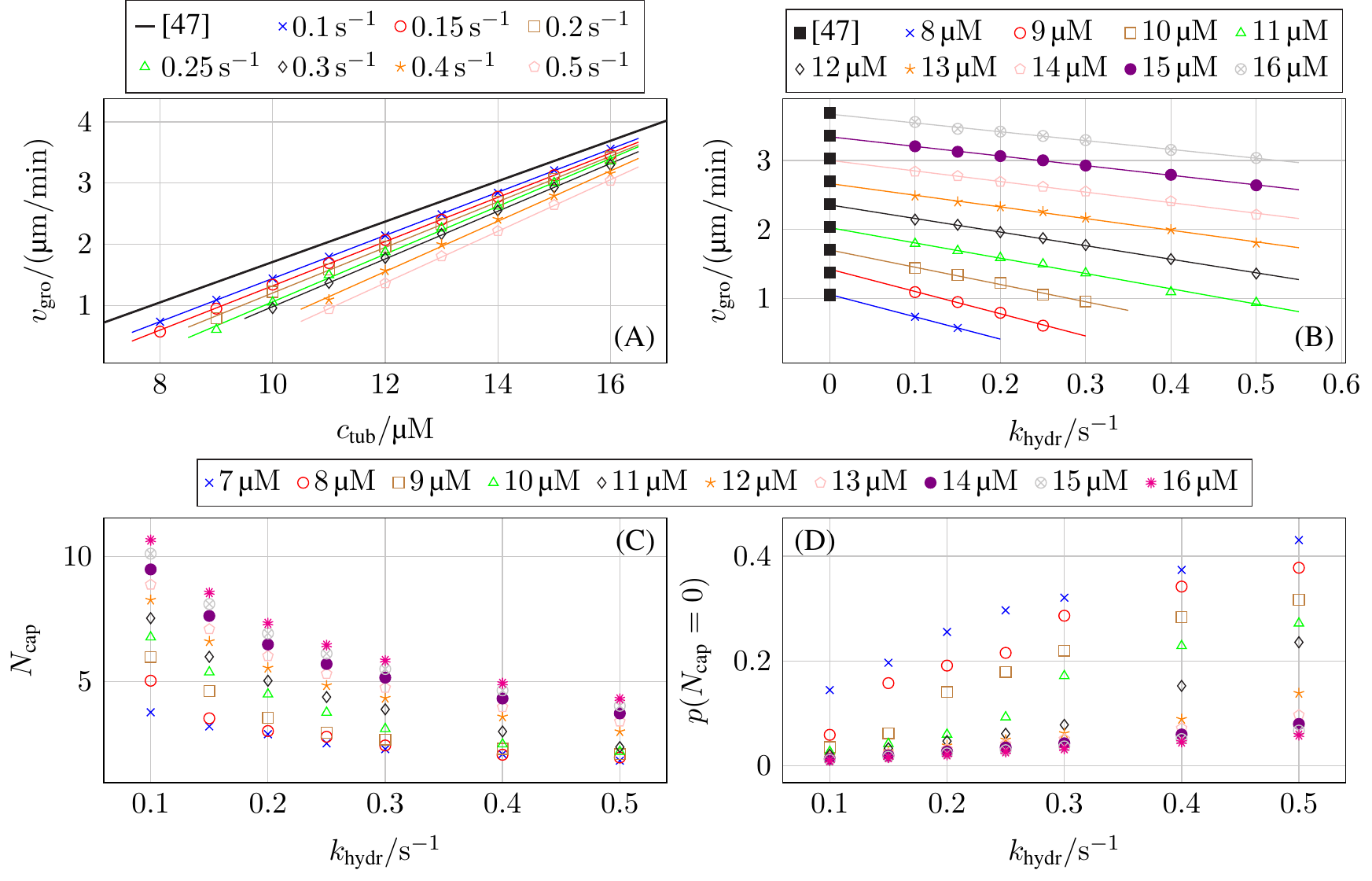}
	\caption{
		Growth velocity $\vgro$ as a function of (A) the free tubulin
		dimer concentration $\ctub$ for different hydrolysis rates
		$\khydr$ and as a function of (B) the hydrolysis rate $\khydr$
		for different free tubulin dimer concentrations $\ctub$ in
		comparison to the experimental data from \cite{Walker1988}. (C)
		Average GTP-tubulin cap length
                $\langle \Ncap \rangle$ of protofilaments and
                (D) fraction of protofilaments without a
		GTP cap as a function of the hydrolysis rate $\khydr$.
		The standard set of parameters from \autoref{tab:parameters} was
		used.
	}
	\label{fig:vgro_of_ctub_with_hydrolysis}
\end{figure}

The last row of \autoref{fig:full_simulation_klat100} indicates and
\autoref{fig:vgro_of_ctub_with_hydrolysis}(B) shows  explicitly 
that increasing the
hydrolysis rate decreases the growth velocity linearly
although the growth reduction
mechanism is indirect via the increased probability of bond rupture for
hydrolyzed GDP-dimers.
Our model parameterization was such that we obtain the
experimentally measured growth velocities by Walker \etal~\cite{Walker1988}
at $\khydrVal{0}$ in \autoref{fig:vgro_of_ctub_with_hydrolysis}(B).
Nevertheless, \autoref{fig:vgro_of_ctub_with_hydrolysis}(A) shows that there is
still a linear relation between the free tubulin dimer concentration $\ctub$ and
the growth velocity $\vgro$
so that it is possible to re-adjust parameters to  reproduce the
growth velocity in the presence of hydrolysis, once a particular hydrolysis
rate can be reliably selected.

Because both the dependence on tubulin concentration in
\autoref{fig:vgro_of_ctub_with_hydrolysis}(A)
remains linear and the reduction by
the hydrolysis rate in \autoref{fig:vgro_of_ctub_with_hydrolysis}(B)
is linear, we
also expect that the individual critical concentration (where $\vgro = 0$ is
reached) increases linearly with the hydrolysis rate beyond the value
$\ctubc \simeq \SI{5}{\micro\molar}$ of Walker \etal~\cite{Walker1988}.
\autoref{fig:full_simulation_klat100} clearly shows that increasing $\khydr$
actually increases the individual critical concentration $\ctubc$.
\footnote{
  The individual critical concentration can be read off from
\autoref{fig:full_simulation_klat100} as the
concentration below which immediate MT
shrinkage sets in.
}

The mechanism of growth velocity reduction by hydrolysis can be further
elucidated by comparing the average GTP-tubulin
cap length $\langle \Ncap \rangle$ of protofilaments (see
\autoref{fig:vgro_of_ctub_with_hydrolysis}(C)),
and the fraction of protofilaments
without a GTP-cap (see \autoref{fig:vgro_of_ctub_with_hydrolysis}(D)):
The higher the hydrolysis rate is, the smaller the GTP-cap and the higher the
fraction of cap-less protofilaments is.
\footnote{
  As the cap lengths shown in \autoref{fig:vgro_of_ctub_with_hydrolysis}(C) are
averaged over the whole duration of the simulations, these cap lengths also
average over growth and shrinkage phases.
As cap lengths are shorter during shrinkage than growth, the cap lengths in
\autoref{fig:vgro_of_ctub_with_hydrolysis}(C) can be regarded as a lower limit
on the average cap length during MT growth.
}
The increase in GDP-tubulin dimers depolymerizing from the protofilament tips
for higher hydrolysis rates is due to an increase in the probability of uncapped
protofilaments with the hydrolysis rate as shown in
\autoref{fig:vgro_of_ctub_with_hydrolysis}(D).
In Ref.\ \cite{Li2010}, dependencies $\langle \Ncap\rangle
\propto \sqrt{\ctub/\khydr}$
and $p(\Ncap=0) \propto \khydr/\ctub$ have been predicted, which are
in agreement with \autoref{fig:vgro_of_ctub_with_hydrolysis}(C) and (D).

\subsection{Detailed dynamics within single catastrophe and rescue events}

\begin{figure}[ht!]
	\centering
        \includegraphics[width=0.9\linewidth]{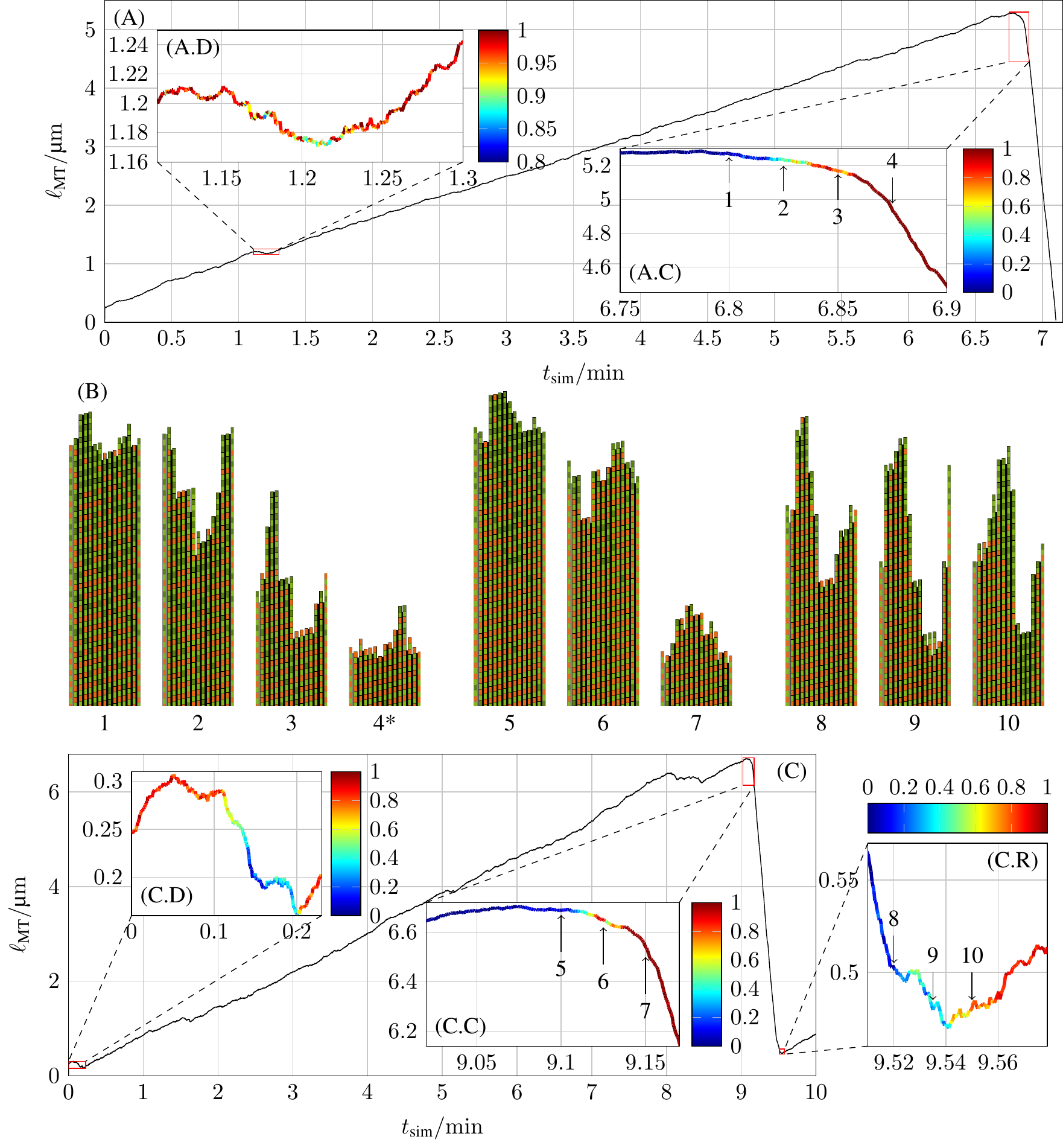}
	\caption{
		Lengths of two MTs as a function of simulation time $\tsim$
		with $\koncVal{4}$, $\GlongVal{-9.3}$,
		$\klat = \SI{100}{\kBT \per\nano\meter\squared}$, and
		(A) $\ctub = \SI{8}{\micro\molar}$ and $\khydrVal{0.1}$
		and (C) $\ctub = \SI{9}{\micro\molar}$ and $\khydrVal{0.2}$.
		The insets highlight parts of the trajectories of interest for
		the dynamics and color-code the probability of the $\Lmt(\tsim)$
		curve to stay quantitatively the same at the relevant point in
		time if new simulations are started with the relevant
		configuration as the initial configuration (for more details,
		refer to the text).
		(B) shows the two-dimensional representations of
		certain MT tip configurations that are marked by arrows
		in the insets of (A) and (C)
                (configuration $4*$ has been shifted towards
                  the MT tip by 24
                  tubulin dimer lengths). The first protofilament is the
		periodic image of $p = 13$ and the last protofilament is the
		periodic image of $p = 1$.
		Lateral bonds are represented by the thick black line between
		protofilaments.
	}
	\label{fig:full_simulations}
\end{figure}

The chemomechanical model reproduces realistic MT dynamics including
catastrophe and rescue events.
\autoref{fig:full_simulations} shows typical MT growth paths featuring two
catastrophe events and a rescue event in subfigure (C).
Moreover, we observe \enquote{dips} in the growth path where a short phase of
shrinking appears, which are similar to \enquote{stutter} events that have been
observed in Ref.\ \cite{Mahserejian2019}.
Videos of these two simulations  with two- and
three-dimensional representations of the MT structure
can be found in the Supplementary Material.

Using our computational model, we can systematically identify the point in a MT
growth path, where a catastrophe becomes structurally unavoidable.
This allows us to search for typical catastrophe-triggering features in MT
growth.
To analyze how probable it is at specific points in the simulation of MT
dynamics that the MT continues a certain growth path, we chose two simulations
with at least one significant event (meaning a catastrophe, rescue, or a
\enquote{dip}/\enquote{stutter}) and took configurations around such events as
starting points for new simulations (similar to \cite{Margolin2012}).
In these new simulations, MTs were allowed to grow (or shrink) for a maximum of
\SI{60}{\second}, a sufficient amount of time to check if the new simulations
show dynamics similar to the original simulation around the significant event.

The MT growth trajectory shown in \autoref{fig:full_simulations}(A) has two
significant events: a dip at $\tsim = \SI{1.2}{\minute}$ and a catastrophe at
$\tsim = \SI{6.85}{\minute}$;
the trajectory
in \autoref{fig:full_simulations}(C) contains three significant events:
 a dip at the
 very beginning, a catastrophe at $\tsim = \SI{9.15}{\minute}$, and a rescue
 at $\tsim = \SI{9.54}{\minute}$.
To determine whether newly run simulations with 
starting points from the initial simulation
qualitatively follow the original simulation, we need
criteria to identify dips, catastrophes, or rescue events.
The exact criteria for these events in  \autoref{fig:full_simulations}(A) and
(C) are stated in the Supplementary Material.
In short, in order to identify whether a new simulation reproduces
a catastrophe, we check  after
a time of \SIrange{10}{15}{\second}
whether the MT is sufficiently short
that a catastrophe must have happened;
for a dip, we check  whether the MT continued to grow without entering
a catastrophe; for a rescue, we check that the MT did not
completely vanish because it continued to shrink.
For each initial configuration, we ran 20 new simulations and calculated the
 fraction of simulations that fulfilled these criteria.
 These fractions are the probabilities for the original growth path at different
 points in time, and they are shown color-coded in all the insets in
 \autoref{fig:full_simulations}.

Both catastrophes and the rescue show that the transition from a high
probability to stay in the current dynamic state to a high probability to switch
into the other dynamic state occurs within a few seconds.
In \autoref{fig:full_simulations}(A) and (C), we first observe
that catastrophes become practically unavoidable (red color code in (A.C) and
(C.C)) after a phase of relatively slow shrinking by
\SIrange{50}{100}{\nano\meter}; similar \enquote{transitional catastrophe}
behavior has been observed in Ref.\ \cite{Mahserejian2019}.
A dip, on the other hand, can only evade a catastrophe (yellow to red color code
in (A.D) and (C.D)) if the MT length shrinks by significantly less than
$\SI{50}{\nano\meter}$.

Because hydrolysis followed by straining and rupture of the lateral bonds is
required before a laterally unbonded dimer can detach, MT shrinking by
$\SI{50}{\nano\meter}$ suggests that roughly 6 dimer layers must hydrolyze in a
row to trigger a catastrophe.
This is, however, not sufficient to remove the entire GTP-cap.
The GTP-cap length averaged over all protofilaments is still $>1$ when the
catastrophe becomes unavoidable
(at points 3 in \autoref{fig:full_simulations}(A)
and 6 in \autoref{fig:full_simulations}(C), see also 
Figure S1 in the Supplementary Material).
As the corresponding MT snapshot insets 3 and 6 reveal, the reason for this
discrepancy is the average over all protofilaments:
it appears that typically only a \enquote{nucleus} of three neighboring
protofilaments shrinks by more than 6 dimers, such that its GTP-cap is removed
and its ends reach into the GDP-body of the MT, when a catastrophe is triggered.
The MT snapshots in \autoref{fig:full_simulations}(B) also suggest that
rescue events require formation of a GTP-cap on almost all 13 protofilaments
(with an average cap length $\sim 4$)
such that nuclei of three neighboring uncapped GDP-protofilaments are avoided.
Further investigation of more catastrophe events will be necessary to definitely
deduce catastrophe- and rescue-triggering structural MT features.

\subsection{Hydrolysis coupled to mechanics changes the cap structure}

Finally, we test how a mechanical feedback onto the hydrolysis rate as
introduced in \eqref{eq:mechanics_hydrolysis_rate} and
\eqref{eq:mechanics_hydrolysis_rate2} changes the cap structure and
dynamic behavior.
In the presence of this mechanical feedback, tubulin dimers in the MT lattice
with larger bending angles tend to hydrolyze preferentially.

\begin{figure}[ht!]
	\centering
	\includegraphics[width=0.99\linewidth]{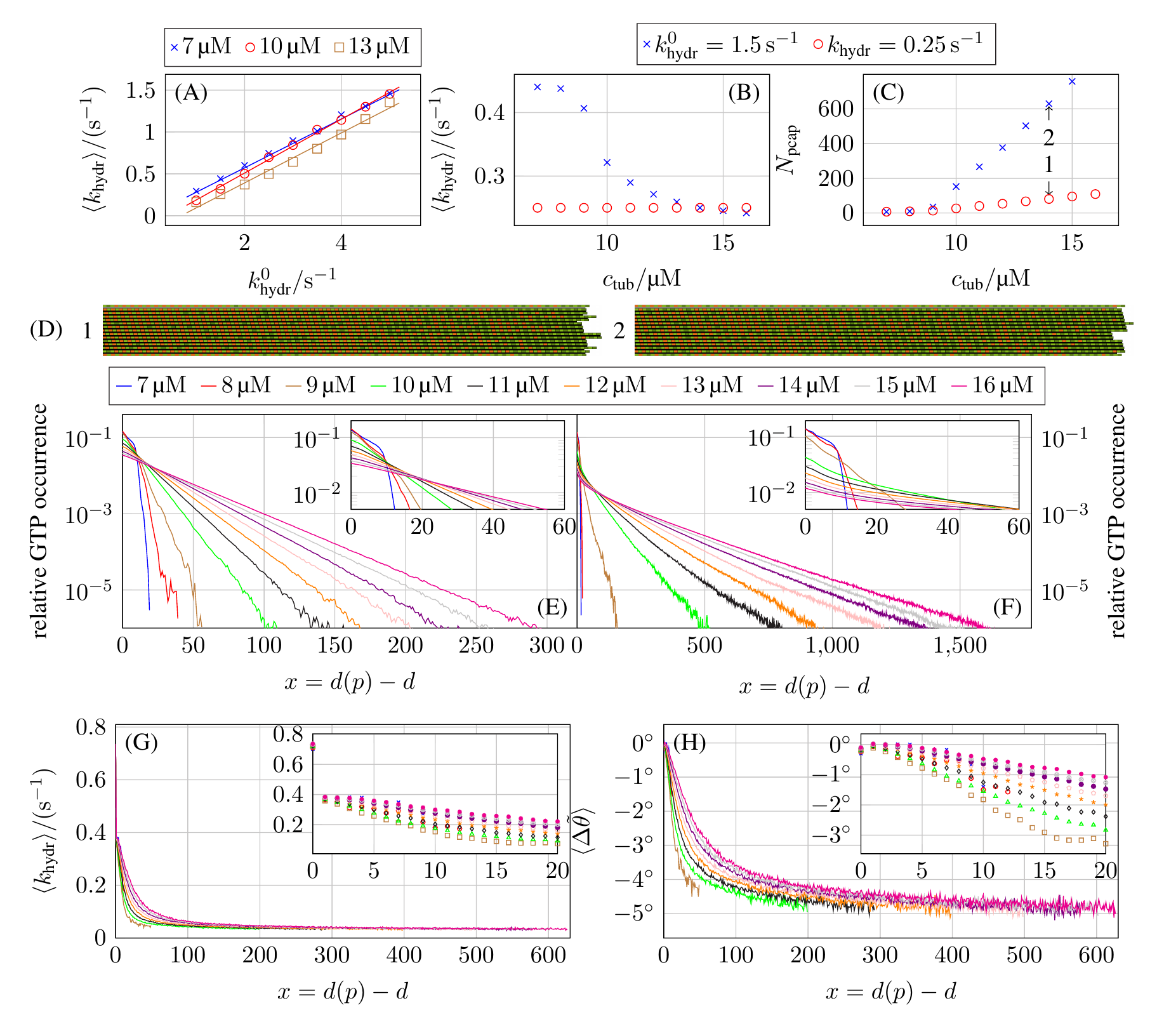}
	\caption{
		(A) Average actual hydrolysis rate $\langle \khydr \rangle$ as a
		function of the constant base hydrolysis rate $\khydrN$.
		Comparison of (B) the average actual hydrolysis rate
		$\langle \khydr \rangle$ and (C) the porous cap length
		$\porousCapLength$ as a function of the free tubulin dimer
		concentration $\ctub$ for hydrolysis coupled to mechanics with
		$\khydrNVal{1.5}$ and a constant hydrolysis rate of
		$\khydrVal{0.25}$.
		(D) shows the two-dimensional representations of
		two MT tip configurations that are marked by arrows in (C) at
		$\tsim = \SI{5}{\min}$.
		The top and bottom protofilaments are periodic images of
		$p = 13$ and $p = 1$, respectively.
		Relative occurrence of GTP tubulin dimers as a function of the
		dimer-based distance from the protofilament tip $d(p) - d$ for
		(E) a constant hydrolysis rate of $\khydrVal{0.25}$ and (F)
		hydrolysis being coupled to mechanics and $\khydrNVal{1.5}$.
		(G) Average hydrolysis rate as a function of distance 
		$d(p) - d$ from the tip and (H)
                the associated average bending angle
		$\langle \Delta \tilde{\theta} \rangle$ for hydrolysis coupled
		to mechanics and  $\khydrNVal{1.5}$.
		All plots are for $\koncVal{4}$, $\GlongVal{-9.3}$, and
		$\klat = \SI{100}{\kBT \per\nano\meter\squared}$.
	}
	\label{fig:mechanical_hydrolysis_results}
\end{figure}

Overall,
we find a linear relation between $\khydrN$ and the average hydrolysis rate
$\langle \khydr \rangle$ (see \autoref{fig:mechanical_hydrolysis_results}(A))
with $\khydrN \gg \langle \khydr \rangle$.
When comparing MT growth with  hydrolysis
coupled to mechanics with average hydrolysis rate
$\langle \khydr \rangle$ to MT growth with constant hydrolysis
rate $\khydr$ (for exmaple in
\autoref{fig:mechanical_hydrolysis_results}(D)-(F)),
we use  \autoref{fig:mechanical_hydrolysis_results}(A)
to choose the base hydrolysis rate $\khydrN$ such that
$\langle \khydr \rangle \approx \khydr$.

\autoref{fig:mechanical_hydrolysis_results}(B) shows the average
hydrolysis rate $\langle \khydr \rangle$ as a function of the free tubulin dimer
concentration $\ctub$ for $\khydrNVal{1.5}$.
Here, we observe a pronounced nonlinear concentration dependence with a decrease
around the individual critical tubulin concentration
$\ctub \simeq \SI{10}{\micro\molar}$.
At the same concentration, also the porous cap length $\porousCapLength$ (see
\autoref{fig:mechanical_hydrolysis_results}(C)),
which is defined as the difference
between the number of tubulin dimers in a protofilament and the value of $d$ of
the first GTP-tubulin dimer counted from the minus end,
starts to increase.
As a result, the porous cap length for hydrolysis
coupled to mechanics is much longer compared to a
constant hydrolysis rate, even if the average effective hydrolysis rate is
roughly the same.
In the following, we argue  that the reason 
for this increase in porous cap length is a decrease of the
hydrolysis rate for GTP-dimers away from  the tip.
Mechanical feedback gives rise to preferential hydrolysis at the tip,
i.e., the average hydrolysis rate $\langle \khydr(x) \rangle$
(over all actually executed
hydrolysis events) is
larger for small layer distances $x \equiv d(p) - d$  from the tip,
as can be seen in  \autoref{fig:mechanical_hydrolysis_results}(G).
This is in line with previous results in Ref.\ \cite{Mueller2014}
from  a much simpler version of our model
with a deterministic hydrolysis kinetics and without dimer attachment and
detachment.

According to \eqref{eq:mechanics_hydrolysis_rate2}, GTP-tubulin dimers with
larger bending angles tend to hydrolyze preferentially.
If a straight GTP-dimer is bent inward ($\bendAngle < \SI{0}{\degree}$),
its hydrolysis rate is reduced according to
\eqref{eq:mechanics_hydrolysis_rate2}; if it is bent outwards
($\bendAngle > \SI{0}{\degree}$) the rate is increased.
From the hydrolysis rates shown in
\autoref{fig:mechanical_hydrolysis_results}(G),
it is possible to calculate the average bending angles using
\eqref{eq:mechanics_hydrolysis_rate} and
\eqref{eq:mechanics_hydrolysis_rate2},
\begin{equation}
	\langle \Delta \tilde{\theta}(p,d) \rangle
	= \frac{1}{\SI{11}{\degree}} \left[
          \frac{1+\delta_{d,d(p)}}{\kappa}
          \ln \left( \frac{\khydr(p,d)}{\khydrN} \right)
          + (\SI{5.5}{\degree})^2 \right] .
        \label{eq:avDelta}
\end{equation}
The results for these bending angles as a function
of the distance $x$ from the top are shown in
\autoref{fig:mechanical_hydrolysis_results}(H).
Surprisingly, almost
all dimers are bent inwards  ($\bendAngle < \SI{0}{\degree}$)
on average apart from dimers close to the tip,
We will try to
interpret these results in the following.

An isolated  GTP-dimer
within the GDP-body can alleviate the bending stress of GDP-dimers
by bending inward ($\bendAngle < \SI{0}{\degree}$), which allows longitudinally
neighboring GDP-dimers to bend outwards (such that
$\bendAngle > \SI{0}{\degree}$)
resulting in an overall decrease of the MT energy
(see Figures S14 and S15   in the Supplementary
Material). 
Therefore,
isolated GTP-dimers deep in the GDP-body hydrolyze with a reduced asymptotic
rate $\langle \khydr \rangle_\infty\ll \khydr$.

We also find that, for several consecutive GTP-dimers in the same protofilament,
GTP-dimers curl inward directly at the GDP/GTP interface resulting in a
reduced hydrolysis rate (see Supplemental Figure S15), 
while GTP-dimers in the center of a GTP-island 
are straight so that they have a higher hydrolysis rate than at the GDP/GTP
interfaces.
Effectively, this hydrolysis rate distribution within a GTP-island results in a
\enquote{anti-vectorial} hydrolysis mechanism with which GTP-islands are
hydrolyzed from the interior in contrast to vectorial hydrolysis where
hydrolysis happens at the GTP/GDP interfaces.

Also for GTP-dimers in layers closer to the MT tip, 
other longitudinally close-by
GTP-dimers cooperate in alleviating bending stresses; then inward bending is
still preferred, but the inward bending angle becomes smaller.
This decrease in inward bending corresponds to an increase of the average
hydrolysis rates $\langle \khydr(x) \rangle$ for GTP-dimers in these layers
compared to GTP-dimers buried deeper in the MT body
(see \autoref{fig:mechanical_hydrolysis_results}(G) and (H)).
For terminal tubulin dimers  ($x = 0$), we observe
a  hydrolysis rate $\langle \khydr(x) \rangle$
\emph{higher} than $\khydr$  (while it is  equal
or lower  than $\khydr$ for all other layers $x > 0$).
Hydrolysis in the first layer  is enhanced because
there are no  tubulin dimers on top, such 
that  hydrolysis has to overcome a smaller energy barrier as pointed
out previously (the  $d+1$-term in 
\eqref{eq:mechanics_hydrolysis_rate2} is missing corresponding to the
$\delta_{d,d(p)}$-contribution in \eqref{eq:avDelta}).
As a result of the hydrolysis bias toward the tip, the spatial GTP-tubulin dimer
distribution also differs.
For concentrations where the MTs are growing only on time scales of
several minutes ($\ctub \ge \SI{11}{\micro\molar}$) for the chosen parameters, a
constant hydrolysis rate leads to the expected exponential distribution of
GTP-dimers shown in \autoref{fig:mechanical_hydrolysis_results}(E)
as observed in \emph{in vivo} experiments \cite{Seetapun2012}.
Using an effective one-dimensional (or single protofilament) model similar to
\cite{Padinhateeri2012} to
calculate the probability of tubulin dimers being GTP-tubulin dimers as a
function of the polymerization rate $\kon$,
effective depolymerization rate $\koffOneD$, and
hydrolysis rate $\khydr$ matches the simulation results for concentrations at
which the MTs can be considered in a steady state of growth (see
Section 4 in the Supplementary Material).
We use an effective depolymerization rate $\koffOneD$ instead of $\koff$,
because we map onto the depolymerization process
of a one-dimensional model so that $\koffOneD$ includes
all effects from lateral bond formation and rupture and the actual
depolymerization process in the full model.

If hydrolysis is coupled to mechanics, the spatial distribution is only
exponential in its tail, has larger values at the MT tip, and 
 GTP-tubulin dimers can be found much deeper in the GDP-body 
 (see \autoref{fig:mechanical_hydrolysis_results}(F)).
These results reflect that the average hydrolysis rate
$\langle \khydr(x) \rangle$ is decreasing towards the GDP-body and reaches a
small limiting value $\langle \khydr \rangle_\infty\ll \khydr$ for distances
$x = d(p)-d > 500$ away from the tip, which governs the exponential tail
(see \autoref{fig:mechanical_hydrolysis_results}(G)).
This can be rationalized by considering the probability $\probGTP(x)$ to
find a GTP-dimer at distance $x$ from the tip in a single protofilament
and continuum approximation.
The balance between attachment/detachment and hydrolysis leads to
\begin{equation}
	0
	= -(\kon - \koffOneD) \frac{\text{d} \probGTP}{\text{d} x}
		- \langle \khydr(x) \rangle \,\probGTP(x)
	\label{eq:pGTPx}
\end{equation}
in the stationary state, which results in
a sharp initial decrease of $\probGTP(x)$ because
$\langle \khydr(0) \rangle$ is large at the tip but a much slower asymptotic
exponential decrease when
$\langle \khydr(x) \rangle \approx \langle \khydr \rangle_\infty\ll \khydr$,
which explains the main features in
\autoref{fig:mechanical_hydrolysis_results}(F).
In Section 4 in the Supplementary Material,
we show that \eqref{eq:pGTPx}
describes simulations with a constant hydrolysis and with hydrolysis coupled to
mechanics equally well.
With $\probGTP(x)$, we can define an \enquote{average cap length} as
$\avgCapLength = \int_0^\infty \text{d}x\, \probGTP(x) x$.
This average cap length $\avgCapLength$ is longer if hydrolysis is coupled to
mechanics compared to a constant hydrolysis rate because $\probGTP(x)$ is much
greater for larger $x$
(see \autoref{fig:mechanical_hydrolysis_results}(E) and (F)).
As $\avgCapLength < \porousCapLength$, this increase in average cap length also
explains the increased porous cap length if hydrolysis is coupled to mechanics.

\begin{figure}[ht!]
	\centering
	\includegraphics[width=0.99\linewidth]{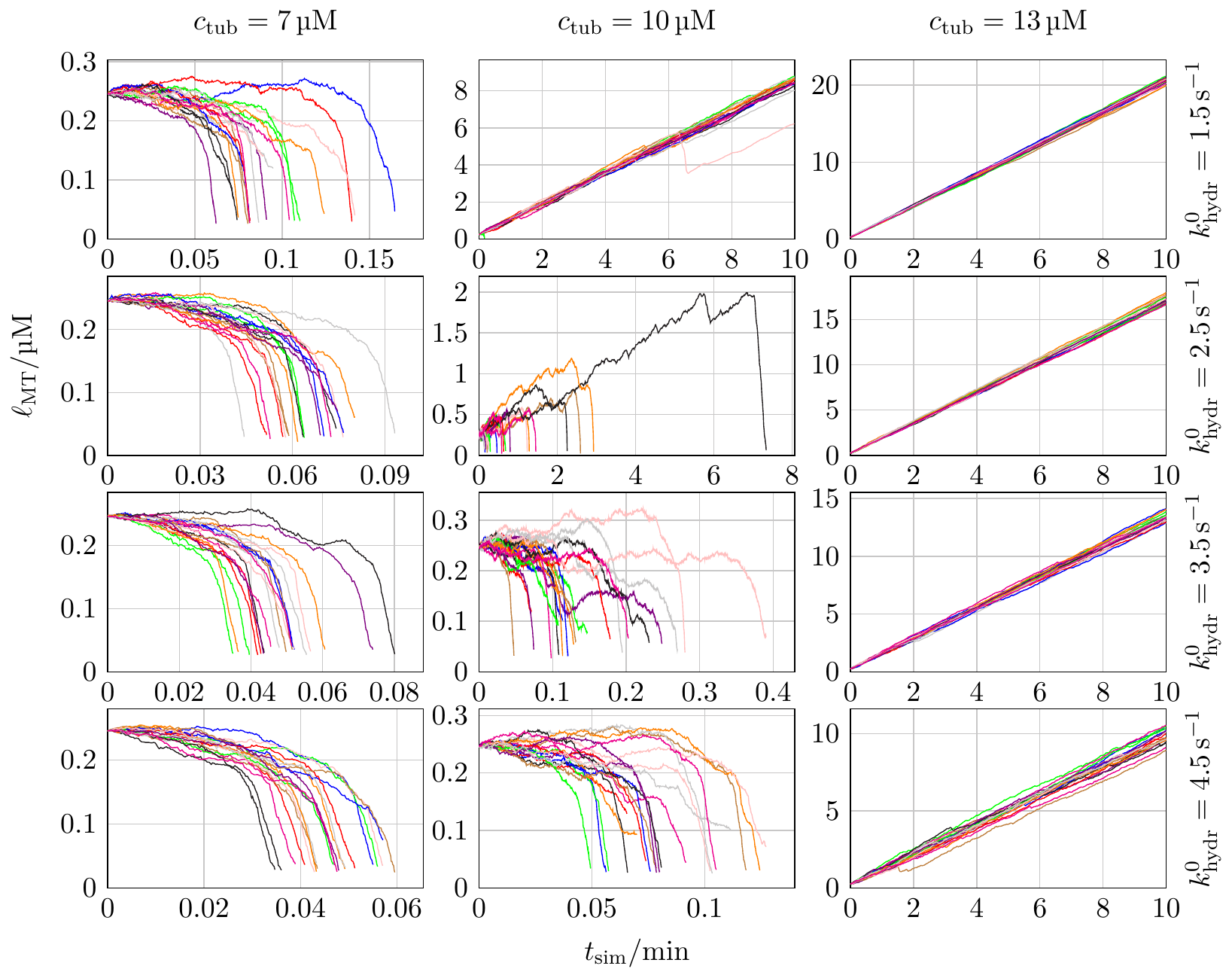}
	\caption{
		MT length $\Lmt$ as a function of the simulation time $\tsim$
		for 20 different simulations with
		$\koncVal{4}$, $\GlongVal{-9.3}$,
		$\klat = \SI{100}{\kBT \per\nano\meter\squared}$, three
		different values of $\ctub$, and four different values of
		$\khydrN$.
	}
	\label{fig:full_simulation_klat100_mechanical_hydrolysis}
\end{figure}

The relative increase of hydrolyzed GDP-dimers at the tip could make MTs more
prone for catastrophes and give rise to an increased catastrophe rate and,
eventually, a more realistic concentration dependence of catastrophe rates.
\autoref{fig:full_simulation_klat100_mechanical_hydrolysis}, however, shows that
this is not the case.
Instead, the same steep dependence on the (base) hydrolysis rate as in
\autoref{fig:full_simulation_klat100} persists.

\begin{figure}[ht!]
	\centering
	\includegraphics[width=0.99\linewidth]{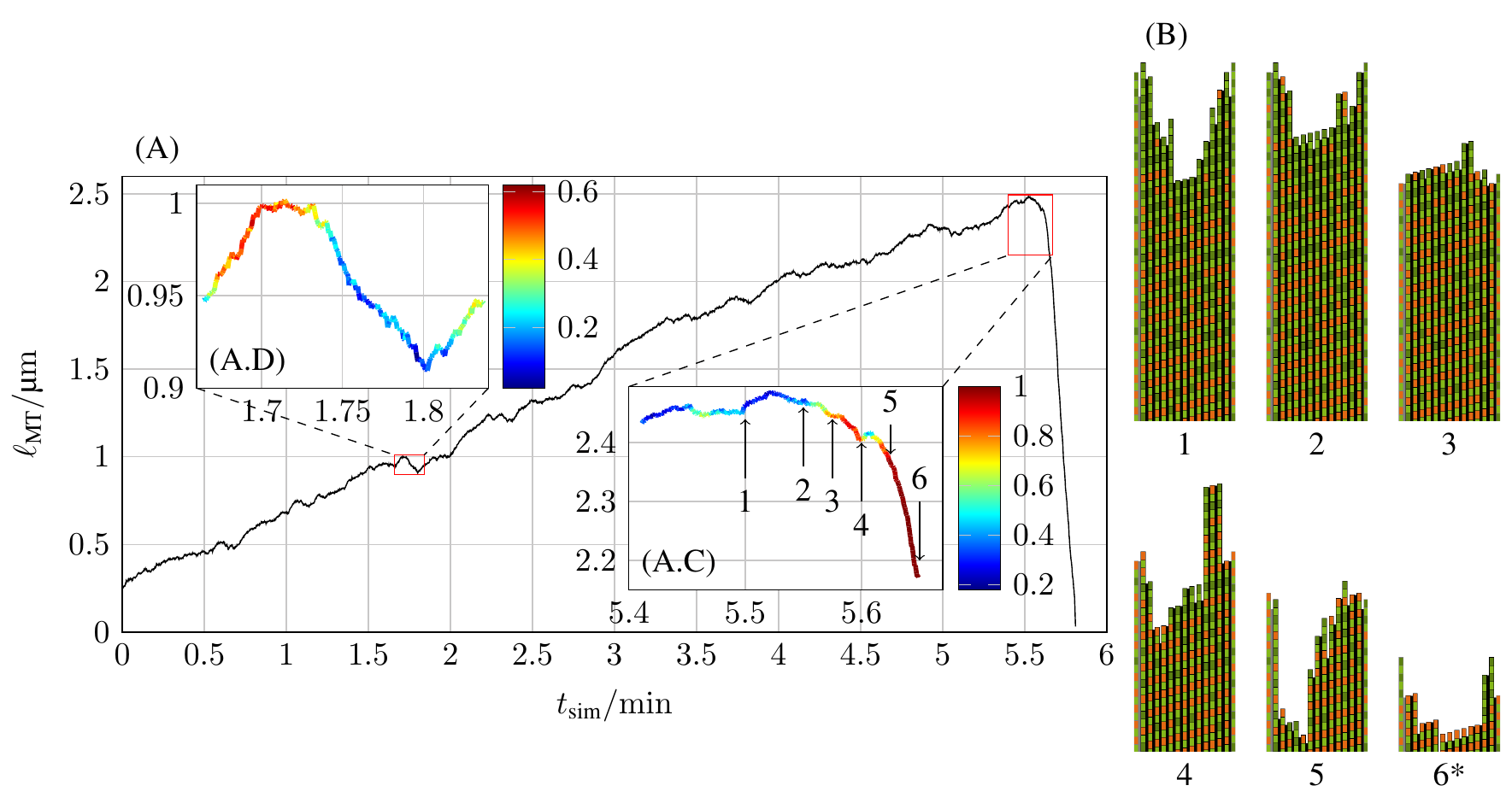}
	\caption{
		(A) Length of a MT $\Lmt$ as a function of the simulation time
		$\tsim$ with $\koncVal{4}$, $\GlongVal{-9.3}$,
		$\klat = \SI{100}{\kBT \per\nano\meter\squared}$,
		$\ctub = \SI{9}{\micro\molar}$ and $\khydrNVal{1.5}$ with
		hydrolysis being coupled to mechanics and (B) the
		two-dimensional representations of certain MT tip
		configurations that are marked by arrows in the inset (A.C)
                (configuration $6*$ has been shifted towards
                  the MT tip by 21
                  tubulin dimer lengths).
	}
	\label{fig:full_simulations_mechanical_hydrolysis}
\end{figure}

In comparison to the MT growth trajectories with a constant
hydrolysis rate shown in \autoref{fig:full_simulations},
\autoref{fig:full_simulations_mechanical_hydrolysis} shows an example of a
MT simulation in which the hydrolysis rate is coupled to mechanics.
To calculate the probabilities shown in the insets, the same criteria as for
\autoref{fig:full_simulations}(A) were used.
At first sight, these trajectories look similar to the corresponding
trajectories for a constant hydrolysis rate \autoref{fig:full_simulations}(A).
There is, however, a significantly increased roughness of the trajectory during
the growth phase, which could be interpreted as increased occurrence of
\enquote{dips} or \enquote{stutter} events.
A high probability of stutter events has also been observed in Ref.\
\cite{Mahserejian2019}, which supports the existence of
a mechanochemical coupling in hydrolysis.
The catastrophe-triggering configuration of a \enquote{nucleus} of several
neighboring protofilaments shrinking by more than 6 dimers is also similar
as snapshots 4 and 5 in \autoref{fig:full_simulations_mechanical_hydrolysis}(B)
show.

\subsection{Dilution experiments}

In dilution experiments, the free tubulin dimer concentration
$\ctub$ is reduced to  $\cdil\ll \ctub$ at a certain point in time
\cite{Voter1991,Walker1991,Duellberg2016}.
If the diluted concentration is sufficiently small or zero,
the GTP-cap stops growing by polymerization (and depolymerizes)
but continues to hydrolyze;
after a characteristic delay time $\Dtdelay$,
the GTP-cap has vanished, a catastrophe is initiated,  and the MT shrinks.
Thus, dilution experiments and their comparison to corresponding
dilution simulations can
give information on the hyrolysis rate.
Simulation results for the delay time are shown in \autoref{fig:dilution}.
In the Supplementary Material, we explain the algorithm that
we used to determine the  delay time $\Dtdelay$ from
MT simulation trajectories in detail.

\begin{figure}[ht!]
        \centering
        \includegraphics[width=0.99\linewidth]{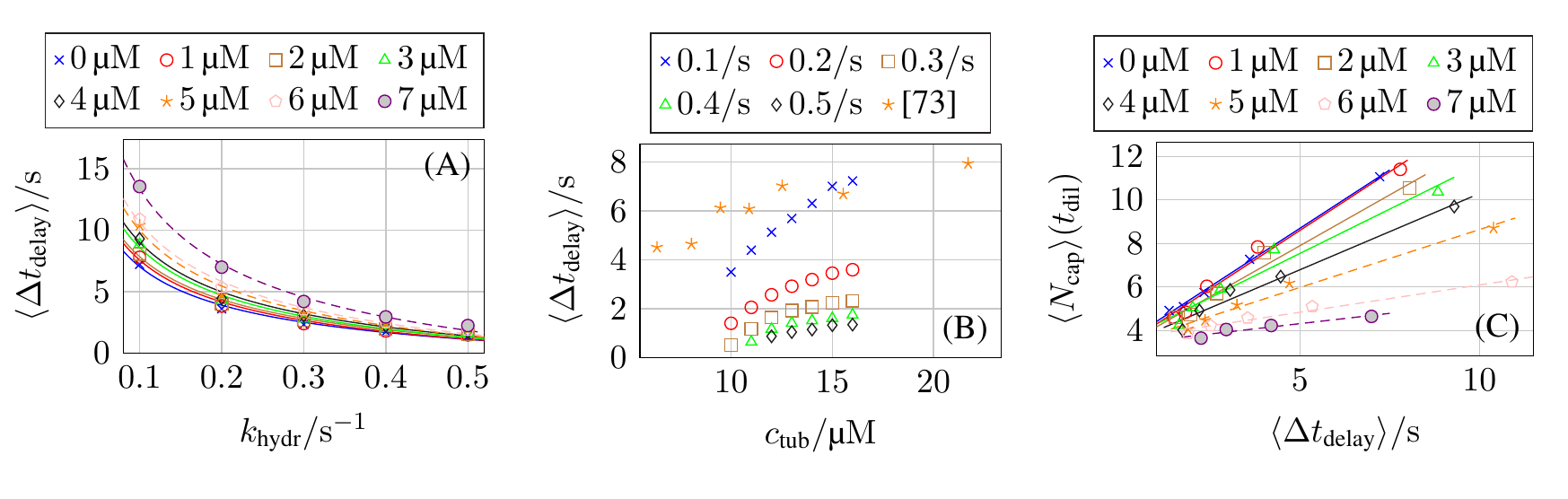}
        \caption{
Average post-dilution delay time $\langle \Dtdelay\rangle$
as a function of
(A) the hydrolysis rate $\khydr$ for $\ctubVal{16}$ and different
post-dilution GTP-tubulin dimer concentrations $\cdil$ and
(B) the pre-dilution GTP-tubulin dimer concentration $\ctub$ for
 $\cdil = \SI{0}{\micro\molar}$ and different hydrolysis rates $\khydr$.
The averaged data from Duellberg \etal~\cite{Duellberg2016}
specified the pre-dilution growth velocity,
which was converted to $\ctub$ for this plot.
(C) Average GTP-cap length $\langle \Ncap \rangle$ at the time of dilution
$\tdil$ as a function of the delay time $\langle \Dtdelay\rangle$ for
$\ctubVal{16}$ and different values of $\cdil$ .
}
        \label{fig:dilution}
\end{figure}

We expect the delay time to be  proportional to the
GTP-cap length, $\Dtdelay \propto \langle \Ncap\rangle$, as
corroborated by \autoref{fig:dilution}(C) and
$\langle \Ncap\rangle\propto
\sqrt{\ctub/\khydr}$  according  to  
Section \ref{sec:vgro_of_ctub_with_hydrolysis} 
(see \autoref{fig:vgro_of_ctub_with_hydrolysis}(C) and (D)) \cite{Li2010}.
This results in  $\Dtdelay \propto\sqrt{\ctub/\khydr}$,
which is in qualitative agreement with our simulation data
in \autoref{fig:dilution}(A) and (B).
The comparison with the experimental dilution data from Ref.\
\cite{Duellberg2016} in \autoref{fig:dilution}(B) shows that delay
times for a hydrolysis rate $\khydrVal{0.1}$
come close to the experimental data but appear to depend too steeply
on $\ctub$.

\section{Discussion}

We introduced, parameterized, and analyzed a chemomechanical model for MT
dynamics in which, in addition to polymerization (attachment of dimers),
depolymerization (detachment of dimers), and hydrolysis of dimers, the rupture
of lateral bonds between monomers in neighboring
protofilaments is explicitly modeled and coupled to the mechanics of the MT.
The basis for this coupling is the allosteric model according to which a
hydrolyzed dimer acquires a more bent configuration, which builds up mechanical
stress in the MT tubular structue via lateral bonds between dimers.

As many model parameters as possible have been determined from the
experimentally measured MT growth and shrinkage velocities measured by Walker
\etal~\cite{Walker1988}.
To determine the values of the model parameters, we use a \enquote{divide and
conquer} approach \cite{VanBuren2002,VanBuren2005}.
We used simulations of growing GTP-only MTs to parameterize longitudinal and
lateral bond energies $\Glong$ and $\Glat$ and the attempt rate $\katt$ for
lateral bond formation.
By requiring a linear concentration dependence of growth
velocity, we can fix all three parameter values for a given value of $\konc$.
We used simulations of shrinking GDP-only MTs to parameterize the bending
constant $\kcurl$ and the spring constant $\klat$ of the lateral bonds.
Here, we can only fix one of the two parameters.
Moreover, the hydrolysis rate $\khydr$ is still a free parameter, for
which we use values in the range
\SIrange[per-mode=reciprocal]{0.1}{0.5}{\per\second} known from experiments
\cite{Melki1996}.

The general philosophy of a divide-and-conquer approach is the successive
fixation of simulation parameters by using first GTP-only growth, then GDP-only
shrinkage and, eventually, catastrophe frequencies or dilution to fix the
hydrolysis rate.
This successive fixation is, however, problematic, as the corresponding
experimental data is influenced by \emph{all} simulation parameters in general.
The problem becomes apparent when considering the hydrolysis rate: changes in
the hydrolysis rate also affect the growth rate over a wide concentration range
because hydrolyzed dimers have an effectively higher
detachment rate, see \autoref{fig:vgro_of_ctub_with_hydrolysis}(B).
Strictly speaking, all simulation parameters in \autoref{tab:parameters} must be
determined at once by fitting several experimental results
simultaneously instead of the successive fixation in the divide-and-conquer
approach or to apply the divide-and-conquer approach iteratively several times
until a self-consistent parameter set is found.
A simultaneous fixation of all parameters has been performed, for example, in
Ref.\ \cite{Piette2009} on a chemical model without bond rupture and, thus, with
only four parameters (on-rate, bond energies, and hydrolysis rate).
Future work on our model should include at least a re-adjustment of the
parameters once a hydrolysis rate is selected such that the growth velocity of
Walker \etal~\cite{Walker1988} is reproduced \emph{in the presence of
hydrolysis}.
If mechanical feedback onto hydrolysis is included, the model has to be
re-parameterized again, in principle.

Our simulation model handles all chemical events, i.e., dimer attachment and
detachment, bond rupture and formation, and hydrolysis using a Gillespie
algorithm. After each chemical event we relax the resulting MT structure
mechanically by mechanical energy minimization based on the assumption that
the microscopic mechanical dynamics is much faster than the chemical steps.
Therefore, mechanical energy minimization is the computationally most demanding
step in the simulation.
This is a common problem in all dimer-based chemomechanical MT models
\cite{VanBuren2005,Coombes2013,Zakharov2015}.
We address this problem by restricting the mechanical energy minimization to
bounded number MT degrees of freedom near the plus end.
We showed that restricting energy minization to a depth of $\dcutoff = 10$
additional layers into the MT (in minus end direction)
from the point of the last chemical event
is an accurate and efficient choice.
Computational efficiency of this procedure is better than performing a
dedicated microscopic Brownian dynamics simulation \cite{Zakharov2015} and
better than random local energy minimization \cite{VanBuren2005,Coombes2013}
(for the same accuracy in energy minimization).
The restricted energy minimization strategy also ensures that the number of
minimization parameters does not scale with the MT length but remains bounded,
which assures that we can simulate arbitrarily long growing MTs at a fixed
minimal computational speed using our approach.

Simulations do not require more than a few hours for \SI{1}{\minute} of
MT dynamics (for a constant hydrolysis rate) using just a single CPU core.
Therefore, we can reach time scales of several minutes of MT dynamics
which is the time scale for repeated catastrophe events for concentrations above
the individual critical concentration, where the dynamic instability can occur.
We performed a first systematic analysis of catastrophe and rescue rates in
\autoref{fig:catastropheRescueRates}, which 
indicates that the decrease of the catastrophe rate
with tubulin concentration is too steep compared to experimental data
\cite{Walker1988,Janson2003}.
It is also much steeper than simulation results of Ref.\ \cite{Zakharov2015} but
these results for the catastrophe rate relied on linear extrapolation from
unrealistically high hydrolysis rates
(\SIrange[per-mode=reciprocal]{3}{11}{\per\second})
down to realistic values
(\SIrange[per-mode=reciprocal]{0.1}{0.5}{\per\second}).
In the future, our computational model can also be used to measure
 the dependence of catastrophe rates on MT lifetime \cite{Gardner2011_kinesins}.

Within our model, we could also study single catastrophe and rescue
events in detail, see \autoref{fig:full_simulations}.
The growth paths appear very similar to experimentally observed catastrophe
and rescue events.
Catastrophes typically feature an initial \enquote{transitional} phase of slow
shrinking by \SIrange{50}{100}{\nano\meter} as also observed in Ref.\ 
\cite{Mahserejian2019}.
Moreover, we observe \enquote{dips} in the growth paths resembling the
\enquote{stutter} events from Ref.\ \cite{Mahserejian2019}.

The most interesting results of chemomechanical models are possible statements
about the typical catastrophe-triggering configurations.
In this respect, our simulations indicate that a catastrophe could be triggered
by a \enquote{nucleus} of three neighboring protofilaments shrinking by more
than 6 dimers, such that its GTP-cap is removed and its ends reach into the
GDP-body of the MT.
To rescue a shrinking MT the GTP-cap has to be re-established on almost all 13
protofilaments such that nuclei of three neighboring uncapped GDP-protofilaments
are avoided.
This shows that mechanical correlations in the dynamics of protofilaments are
important in triggering catastrophe events.
This is an aspect which is absent in the calculation of catastrophe frequencies
based on simplified purely chemical models such as in Ref.\ 
\cite{Flyvbjerg1996}, where protofilaments are regarded as effectively
independent and uncorrelated.

Our model can achieve qualitative agreement with experimental
data on dilution experiments (see \autoref{fig:dilution}(C)) from Ref.\
\cite{Duellberg2016} for relatively low hydrolysis rates of
$\khydrVal{0.1}$, which is an indication that the catastrophe
mechanism is correctly captured by our chemomechanical model. This also
constrains  the hydrolysis rate, which is still a free parameter
in our model, to  lower values around $\khydrVal{0.1}$.

Finally, we
explored the consequences of a mechanochemical coupling in 
the hydrolysis of tubulin dimers.
Because hydrolysis gives rise to bending of the GTP-dimers, we argue that
mechanical forces on a dimer that increase its bending angle should also
lead to higher hydrolysis rates, see \eqref{eq:mechanics_hydrolysis_rate} and
\eqref{eq:mechanics_hydrolysis_rate2}.
In the presence of  mechanical feedback, hydrolysis gets a bias towards the
MT plus end which, in turn, also causes an increase in porous cap length.
At the same average hydrolysis rate,  hydrolysis in the
immediate tip of the GTP-cap is more likely while
it is less likely in the remaining part of the
cap such that GTP-tubulin dimers can be found much deeper in the GDP-body,
see \autoref{fig:mechanical_hydrolysis_results}.
Individual catastrophe and rescue events (see
\autoref{fig:full_simulations_mechanical_hydrolysis})
look qualitatively similar in
the presence of mechanical feedback but the probability of \enquote{dips} or
\enquote{stutter} events is increased in agreement with Ref.\ 
\cite{Mahserejian2019}.
The coupling of hydrolysis to mechanics does not increase catastrophe rates
significantly such that the steep decrease of the catastrophe rate with tubulin
concentration persists.

The main problem of our model appears to be the steep decrease of catastrophe
rate with tubulin concentration, which could hint to a failure of basic
assumptions.
One possibility is that a direct effect of the hydrolysis state of the dimer
onto the off-rate (as also suggested by atomistic simulations
\cite{Grafmueller2013}) is relevant and not included in the model.
Another possibility is a failure of allosteric models in general.
   The steep decline of catastrophe rates with tubulin concentration
  gives a hint that MTs are structurally too stable for GTP-rich
  caps. This might provide evidence for a shortcoming
  of the underlying allosteric model, which inserts GTP-dimers
  in a straight configuration that is  more prone to
  form stable lateral bonds than a curved configuration.
An alternative are so-called lattice models \cite{Buey2006,Rice2008},
according to which dimers are
always bent but hydrolysis affects lateral and longitudinal dimer interaction
energies.
A systematic comparison of allosteric and lattice models towards the resulting
concentration dependence of catastrophe rates within the framework provided here
 could help decide which class of models is more appropriate.

 So far, almost all chemomechanical modelling approaches were based on
  the allosteric model
  \cite{Molodtsov2005,VanBuren2005,Coombes2013,Mueller2014,Zakharov2015,Jain2015} but
  recent experimental advancements in the analysis of the structure of
  MT tips \cite{McIntosh2018} demonstrated that both
  growing and shrinking MTs have
  bent protofilament ends supporting similar earlier results
  \cite{Hoog2011,Kukulski2011,Nawrotek2011,Pecqueur2012}.  Additionally,
  calculations using MT structures with different nucleotide content in the
  beta-tubulin \cite{Manka2018} and all-atom MD
  simulations of
  GTP- and GDP-only MTs \cite{Ayoub2015,Fedorov2019}
  revealed that hydrolysis weakens
  lateral bonds and strengthens longitudinal bonds.  Both aspects
  support the  lattice model for the influence of hydrolysis
  on MT mechanics.
  There is, however, also evidence from MD simulations for
  intermediate models, where hydrolysis
   affects interactions and also
  leads to a  much lower GDP-tubulin flexibility \cite{Igaev2018}.
 Independent of  these findings, our study based on the
  allosteric model is valuable 
   for the following reasons:
   (i)
   In both the allosteric and the lattice model, catastrophes are
   cascades of lateral bond rupture and in both models,
   the bent shape of GDP-dimers is the dominating cause of mechanical
   strain  in the MT structure.
   In the allosteric model, bending and mechanical strain
     is directly generated by the hydrolysis of GTP-dimers, 
   whereas in the lattice model, the tubulin dimers are always bent
   but hydrolysis weakens lateral bonds.
   In both models, the result is an increased lateral bond rupture
     rate of mechanically strained bonds after  hydrolysis. 
   Therefore, an explicit modelling
   approach for lateral bond rupture as a stoachstic process under
   force generated by the bending of GDP-dimers will also be important
   in all future chemomechanical models based on the lattice model.
   So far explicit stochastic models of  lateral bond rupture
   have only been included into  two-dimensional models
   lacking explicit  mechanics \cite{Margolin2011,Margolin2012,Li2014}
   or with heavy computational cost by explicitly simulating the Brownian
   dynamics of dimers and bonds \cite{Zakharov2015}.
   (ii)
  The importance of lateral bond rupture 
   becomes particularly clear for shrinking MTs or MTs
   entering a catastrophe. In these phases of the dynamic instability,
   GDP-tubulin dimers are
   significantly more relavant than GTP-tubulin dimers.
   As GDP-dimers are bent in both models and this bending gives rise
   to lateral bond stretching, we believe that both types
   of models will display a very similar behavior in these phases. 
  The only difference in this scenario is
  that in the lattice model, the lateral bond energy $\Glat$ in the rupture
  rate \eqref{eq:krup} will depend  on the nucleotide type of the bonded tubulin
  monomers, which also makes $\krup$  an explicit function of the nucleotide
  state. Because the nucleotide state is predominantly GDP, 
  the results for properly parameterized models will be very similar. 
  (iii)
  We also introduced a computationally efficient scheme to relax the
  mechanical energy between chemical events, which can also be employed
  in  future  chemomechanical lattice models.
  Within the allosteric model, we 
  achieve a better mechanical energy relaxation than
  previous models  \cite{VanBuren2005}
  with significantly less computational steps
  than a full Brownian dynamics simulation requires \cite{Zakharov2015}.
  (iv) The idea of a feedback of mechanical forces onto the
  hydrolysis rate can also be applied in future  chemomechanical lattice
  models: if hydrolysis leads to a weakening of lateral  bonds, one could
  expect mechanical strains that favor weakening of lateral bonds
  also to favor hydrolysis. 

In the future, our model could be extended to also include regulating TIP+
proteins \cite{Akhmanova2008} for which different mechanisms of how they
influence MTs could be implemented.
Comparing the results of such simulations with experimental data could help
to develop a mechanistic picture of the action of these proteins.
Another future extension is MT polymerization under force
\cite{Dogterom1997,Dogterom2005}.
So far, polymerization under force has been investigated  using chemical models
\cite{VanDoorn2000,Kolomeisky2001,Stukalin2004,Ranjith2009,Krawczyk2011};
the influence of an external force on the microscopic level, in particular the
detailed dynamics of catastrophe events and the catastrophe-triggering
configurations is unknown.

\section{Funding}

We acknowledge funding from the German Research Foundation (DFG, www.dfg.de)
through Grant number KI 662/9-1.
The authors gratefully acknowledge computing time provided on
the Linux HPC cluster at Technical University Dortmund (LiDO3),
partially funded in the course of the Large-Scale Equipment
Initiative by the German Research Foundation (DFG) as project
271512359.

\bibliographystyle{naturemag-doi}
\bibliography{paper}

\begin{thebibliography}{10}
\urlstyle{rm}
\expandafter\ifx\csname url\endcsname\relax
  \def\url#1{\texttt{#1}}\fi
\expandafter\ifx\csname urlprefix\endcsname\relax\def\urlprefix{URL }\fi
\expandafter\ifx\csname doiprefix\endcsname\relax\def\doiprefix{DOI: }\fi
\providecommand{\bibinfo}[2]{#2}
\providecommand{\eprint}[2][]{\url{#2}}

\bibitem{McIntosh2002}
\bibinfo{author}{McIntosh, J.~R.}, \bibinfo{author}{Grishchuk, E.~L.} \&
  \bibinfo{author}{West, R.~R.}
\newblock \bibinfo{journal}{\bibinfo{title}{{Chromosome-Microtubule
  Interactions During Mitosis}}}.
\newblock {\emph{\JournalTitle{Annu. Rev. Cell Dev. Biol.}}}
  \textbf{\bibinfo{volume}{18}}, \bibinfo{pages}{193--219},
  \doiprefix\url{10.1146/annurev.cellbio.18.032002.132412}
  (\bibinfo{year}{2002}).

\bibitem{Siegrist2007}
\bibinfo{author}{Siegrist, S.~E.} \& \bibinfo{author}{Doe, C.~Q.}
\newblock \bibinfo{journal}{\bibinfo{title}{Microtubule-induced cortical cell
  polarity}}.
\newblock {\emph{\JournalTitle{Genes Dev.}}} \textbf{\bibinfo{volume}{21}},
  \bibinfo{pages}{483--496}, \doiprefix\url{10.1101/gad.1511207}
  (\bibinfo{year}{2007}).

\bibitem{Mitchison1984}
\bibinfo{author}{Mitchison, T.} \& \bibinfo{author}{Kirschner, M.}
\newblock \bibinfo{journal}{\bibinfo{title}{Dynamic instability of microtubule
  growth}}.
\newblock {\emph{\JournalTitle{Nature}}} \textbf{\bibinfo{volume}{312}},
  \bibinfo{pages}{237--242}, \doiprefix\url{10.1038/312237a0}
  (\bibinfo{year}{1984}).

\bibitem{Carlier1984}
\bibinfo{author}{Carlier, M.-F.}, \bibinfo{author}{Hill, T.~L.} \&
  \bibinfo{author}{Chen, Y.-d.}
\newblock \bibinfo{journal}{\bibinfo{title}{{Interference of GTP hydrolysis in
  the mechanism of microtubule assembly: an experimental study}}}.
\newblock {\emph{\JournalTitle{Proc. Natl. Acad. Sci. USA}}}
  \textbf{\bibinfo{volume}{81}}, \bibinfo{pages}{771--775},
  \doiprefix\url{10.1073/pnas.81.3.771} (\bibinfo{year}{1984}).

\bibitem{Walker1991}
\bibinfo{author}{Walker, R.~A.}, \bibinfo{author}{Pryer, N.~K.} \&
  \bibinfo{author}{Salmon, E.~D.}
\newblock \bibinfo{journal}{\bibinfo{title}{Dilution of individual microtubules
  observed in real time in vitro: evidence that cap size is small and
  independent of elongation rate.}}
\newblock {\emph{\JournalTitle{J. Cell Biol}}} \textbf{\bibinfo{volume}{114}},
  \bibinfo{pages}{73--81}, \doiprefix\url{10.1083/jcb.114.1.73}
  (\bibinfo{year}{1991}).

\bibitem{Howard2009}
\bibinfo{author}{Howard, J.} \& \bibinfo{author}{Hyman, A.~A.}
\newblock \bibinfo{journal}{\bibinfo{title}{Growth, fluctuation and switching
  at microtubule plus ends}}.
\newblock {\emph{\JournalTitle{Nat. Rev. Mol. Cell Biol.}}}
  \textbf{\bibinfo{volume}{10}}, \bibinfo{pages}{569--574},
  \doiprefix\url{10.1038/nrm2713} (\bibinfo{year}{2009}).

\bibitem{VanHaren2019}
\bibinfo{author}{van Haren, J.} \& \bibinfo{author}{Wittmann, T.}
\newblock \bibinfo{journal}{\bibinfo{title}{{Microtubule Plus End Dynamics --
  Do We Know How Microtubules Grow?}}}
\newblock {\emph{\JournalTitle{Bioessays}}} \textbf{\bibinfo{volume}{41}},
  \bibinfo{pages}{1800194}, \doiprefix\url{10.1002/bies.201800194}
  (\bibinfo{year}{2019}).

\bibitem{Mandelkow1991}
\bibinfo{author}{Mandelkow, E.~M.}, \bibinfo{author}{Mandelkow, E.} \&
  \bibinfo{author}{Milligan, R.~A.}
\newblock \bibinfo{journal}{\bibinfo{title}{{Microtubule Dynamics and
  Microtubule Caps: A Time-resolved Cryo-Electron Microscopy Study}}}.
\newblock {\emph{\JournalTitle{J. Cell Biol.}}} \textbf{\bibinfo{volume}{114}},
  \bibinfo{pages}{977--991}, \doiprefix\url{10.1083/jcb.114.5.977}
  (\bibinfo{year}{1991}).

\bibitem{McIntosh2010}
\bibinfo{author}{McIntosh, J.~R.}, \bibinfo{author}{Volkov, V.},
  \bibinfo{author}{Ataullakhanov, F.~I.} \& \bibinfo{author}{Grishchuk, E.~L.}
\newblock \bibinfo{journal}{\bibinfo{title}{Tubulin depolymerization may be an
  ancient biological motor}}.
\newblock {\emph{\JournalTitle{J. Cell Sci.}}} \textbf{\bibinfo{volume}{123}},
  \bibinfo{pages}{3425--3434}, \doiprefix\url{10.1242/jcs.067611}
  (\bibinfo{year}{2010}).

\bibitem{Mueller-Reichert1998}
\bibinfo{author}{Müller-Reichert, T.}, \bibinfo{author}{Chrétien, D.},
  \bibinfo{author}{Severin, F.} \& \bibinfo{author}{Hyman, A.~A.}
\newblock \bibinfo{journal}{\bibinfo{title}{{Structural changes at microtubule
  ends accompanying GTP hydrolysis: Information from a slowly hydrolyzable
  analogue of GTP, guanylyl ($\alpha$,$\beta$)methylenediphosphonate}}}.
\newblock {\emph{\JournalTitle{Proc. Natl. Acad. Sci. USA}}}
  \textbf{\bibinfo{volume}{95}}, \bibinfo{pages}{3661--3666}
  (\bibinfo{year}{1998}).

\bibitem{Downing1998}
\bibinfo{author}{Downing, K.~H.} \& \bibinfo{author}{Eva, N.}
\newblock \bibinfo{journal}{\bibinfo{title}{Tubulin and microtubule
  structure}}.
\newblock {\emph{\JournalTitle{Curr. Opin. Cell Biol.}}}
  \textbf{\bibinfo{volume}{10}}, \bibinfo{pages}{16--22},
  \doiprefix\url{10.1016/S0955-0674(98)80082-3} (\bibinfo{year}{1998}).

\bibitem{Nogales2006}
\bibinfo{author}{Nogales, E.} \& \bibinfo{author}{Wang, H.-W.}
\newblock \bibinfo{journal}{\bibinfo{title}{{Structural mechanisms underlying
  nucleotide-dependent self-assembly of tubulin and its relatives.}}}
\newblock {\emph{\JournalTitle{Curr. Opin. Struct. Biol.}}}
  \textbf{\bibinfo{volume}{16}}, \bibinfo{pages}{221--229},
  \doiprefix\url{10.1016/j.sbi.2006.03.005} (\bibinfo{year}{2006}).

\bibitem{Nogales1999}
\bibinfo{author}{Nogales, E.}, \bibinfo{author}{Whittaker, M.},
  \bibinfo{author}{Milligan, R.~A.} \& \bibinfo{author}{Downing, K.~H.}
\newblock \bibinfo{journal}{\bibinfo{title}{{High-Resolution Model of the
  Microtubule}}}.
\newblock {\emph{\JournalTitle{Cell}}} \textbf{\bibinfo{volume}{96}},
  \bibinfo{pages}{79--88}, \doiprefix\url{10.1016/S0092-8674(00)80961-7}
  (\bibinfo{year}{1999}).

\bibitem{Molodtsov2005}
\bibinfo{author}{Molodtsov, M.~I.} \emph{et~al.}
\newblock \bibinfo{journal}{\bibinfo{title}{{A Molecular-Mechanical Model of
  the Microtubule}}}.
\newblock {\emph{\JournalTitle{Biophys. J.}}} \textbf{\bibinfo{volume}{88}},
  \bibinfo{pages}{3167--3179}, \doiprefix\url{10.1529/biophysj.104.051789}
  (\bibinfo{year}{2005}).

\bibitem{VanBuren2005}
\bibinfo{author}{VanBuren, V.}, \bibinfo{author}{Cassimeris, L.} \&
  \bibinfo{author}{Odde, D.~J.}
\newblock \bibinfo{journal}{\bibinfo{title}{{Mechanochemical Model of
  Microtubule Structure and Self-Assembly Kinetics}}}.
\newblock {\emph{\JournalTitle{Biophys. J.}}} \textbf{\bibinfo{volume}{89}},
  \bibinfo{pages}{2911--2926}, \doiprefix\url{10.1529/biophysj.105.060913}
  (\bibinfo{year}{2005}).

\bibitem{Coombes2013}
\bibinfo{author}{Coombes, C.}, \bibinfo{author}{Yamamoto, A.},
  \bibinfo{author}{Kenzie, M.}, \bibinfo{author}{Odde, D.} \&
  \bibinfo{author}{Gardner, M.}
\newblock \bibinfo{journal}{\bibinfo{title}{{Evolving Tip Structures Can
  Explain Age-Dependent Microtubule Catastrophe}}}.
\newblock {\emph{\JournalTitle{Curr. Biol.}}} \textbf{\bibinfo{volume}{23}},
  \bibinfo{pages}{1342--1348}, \doiprefix\url{10.1016/j.cub.2013.05.059}
  (\bibinfo{year}{2013}).

\bibitem{Mueller2014}
\bibinfo{author}{Müller, N.} \& \bibinfo{author}{Kierfeld, J.}
\newblock \bibinfo{journal}{\bibinfo{title}{Effects of microtubule mechanics on
  hydrolysis and catastrophes}}.
\newblock {\emph{\JournalTitle{Phys. Biol.}}} \textbf{\bibinfo{volume}{11}},
  \bibinfo{pages}{046001}, \doiprefix\url{10.1088/1478-3975/11/4/046001}
  (\bibinfo{year}{2014}).

\bibitem{Zakharov2015}
\bibinfo{author}{Zakharov, P.} \emph{et~al.}
\newblock \bibinfo{journal}{\bibinfo{title}{{Molecular and Mechanical Causes of
  Microtubule Catastrophe and Aging}}}.
\newblock {\emph{\JournalTitle{Biophys. J.}}} \textbf{\bibinfo{volume}{109}},
  \bibinfo{pages}{2574--2591}, \doiprefix\url{10.1016/j.bpj.2015.10.048}
  (\bibinfo{year}{2015}).

\bibitem{Jain2015}
\bibinfo{author}{Jain, I.}, \bibinfo{author}{Inamdar, M.~M.} \&
  \bibinfo{author}{Padinhateeri, R.}
\newblock \bibinfo{journal}{\bibinfo{title}{{Statistical Mechanics Provides
  Novel Insights into Microtubule Stability and Mechanism of Shrinkage}}}.
\newblock {\emph{\JournalTitle{PLoS Comput. Biol.}}}
  \textbf{\bibinfo{volume}{11}}, \bibinfo{pages}{1--23},
  \doiprefix\url{10.1371/journal.pcbi.1004099} (\bibinfo{year}{2015}).

\bibitem{Buey2006}
\bibinfo{author}{Buey, R.~M.}, \bibinfo{author}{Díaz, J.~F.} \&
  \bibinfo{author}{Andreu, J.~M.}
\newblock \bibinfo{journal}{\bibinfo{title}{{The Nucleotide Switch of Tubulin
  and Microtubule Assembly: A Polymerization-Driven Structural Change}}}.
\newblock {\emph{\JournalTitle{Biochemistry}}} \textbf{\bibinfo{volume}{45}},
  \bibinfo{pages}{5933--5938}, \doiprefix\url{10.1021/bi060334m}
  (\bibinfo{year}{2006}).

\bibitem{Rice2008}
\bibinfo{author}{Rice, L.~M.}, \bibinfo{author}{Montabana, E.~A.} \&
  \bibinfo{author}{Agard, D.~A.}
\newblock \bibinfo{journal}{\bibinfo{title}{{The lattice as allosteric
  effector: Structural studies of $\alpha\beta$- and $\gamma$-tubulin clarify
  the role of GTP in microtubule assembly}}}.
\newblock {\emph{\JournalTitle{Proc. Natl. Acad. Sci. USA}}}
  \textbf{\bibinfo{volume}{105}}, \bibinfo{pages}{5378--5383},
  \doiprefix\url{10.1073/pnas.0801155105} (\bibinfo{year}{2008}).

\bibitem{Alushin2014}
\bibinfo{author}{Alushin, G.~M.} \emph{et~al.}
\newblock \bibinfo{journal}{\bibinfo{title}{{High-Resolution microtubule
  structures reveal the structural transitions in $\alpha$$\beta$-tubulin upon
  GTP hydrolysis}}}.
\newblock {\emph{\JournalTitle{Cell}}} \textbf{\bibinfo{volume}{157}},
  \bibinfo{pages}{1117--1129}, \doiprefix\url{10.1016/j.cell.2014.03.053}
  (\bibinfo{year}{2014}).

\bibitem{Manka2018}
\bibinfo{author}{Manka, S.~W.} \& \bibinfo{author}{Moores, C.~A.}
\newblock \bibinfo{journal}{\bibinfo{title}{The role of tubulin–tubulin
  lattice contacts in the mechanism of microtubule dynamic instability}}.
\newblock {\emph{\JournalTitle{Nat. Struct. Mol. Biol.}}}
  \textbf{\bibinfo{volume}{25}}, \bibinfo{pages}{607--615},
  \doiprefix\url{10.1038/s41594-018-0087-8} (\bibinfo{year}{2018}).

\bibitem{Ayoub2015}
\bibinfo{author}{Ayoub, A.~T.}, \bibinfo{author}{Klobukowski, M.} \&
  \bibinfo{author}{Tuszynski, J.~A.}
\newblock \bibinfo{journal}{\bibinfo{title}{Detailed per-residue energetic
  analysis explains the driving force for microtubule disassembly}}.
\newblock {\emph{\JournalTitle{PLoS Comput. Biol.}}}
  \textbf{\bibinfo{volume}{11}}, \bibinfo{pages}{1--21},
  \doiprefix\url{10.1371/journal.pcbi.1004313} (\bibinfo{year}{2015}).

\bibitem{Fedorov2019}
\bibinfo{author}{Fedorov, V.~A.} \emph{et~al.}
\newblock \bibinfo{journal}{\bibinfo{title}{{Mechanical properties of tubulin
  intra- and inter-dimer interfaces and their implications for microtubule
  dynamic instability}}}.
\newblock {\emph{\JournalTitle{PLOS Comput. Biol.}}}
  \textbf{\bibinfo{volume}{15}}, \bibinfo{pages}{e1007327},
  \doiprefix\url{10.1371/journal.pcbi.1007327} (\bibinfo{year}{2019}).

\bibitem{McIntosh2018}
\bibinfo{author}{McIntosh, J.~R.} \emph{et~al.}
\newblock \bibinfo{journal}{\bibinfo{title}{Microtubules grow by the addition
  of bent guanosine triphosphate tubulin to the tips of curved
  protofilaments}}.
\newblock {\emph{\JournalTitle{J. Cell Biol.}}} \textbf{\bibinfo{volume}{217}},
  \bibinfo{pages}{2691--2708}, \doiprefix\url{10.1083/jcb.201802138}
  (\bibinfo{year}{2018}).

\bibitem{Igaev2018}
\bibinfo{author}{Igaev, M.} \& \bibinfo{author}{Grubm{\"{u}}ller, H.}
\newblock \bibinfo{journal}{\bibinfo{title}{{Microtubule assembly governed by
  tubulin allosteric gain in flexibility and lattice induced fit}}}.
\newblock {\emph{\JournalTitle{Elife}}} \textbf{\bibinfo{volume}{7}},
  \bibinfo{pages}{e34353}, \doiprefix\url{10.7554/eLife.34353}
  (\bibinfo{year}{2018}).

\bibitem{Zakharov2016}
\bibinfo{author}{Zakharov, P.~N.}, \bibinfo{author}{Arzhanik, V.~K.},
  \bibinfo{author}{Ulyanov, E.~V.}, \bibinfo{author}{Gudimchuk, N.~B.} \&
  \bibinfo{author}{Ataullakhanov, F.~I.}
\newblock \bibinfo{journal}{\bibinfo{title}{Microtubules: dynamically unstable
  stochastic phase-switching polymers}}.
\newblock {\emph{\JournalTitle{Physics-Uspekhi}}}
  \textbf{\bibinfo{volume}{59}}, \bibinfo{pages}{773--786},
  \doiprefix\url{10.3367/ufne.2016.04.037779} (\bibinfo{year}{2016}).

\bibitem{Dogterom1993}
\bibinfo{author}{Dogterom, M.} \& \bibinfo{author}{Leibler, S.}
\newblock \bibinfo{journal}{\bibinfo{title}{Physical aspects of the growth and
  regulation of microtubule structures}}.
\newblock {\emph{\JournalTitle{Phys. Rev. Lett.}}}
  \textbf{\bibinfo{volume}{70}}, \bibinfo{pages}{1347--1350},
  \doiprefix\url{10.1103/PhysRevLett.70.1347} (\bibinfo{year}{1993}).

\bibitem{VanBuren2002}
\bibinfo{author}{VanBuren, V.}, \bibinfo{author}{Odde, D.~J.} \&
  \bibinfo{author}{Cassimeris, L.}
\newblock \bibinfo{journal}{\bibinfo{title}{Estimates of lateral and
  longitudinal bond energies within the microtubule lattice}}.
\newblock {\emph{\JournalTitle{Proc. Natl. Acad. Sci. USA}}}
  \textbf{\bibinfo{volume}{99}}, \bibinfo{pages}{6035--6040},
  \doiprefix\url{10.1073/pnas.092504999} (\bibinfo{year}{2002}).

\bibitem{Piette2009}
\bibinfo{author}{Piette, B. M. A.~G.} \emph{et~al.}
\newblock \bibinfo{journal}{\bibinfo{title}{{A Thermodynamic Model of
  Microtubule Assembly and Disassembly}}}.
\newblock {\emph{\JournalTitle{PLoS ONE}}} \textbf{\bibinfo{volume}{4}},
  \bibinfo{pages}{1--11}, \doiprefix\url{10.1371/journal.pone.0006378}
  (\bibinfo{year}{2009}).

\bibitem{Margolin2011}
\bibinfo{author}{Margolin, G.}, \bibinfo{author}{Goodson, H.~V.} \&
  \bibinfo{author}{Alber, M.~S.}
\newblock \bibinfo{journal}{\bibinfo{title}{Mean-field study of the role of
  lateral cracks in microtubule dynamics}}.
\newblock {\emph{\JournalTitle{Phys. Rev. E}}} \textbf{\bibinfo{volume}{83}},
  \bibinfo{pages}{041905}, \doiprefix\url{10.1103/PhysRevE.83.041905}
  (\bibinfo{year}{2011}).

\bibitem{Margolin2012}
\bibinfo{author}{Margolin, G.} \emph{et~al.}
\newblock \bibinfo{journal}{\bibinfo{title}{The mechanisms of microtubule
  catastrophe and rescue: implications from analysis of a dimer-scale
  computational model}}.
\newblock {\emph{\JournalTitle{Mol. Biol. Cell}}}
  \textbf{\bibinfo{volume}{23}}, \bibinfo{pages}{642--656},
  \doiprefix\url{10.1091/mbc.E11-08-0688} (\bibinfo{year}{2012}).

\bibitem{Li2014}
\bibinfo{author}{Li, C.}, \bibinfo{author}{Li, J.}, \bibinfo{author}{Goodson,
  H.~V.} \& \bibinfo{author}{Alber, M.~S.}
\newblock \bibinfo{journal}{\bibinfo{title}{Microtubule dynamic instability:
  the role of cracks between protofilaments}}.
\newblock {\emph{\JournalTitle{Soft Matter}}} \textbf{\bibinfo{volume}{10}},
  \bibinfo{pages}{2069--2080}, \doiprefix\url{10.1039/C3SM52892H}
  (\bibinfo{year}{2014}).

\bibitem{Mahserejian2019}
\bibinfo{author}{Mahserejian, S.~M.} \emph{et~al.}
\newblock \bibinfo{journal}{\bibinfo{title}{{Stutter: a Transient Dynamic
  Instability Phase that is Strongly Associated with Catastrophe}}}.
\newblock {\emph{\JournalTitle{bioRxiv}}}
  \doiprefix\url{10.1101/2019.12.16.878603} (\bibinfo{year}{2019}).

\bibitem{Wang2005}
\bibinfo{author}{Wang, H.-W.} \& \bibinfo{author}{Nogales, E.}
\newblock \bibinfo{journal}{\bibinfo{title}{Nucleotide-dependent bending
  flexibility of tubulin regulates microtubule assembly}}.
\newblock {\emph{\JournalTitle{Nature}}} \textbf{\bibinfo{volume}{435}},
  \bibinfo{pages}{911--915}, \doiprefix\url{10.1038/nature03606}
  (\bibinfo{year}{2005}).

\bibitem{VanBuren2004}
\bibinfo{author}{VanBuren, V.}, \bibinfo{author}{Odde, D.~J.} \&
  \bibinfo{author}{Cassimeris, L.}
\newblock \bibinfo{journal}{\bibinfo{title}{{errata for VanBuren et al.,
  Estimates of lateral and longitudinal bond energies within the microtubule
  lattice, PNAS 2002 99:6035-6040}}}.
\newblock {\emph{\JournalTitle{Proc. Natl. Acad. Sci. USA}}}
  \textbf{\bibinfo{volume}{101}}, \bibinfo{pages}{14989},
  \doiprefix\url{10.1073/pnas.0406393101} (\bibinfo{year}{2004}).

\bibitem{Harris2018}
\bibinfo{author}{Harris, B.~J.}, \bibinfo{author}{Ross, J.~L.} \&
  \bibinfo{author}{Hawkins, T.~L.}
\newblock \bibinfo{journal}{\bibinfo{title}{{Microtubule seams are not
  mechanically weak defects}}}.
\newblock {\emph{\JournalTitle{Phys. Rev. E}}} \textbf{\bibinfo{volume}{97}},
  \bibinfo{pages}{062408}, \doiprefix\url{10.1103/PhysRevE.97.062408}
  (\bibinfo{year}{2018}).

\bibitem{ElieCaille2007}
\bibinfo{author}{Elie-Caille, C.} \emph{et~al.}
\newblock \bibinfo{journal}{\bibinfo{title}{{Straight GDP-Tubulin
  Protofilaments Form in the Presence of Taxol}}}.
\newblock {\emph{\JournalTitle{Curr. Biol.}}} \textbf{\bibinfo{volume}{17}},
  \bibinfo{pages}{1765--1770}, \doiprefix\url{10.1016/j.cub.2007.08.063}
  (\bibinfo{year}{2007}).

\bibitem{Kroy1997}
\bibinfo{author}{Kroy, K.} \& \bibinfo{author}{Frey, E.}
\newblock \bibinfo{journal}{\bibinfo{title}{{Dynamic scattering from solutions
  of semiflexible polymers}}}.
\newblock {\emph{\JournalTitle{Phys. Rev. E}}} \textbf{\bibinfo{volume}{55}},
  \bibinfo{pages}{3092--3101}, \doiprefix\url{10.1103/PhysRevE.55.3092}
  (\bibinfo{year}{1997}).

\bibitem{Ulyanov2021}
\bibinfo{author}{Ulyanov, E.~V.}, \bibinfo{author}{Vinogradov, D.~S.},
  \bibinfo{author}{McIntosh, J.~R.} \& \bibinfo{author}{Gudimchuk, N.~B.}
\newblock \bibinfo{journal}{\bibinfo{title}{{Brownian dynamics simulation of
  protofilament relaxation during rapid freezing}}}.
\newblock {\emph{\JournalTitle{PLoS One}}} \textbf{\bibinfo{volume}{16}},
  \bibinfo{pages}{e0247022}, \doiprefix\url{10.1371/journal.pone.0247022}
  (\bibinfo{year}{2021}).

\bibitem{Gardner2011}
\bibinfo{author}{Gardner, M.} \emph{et~al.}
\newblock \bibinfo{journal}{\bibinfo{title}{{Rapid Microtubule Self-Assembly
  Kinetics}}}.
\newblock {\emph{\JournalTitle{Cell}}} \textbf{\bibinfo{volume}{146}},
  \bibinfo{pages}{582--592}, \doiprefix\url{10.1016/j.cell.2011.06.053}
  (\bibinfo{year}{2011}).

\bibitem{Kononova2014}
\bibinfo{author}{Kononova, O.} \emph{et~al.}
\newblock \bibinfo{journal}{\bibinfo{title}{{Tubulin Bond Energies and
  Microtubule Biomechanics Determined from Nanoindentation in Silico}}}.
\newblock {\emph{\JournalTitle{J. Am. Chem. Soc.}}}
  \textbf{\bibinfo{volume}{136}}, \bibinfo{pages}{17036--17045},
  \doiprefix\url{10.1021/ja506385p} (\bibinfo{year}{2014}).

\bibitem{Bell1978}
\bibinfo{author}{Bell, G.~I.}
\newblock \bibinfo{journal}{\bibinfo{title}{Models for the specific adhesion of
  cells to cells}}.
\newblock {\emph{\JournalTitle{Science}}} \textbf{\bibinfo{volume}{200}},
  \bibinfo{pages}{618--627}, \doiprefix\url{10.1126/science.347575}
  (\bibinfo{year}{1978}).

\bibitem{Evans1997}
\bibinfo{author}{Evans, E.} \& \bibinfo{author}{Ritchie, K.}
\newblock \bibinfo{journal}{\bibinfo{title}{{Dynamic Strength of Molecular
  Adhesion Bonds}}}.
\newblock {\emph{\JournalTitle{Biophys. J.}}} \textbf{\bibinfo{volume}{72}},
  \bibinfo{pages}{1541--1555}, \doiprefix\url{10.1016/S0006-3495(97)78802-7}
  (\bibinfo{year}{1997}).

\bibitem{Gillespie1976}
\bibinfo{author}{Gillespie, D.~T.}
\newblock \bibinfo{journal}{\bibinfo{title}{A general method for numerically
  simulating the stochastic time evolution of coupled chemical reactions}}.
\newblock {\emph{\JournalTitle{J. Comput. Phys.}}}
  \textbf{\bibinfo{volume}{22}}, \bibinfo{pages}{403--434},
  \doiprefix\url{10.1016/0021-9991(76)90041-3} (\bibinfo{year}{1976}).

\bibitem{Walker1988}
\bibinfo{author}{Walker, R.~A.} \emph{et~al.}
\newblock \bibinfo{journal}{\bibinfo{title}{Dynamic instability of individual
  microtubules analyzed by video light microscopy: rate constants and
  transition frequencies}}.
\newblock {\emph{\JournalTitle{J. Cell Biol.}}} \textbf{\bibinfo{volume}{107}},
  \bibinfo{pages}{1437--1448}, \doiprefix\url{10.1083/jcb.107.4.1437}
  (\bibinfo{year}{1988}).

\bibitem{OBrien1990}
\bibinfo{author}{O'Brien, E.~T.}, \bibinfo{author}{Salmon, E.~D.},
  \bibinfo{author}{Walker, R.~A.} \& \bibinfo{author}{Erickson, H.~P.}
\newblock \bibinfo{journal}{\bibinfo{title}{Effects of magnesium on the dynamic
  instability of individual microtubules}}.
\newblock {\emph{\JournalTitle{Biochemistry}}} \textbf{\bibinfo{volume}{29}},
  \bibinfo{pages}{6648--6656}, \doiprefix\url{10.1021/bi00480a014}
  (\bibinfo{year}{1990}).

\bibitem{Drechsel1992}
\bibinfo{author}{Drechsel, D.~N.}, \bibinfo{author}{Hyman, A.~A.},
  \bibinfo{author}{Cobb, M.~H.} \& \bibinfo{author}{Kirschner, M.~W.}
\newblock \bibinfo{journal}{\bibinfo{title}{Modulation of the dynamic
  instability of tubulin assembly by the microtubule-associated protein tau}}.
\newblock {\emph{\JournalTitle{Mol. Biol. Cell}}} \textbf{\bibinfo{volume}{3}},
  \bibinfo{pages}{1141--1154}, \doiprefix\url{10.1091/mbc.3.10.1141}
  (\bibinfo{year}{1992}).

\bibitem{Trinczek1993}
\bibinfo{author}{Trinczek, B.}, \bibinfo{author}{Marx, A.},
  \bibinfo{author}{Mandelkow, E.~M.}, \bibinfo{author}{Murphy, D.~B.} \&
  \bibinfo{author}{Mandelkow, E.}
\newblock \bibinfo{journal}{\bibinfo{title}{{Dynamics of microtubules from
  erythrocyte marginal bands}}}.
\newblock {\emph{\JournalTitle{Mol. Biol. Cell}}} \textbf{\bibinfo{volume}{4}},
  \bibinfo{pages}{323--335}, \doiprefix\url{10.1091/mbc.4.3.323}
  (\bibinfo{year}{1993}).

\bibitem{Chretien1995}
\bibinfo{author}{Chrétien, D.}, \bibinfo{author}{Fuller, S.~D.} \&
  \bibinfo{author}{Karsenti, E.}
\newblock \bibinfo{journal}{\bibinfo{title}{{Structure of Growing Microtubule
  Ends: Two-Dimensional Sheets Close Into Tubes at Variable Rates}}}.
\newblock {\emph{\JournalTitle{J. Cell Biol.}}} \textbf{\bibinfo{volume}{129}},
  \bibinfo{pages}{1311--1328}, \doiprefix\url{10.1083/jcb.129.5.1311}
  (\bibinfo{year}{1995}).

\bibitem{Pedigo2002}
\bibinfo{author}{Pedigo, S.} \& \bibinfo{author}{Williams, R.~C., Jr.}
\newblock \bibinfo{journal}{\bibinfo{title}{{Concentration Dependence of
  Variability in Growth Rates of Microtubules}}}.
\newblock {\emph{\JournalTitle{Biophys. J.}}} \textbf{\bibinfo{volume}{83}},
  \bibinfo{pages}{1809--1819}, \doiprefix\url{10.1016/S0006-3495(02)73946-5}
  (\bibinfo{year}{2002}).

\bibitem{GSL}
\bibinfo{author}{Galassi, M.} \emph{et~al.}
\newblock \bibinfo{title}{{GNU Scientific Library Reference Manual}}.

\bibitem{Ayaz2014}
\bibinfo{author}{Ayaz, P.} \emph{et~al.}
\newblock \bibinfo{journal}{\bibinfo{title}{A tethered delivery mechanism
  explains the catalytic action of a microtubule polymerase}}.
\newblock {\emph{\JournalTitle{eLife}}} \textbf{\bibinfo{volume}{3}},
  \bibinfo{pages}{e03069}, \doiprefix\url{10.7554/eLife.03069}
  (\bibinfo{year}{2014}).

\bibitem{Mickolajczyk2019}
\bibinfo{author}{Mickolajczyk, K.~J.}, \bibinfo{author}{Geyer, E.~A.},
  \bibinfo{author}{Kim, T.}, \bibinfo{author}{Rice, L.~M.} \&
  \bibinfo{author}{Hancock, W.~O.}
\newblock \bibinfo{journal}{\bibinfo{title}{Direct observation of individual
  tubulin dimers binding to growing microtubules}}.
\newblock {\emph{\JournalTitle{Proc. Natl. Acad. Sci. USA}}}
  \textbf{\bibinfo{volume}{116}}, \bibinfo{pages}{7314--7322},
  \doiprefix\url{10.1073/pnas.1815823116} (\bibinfo{year}{2019}).

\bibitem{Stukalin2004}
\bibinfo{author}{Stukalin, E.~B.} \& \bibinfo{author}{Kolomeisky, A.~B.}
\newblock \bibinfo{journal}{\bibinfo{title}{Simple growth models of rigid
  multifilament biopolymers}}.
\newblock {\emph{\JournalTitle{J. Chem. Phys.}}}
  \textbf{\bibinfo{volume}{121}}, \bibinfo{pages}{1097--1104},
  \doiprefix\url{10.1063/1.1759316} (\bibinfo{year}{2004}).

\bibitem{Sim2013}
\bibinfo{author}{Sim, H.} \& \bibinfo{author}{Sept, D.}
\newblock \bibinfo{journal}{\bibinfo{title}{Properties of microtubules with
  isotropic and anisotropic mechanics}}.
\newblock {\emph{\JournalTitle{Cell Mol. Bioeng.}}}
  \textbf{\bibinfo{volume}{6}}, \bibinfo{pages}{361--368},
  \doiprefix\url{10.1007/s12195-013-0302-y} (\bibinfo{year}{2013}).

\bibitem{Driver2017}
\bibinfo{author}{Driver, J.~W.}, \bibinfo{author}{Geyer, E.~A.},
  \bibinfo{author}{Bailey, M.~E.}, \bibinfo{author}{Rice, L.~M.} \&
  \bibinfo{author}{Asbury, C.~L.}
\newblock \bibinfo{journal}{\bibinfo{title}{Direct measurement of
  conformational strain energy in protofilaments curling outward from
  disassembling microtubule tips}}.
\newblock {\emph{\JournalTitle{eLife}}} \textbf{\bibinfo{volume}{6}},
  \bibinfo{pages}{e28433}, \doiprefix\url{10.7554/eLife.28433}
  (\bibinfo{year}{2017}).

\bibitem{Deriu2007}
\bibinfo{author}{Deriu, M.~A.}, \bibinfo{author}{Enemark, S.},
  \bibinfo{author}{Soncini, M.}, \bibinfo{author}{Montevecchi, F.~M.} \&
  \bibinfo{author}{Redaelli, A.}
\newblock \bibinfo{journal}{\bibinfo{title}{Tubulin: from atomistic structure
  to supramolecular mechanical properties}}.
\newblock {\emph{\JournalTitle{J. Mater. Sci.}}} \textbf{\bibinfo{volume}{42}},
  \bibinfo{pages}{8864--8872}, \doiprefix\url{10.1007/s10853-007-1784-6}
  (\bibinfo{year}{2007}).

\bibitem{Grafmueller2011}
\bibinfo{author}{Grafmüller, A.} \& \bibinfo{author}{Voth, G.}
\newblock \bibinfo{journal}{\bibinfo{title}{Intrinsic bending of microtubule
  protofilaments}}.
\newblock {\emph{\JournalTitle{Structure}}} \textbf{\bibinfo{volume}{19}},
  \bibinfo{pages}{409--417}, \doiprefix\url{10.1016/j.str.2010.12.020}
  (\bibinfo{year}{2011}).

\bibitem{Melki1996}
\bibinfo{author}{Melki, R.}, \bibinfo{author}{Fievez, S.} \&
  \bibinfo{author}{Carlier, M.-F.}
\newblock \bibinfo{journal}{\bibinfo{title}{{Continuous Monitoring of
  P${}_\text{i}$ Release Following Nucleotide Hydrolysis in Actin or Tubulin
  Assembly Using 2-Amino-6-mer-capto-7-methylpurine Ribonucleoside and
  Purine-Nucleoside Phosphorylase as an Enzyme-Linked Assay}}}.
\newblock {\emph{\JournalTitle{Biochemistry}}} \textbf{\bibinfo{volume}{35}},
  \bibinfo{pages}{12038--12045}, \doiprefix\url{10.1021/bi961325o}
  (\bibinfo{year}{1996}).

\bibitem{Aparna2017}
\bibinfo{author}{Aparna, J.~S.}, \bibinfo{author}{Padinhateeri, R.} \&
  \bibinfo{author}{Das, D.}
\newblock \bibinfo{journal}{\bibinfo{title}{Signatures of a macroscopic
  switching transition for a dynamic microtubule}}.
\newblock {\emph{\JournalTitle{Sci. Rep.}}} \textbf{\bibinfo{volume}{7}},
  \bibinfo{pages}{45747}, \doiprefix\url{10.1038/srep45747}
  (\bibinfo{year}{2017}).

\bibitem{Padinhateeri2012}
\bibinfo{author}{Padinhateeri, R.}, \bibinfo{author}{Kolomeisky, A.} \&
  \bibinfo{author}{Lacoste, D.}
\newblock \bibinfo{journal}{\bibinfo{title}{{Random Hydrolysis Controls the
  Dynamic Instability of Microtubules}}}.
\newblock {\emph{\JournalTitle{Biophys. J.}}} \textbf{\bibinfo{volume}{102}},
  \bibinfo{pages}{1274--1283}, \doiprefix\url{10.1016/j.bpj.2011.12.059}
  (\bibinfo{year}{2012}).

\bibitem{Bowne-Anderson2013}
\bibinfo{author}{Bowne-Anderson, H.}, \bibinfo{author}{Zanic, M.},
  \bibinfo{author}{Kauer, M.} \& \bibinfo{author}{Howard, J.}
\newblock \bibinfo{journal}{\bibinfo{title}{{Microtubule dynamic instability: A
  new model with coupled GTP hydrolysis and multistep catastrophe}}}.
\newblock {\emph{\JournalTitle{Bioessays}}} \textbf{\bibinfo{volume}{35}},
  \bibinfo{pages}{452--461}, \doiprefix\url{10.1002/bies.201200131}
  (\bibinfo{year}{2013}).

\bibitem{Piedra2016}
\bibinfo{author}{Piedra, F.-A.} \emph{et~al.}
\newblock \bibinfo{journal}{\bibinfo{title}{{GDP-to-GTP exchange on the
  microtubule end can contribute to the frequency of catastrophe}}}.
\newblock {\emph{\JournalTitle{Molecular Biology of the Cell}}}
  \textbf{\bibinfo{volume}{27}}, \bibinfo{pages}{3515--3525},
  \doiprefix\url{10.1091/mbc.E16-03-0199} (\bibinfo{year}{2016}).

\bibitem{Gardner2011_kinesins}
\bibinfo{author}{Gardner, M.}, \bibinfo{author}{Zanic, M.},
  \bibinfo{author}{Gell, C.}, \bibinfo{author}{Bormuth, V.} \&
  \bibinfo{author}{Howard, J.}
\newblock \bibinfo{journal}{\bibinfo{title}{Depolymerizing kinesins kip3 and
  mcak shape cellular microtubule architecture by differential control of
  catastrophe}}.
\newblock {\emph{\JournalTitle{Cell}}} \textbf{\bibinfo{volume}{147}},
  \bibinfo{pages}{1092--1103}, \doiprefix\url{10.1016/j.cell.2011.10.037}
  (\bibinfo{year}{2011}).

\bibitem{Flyvbjerg1996}
\bibinfo{author}{Flyvbjerg, H.}, \bibinfo{author}{Holy, T.~E.} \&
  \bibinfo{author}{Leibler, S.}
\newblock \bibinfo{journal}{\bibinfo{title}{{Microtubule dynamics: Caps,
  catastrophes, and coupled hydrolysis}}}.
\newblock {\emph{\JournalTitle{Phys. Rev. E}}} \textbf{\bibinfo{volume}{54}},
  \bibinfo{pages}{5538--5560}, \doiprefix\url{10.1103/PhysRevE.54.5538}
  (\bibinfo{year}{1996}).

\bibitem{Zelinski2013}
\bibinfo{author}{Zelinski, B.} \& \bibinfo{author}{Kierfeld, J.}
\newblock \bibinfo{journal}{\bibinfo{title}{{Cooperative dynamics of
  microtubule ensembles: Polymerization forces and rescue-induced
  oscillations}}}.
\newblock {\emph{\JournalTitle{Phys. Rev. E}}} \textbf{\bibinfo{volume}{87}},
  \bibinfo{pages}{012703}, \doiprefix\url{10.1103/PhysRevE.87.012703}
  (\bibinfo{year}{2013}).

\bibitem{Janson2003}
\bibinfo{author}{Janson, M.~E.}, \bibinfo{author}{de~Dood, M.~E.} \&
  \bibinfo{author}{Dogterom, M.}
\newblock \bibinfo{journal}{\bibinfo{title}{Dynamic instability of microtubules
  is regulated by force}}.
\newblock {\emph{\JournalTitle{J. Cell Biol.}}} \textbf{\bibinfo{volume}{161}},
  \bibinfo{pages}{1029--1034}, \doiprefix\url{10.1083/jcb.200301147}
  (\bibinfo{year}{2003}).

\bibitem{Li2010}
\bibinfo{author}{Li, X.}, \bibinfo{author}{Lipowsky, R.} \&
  \bibinfo{author}{Kierfeld, J.}
\newblock \bibinfo{journal}{\bibinfo{title}{{Coupling of actin hydrolysis and
  polymerization: Reduced description with two nucleotide states}}}.
\newblock {\emph{\JournalTitle{EPL}}} \textbf{\bibinfo{volume}{89}},
  \bibinfo{pages}{38010}, \doiprefix\url{10.1209/0295-5075/89/38010}
  (\bibinfo{year}{2011}).

\bibitem{Seetapun2012}
\bibinfo{author}{Seetapun, D.}, \bibinfo{author}{Castle, B.},
  \bibinfo{author}{McIntyre, A.}, \bibinfo{author}{Tran, P.} \&
  \bibinfo{author}{Odde, D.}
\newblock \bibinfo{journal}{\bibinfo{title}{{Estimating the Microtubule GTP Cap
  Size In Vivo}}}.
\newblock {\emph{\JournalTitle{Curr. Biol.}}} \textbf{\bibinfo{volume}{22}},
  \bibinfo{pages}{1681--1687}, \doiprefix\url{10.1016/j.cub.2012.06.068}
  (\bibinfo{year}{2012}).

\bibitem{Voter1991}
\bibinfo{author}{Voter, W.~A.}, \bibinfo{author}{O'Brien, E.~T.} \&
  \bibinfo{author}{Erickson, H.~P.}
\newblock \bibinfo{journal}{\bibinfo{title}{{Dilution-Induced Disassembly of
  Microtubules: Relation to Dynamic Instability and the GTP Cap}}}.
\newblock {\emph{\JournalTitle{Cell Motil. Cytoskeleton}}}
  \textbf{\bibinfo{volume}{18}}, \bibinfo{pages}{55--62},
  \doiprefix\url{10.1002/cm.970180106} (\bibinfo{year}{1991}).

\bibitem{Duellberg2016}
\bibinfo{author}{Duellberg, C.}, \bibinfo{author}{Cade, N.~I.},
  \bibinfo{author}{Holmes, D.} \& \bibinfo{author}{Surrey, T.}
\newblock \bibinfo{journal}{\bibinfo{title}{The size of the eb cap determines
  instantaneous microtubule stability}}.
\newblock {\emph{\JournalTitle{eLife}}} \textbf{\bibinfo{volume}{5}},
  \bibinfo{pages}{e13470}, \doiprefix\url{10.7554/eLife.13470}
  (\bibinfo{year}{2016}).

\bibitem{Grafmueller2013}
\bibinfo{author}{Grafmüller, A.}, \bibinfo{author}{Noya, E.~G.} \&
  \bibinfo{author}{Voth, G.~A.}
\newblock \bibinfo{journal}{\bibinfo{title}{{Nucleotide-Dependent Lateral and
  Longitudinal Interactions in Microtubules}}}.
\newblock {\emph{\JournalTitle{J. Mol. Biol.}}} \textbf{\bibinfo{volume}{425}},
  \bibinfo{pages}{2232--2246}, \doiprefix\url{10.1016/j.jmb.2013.03.029}
  (\bibinfo{year}{2013}).

\bibitem{Hoog2011}
\bibinfo{author}{H{\"o}{\"o}g, J.~L.} \emph{et~al.}
\newblock \bibinfo{journal}{\bibinfo{title}{Electron tomography reveals a
  flared morphology on growing microtubule ends}}.
\newblock {\emph{\JournalTitle{J. Cell Sci.}}} \textbf{\bibinfo{volume}{124}},
  \bibinfo{pages}{693--698}, \doiprefix\url{10.1242/jcs.072967}
  (\bibinfo{year}{2011}).

\bibitem{Kukulski2011}
\bibinfo{author}{Kukulski, W.} \emph{et~al.}
\newblock \bibinfo{journal}{\bibinfo{title}{Correlated fluorescence and 3d
  electron microscopy with high sensitivity and spatial precision}}.
\newblock {\emph{\JournalTitle{J. Cell Biol.}}} \textbf{\bibinfo{volume}{192}},
  \bibinfo{pages}{111--119}, \doiprefix\url{10.1083/jcb.201009037}
  (\bibinfo{year}{2011}).

\bibitem{Nawrotek2011}
\bibinfo{author}{Nawrotek, A.}, \bibinfo{author}{Knossow, M.} \&
  \bibinfo{author}{Gigant, B.}
\newblock \bibinfo{journal}{\bibinfo{title}{The determinants that govern
  microtubule assembly from the atomic structure of gtp-tubulin}}.
\newblock {\emph{\JournalTitle{J. Mol. Biol.}}} \textbf{\bibinfo{volume}{412}},
  \bibinfo{pages}{35--42}, \doiprefix\url{10.1016/j.jmb.2011.07.029}
  (\bibinfo{year}{2011}).

\bibitem{Pecqueur2012}
\bibinfo{author}{Pecqueur, L.} \emph{et~al.}
\newblock \bibinfo{journal}{\bibinfo{title}{A designed ankyrin repeat protein
  selected to bind to tubulin caps the microtubule plus end}}.
\newblock {\emph{\JournalTitle{Proc. Natl. Acad. Sci. USA}}}
  \textbf{\bibinfo{volume}{109}}, \bibinfo{pages}{12011--12016},
  \doiprefix\url{10.1073/pnas.1204129109} (\bibinfo{year}{2012}).

\bibitem{Akhmanova2008}
\bibinfo{author}{Akhmanova, A.} \& \bibinfo{author}{Steinmetz, M.~O.}
\newblock \bibinfo{journal}{\bibinfo{title}{Tracking the ends: a dynamic
  protein network controls the fate of microtubule tips}}.
\newblock {\emph{\JournalTitle{Nat. Rev. Mol. Cell Biol.}}}
  \textbf{\bibinfo{volume}{9}}, \bibinfo{pages}{309--322},
  \doiprefix\url{10.1038/nrm2369} (\bibinfo{year}{2008}).

\bibitem{Dogterom1997}
\bibinfo{author}{Dogterom, M.} \& \bibinfo{author}{Yurke, B.}
\newblock \bibinfo{journal}{\bibinfo{title}{{Measurement of the Force-Velocity
  Relation for Growing Microtubules}}}.
\newblock {\emph{\JournalTitle{Science}}} \textbf{\bibinfo{volume}{278}},
  \bibinfo{pages}{856--860}, \doiprefix\url{10.1126/science.278.5339.856}
  (\bibinfo{year}{1997}).

\bibitem{Dogterom2005}
\bibinfo{author}{Dogterom, M.}, \bibinfo{author}{Kerssemakers, J. W.~J.},
  \bibinfo{author}{Romet-Lemonne, G.} \& \bibinfo{author}{Janson, M.~E.}
\newblock \bibinfo{journal}{\bibinfo{title}{Force generation by dynamic
  microtubules}}.
\newblock {\emph{\JournalTitle{Curr. Opin. Cell Biol.}}}
  \textbf{\bibinfo{volume}{17}}, \bibinfo{pages}{67--74},
  \doiprefix\url{10.1016/j.ceb.2004.12.011} (\bibinfo{year}{2005}).

\bibitem{VanDoorn2000}
\bibinfo{author}{van Doorn, G.~S.}, \bibinfo{author}{Tănase, C.},
  \bibinfo{author}{Mulder, B.~M.} \& \bibinfo{author}{Dogterom, M.}
\newblock \bibinfo{journal}{\bibinfo{title}{On the stall force for growing
  microtubules}}.
\newblock {\emph{\JournalTitle{Eur. Biophys. J.}}}
  \textbf{\bibinfo{volume}{29}}, \bibinfo{pages}{2--6},
  \doiprefix\url{10.1007/s002490050245} (\bibinfo{year}{2000}).

\bibitem{Kolomeisky2001}
\bibinfo{author}{Kolomeisky, A.~B.} \& \bibinfo{author}{Fisher, M.~E.}
\newblock \bibinfo{journal}{\bibinfo{title}{{Force-Velocity Relation for
  Growing Microtubules}}}.
\newblock {\emph{\JournalTitle{Biophys. J.}}} \textbf{\bibinfo{volume}{80}},
  \bibinfo{pages}{149--154}, \doiprefix\url{10.1016/S0006-3495(01)76002-X}
  (\bibinfo{year}{2001}).

\bibitem{Ranjith2009}
\bibinfo{author}{Ranjith, P.}, \bibinfo{author}{Lacoste, D.},
  \bibinfo{author}{Mallick, K.} \& \bibinfo{author}{Joanny, J.-F.}
\newblock \bibinfo{journal}{\bibinfo{title}{{Nonequilibrium Self-Assembly of a
  Filament Coupled to ATP/GTP Hydrolysis}}}.
\newblock {\emph{\JournalTitle{Biophys. J.}}} \textbf{\bibinfo{volume}{96}},
  \bibinfo{pages}{2146--2159}, \doiprefix\url{10.1016/j.bpj.2008.12.3920}
  (\bibinfo{year}{2009}).

\bibitem{Krawczyk2011}
\bibinfo{author}{Krawczyk, J.} \& \bibinfo{author}{Kierfeld, J.}
\newblock \bibinfo{journal}{\bibinfo{title}{Stall force of polymerizing
  microtubules and filament bundles}}.
\newblock {\emph{\JournalTitle{Europhys. Lett.}}}
  \textbf{\bibinfo{volume}{93}}, \bibinfo{pages}{28006},
  \doiprefix\url{10.1209/0295-5075/93/28006} (\bibinfo{year}{2011}).

\end{thebibliography}

\newpage
\section*{Supplementary Material}

\subsection*{Microtubule structure and energy}

At the minus end of the MT, each  protofilament $p$ starts with
an alpha-tubulin at 
\begin{equation}
	\vec{m}(p,1,1)
	= \begin{pmatrix}
		\Rmt \cos \phi(p) \\
		- \Rmt \sin \phi(p) \\
		3 \Ltub (p - 1) / 13
	\end{pmatrix}
\end{equation}
with the mean MT radius $\Rmt = \SI{10.5}{\nano\meter}$, such that
the seam is located
between the 13th and the 1st protofilament.

Using the direction vectors
\begin{equation}
	\vec{d}(p,d,t)
	= \vec{p}(p,d,t) - \vec{m}(p,d,t)
	= \Ltub \begin{pmatrix}
		\cos \phi(p) \sin \theta(p,d,t) \\
		- \sin \phi(p) \sin \theta(p,d,t) \\
		\cos \theta(p,d,t)
	\end{pmatrix},
\end{equation}
the plus end position $\vec{p}(p,d,t)$ of any tubulin monomer can be calculated
by adding all direction vectors to the minus end vector, 
\begin{equation}
	\vec{p}(p,d,t)
	= \vec{m}(p,1,1) + \sum_{d' = 1}^{d} \sum_{t' = 1}^2 \vec{d}(p,d',t')
		- \delta_{t,1} \vec{d}(p,d,2) .
\end{equation}
The protofilament length that will be used to
calculate the growth and shrinkage
velocities is the maximum $z$-coordinate of all tubulin
monomers within the protofilament:
\begin{equation}
  \lmax(p)
  = \max_{d,t} \left( \vec{p}(p,d,t) \cdot \vec{e}_z \right).
\end{equation}
For straight and slightly curved protofilaments,
$\lmax(p) = \vec{p}(p,d(p),2) \cdot \vec{e}_z$ is the position of the plus end
of the protofilament.
For strongly curved protofilaments exhibiting a ram's horn and curling
backwards, the length can exceed  the
$z$-coordinate of the terminal beta-tubulin,
$\lmax(p) > \vec{p}(p,d(p),2) \cdot \vec{e}_z$.

In order to define  the lateral bond energies between tubulin dimers 
in neighboring protofilaments, we need to introduce
interaction points, where the harmonic springs of the lateral bonds attach.
The lateral interaction points are located at the edge of the upper base
at $\vec{p}(p,d,t) + \vec{c}(p,d,t)$ and
$\vec{p}(p,d,t) - \vec{c}(p,d,t)$ with the connection vector
\begin{equation}
	\vec{c}(p,d,t)
	= \begin{pmatrix}
		- \Rtub \sin \phi(p) \\
		- \Rtub \cos \phi(p) \\
		0
	\end{pmatrix} .
\end{equation}
The connection vector
\begin{equation}
  \vec{s}(p,d,t)  = (\vec{p}(p+1,d,t) - \vec{c}(p+1,d,t)) -
  (\vec{p}(p,d,t) + \vec{c}(p,d,t))
\end{equation}
points from the edge of the upper base of
tubulin monomer $(p, d, t)$ to its
next neighbor $(p + 1, d, t)$ and is used
to define the harmonic spring energies of the lateral bonds.

\subsection*{Detailed dynamics of  single catastrophes, rescues and dips}

The MT growth trajectory shown in Fig.\ 10(A) in the main text has two
significant events: a dip at $\tsim = \SI{1.2}{\minute}$ and a catastrophe at
$\tsim = \SI{6.85}{\minute}$.
To determine whether newly run simulations with configurations from the initial
simulation as starting points qualitatively follow the original simulation, we
chose the following criteria for this particular simulation:
\begin{itemize}
	\item	To identify whether a new simulation reproduces the dip, we
		checked that at the end of the simulation, the
		MT length is at most \SI{400}{\nano\meter} shorter than
		the original simulation.
		As the relevant question for the dip is whether it could
		actually result in a catastrophe, we are only interested in the
	new simulations being \emph{shorter} than the original simulation,
		hence there is no upper limit on the MT length difference.
	\item	To identify whether a new simulation reproduces the catastrophe,
		we took the last entry in the MT length log and checked
		if its time $\tsim$ differs less than \SI{10}{\second} from the
		original simulation end and if $\Lmt < \SI{200}{\nano\meter}$,
		i.e., if the MT continued shrinking and depolymerized (almost)
		completely.
\end{itemize}
For each initial configuration, we ran 20 new simulations and calculated the
fraction of simulations that fulfilled the criteria above.
These fractions are the probabilities for the original growth path at different
 points in time and they are shown color-coded in the insets (A.D) and (A.C) in
 Fig.\ 10(A).

 For the analysis of Fig.\ 13(A) in the main text, where
 the hydrolysis rate is coupled to mechanics, we used the same criteria.

 Figure 10(C) in the main text contains three significant events:
 a dip at the
 very beginning, a catastrophe at $\tsim = \SI{9.15}{\minute}$, and a rescue
 at $\tsim = \SI{9.54}{\minute}$.
 For this simulation, we used the following criteria:
 \begin{itemize}
 	\item	For the dip, we used the same criterion as before.
 	\item	For the catastrophe, we checked that after \SI{15}{\second}
 		(or \SI{10}{\second} for configurations after the catastrophe),
 		the MT was at most $\SI{400}{\nano\meter}$ longer than
 		the original MT at the same point in time.
 	\item	For the rescue, we took the last entry in the MT length
 		log of the new simulations and checked if
 		$\tsim > \SI{59}{\second}$, i.e., if the new simulation finished
 		due to the time constraint and not due to the MT having vanished
 		meaning that the rescue actually happened.
 \end{itemize}
 Again, we ran 20 new simulations for each of the three initial configurations
 and obtained the probabilities for the simulation to follow the original growth
path, which are shown as color code in the insets (C.D), (C.C), and (C.R) in
 Fig.\ 10(C).
The color coding reveals that the dip in (C.D) is actually better characterized
 as a catastrophe immediately followed by a rescue.
 In addition, Fig.\ 10(B) shows snapshots of the MT
 configuration at characteristic points in the growth path, for example, before,
 in the middle, and after the catastrophe and rescue events.

\begin{figure}[ht!]
	\centering
	\includegraphics[width=0.99\linewidth]{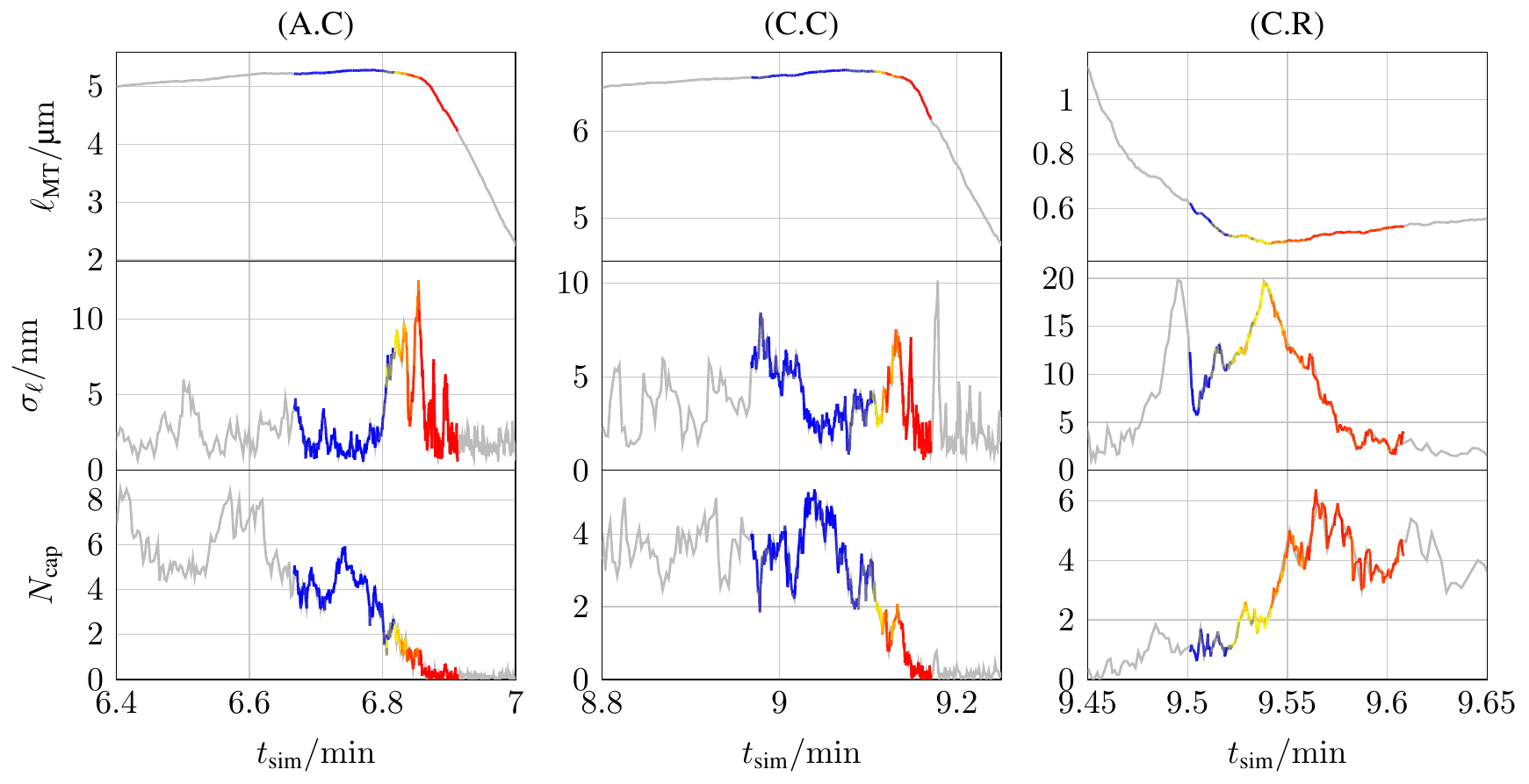}
	\caption{
		The standard deviation $\sigma_\ell$ of the individual
		protofilament lengths from the mean protofilament length $\Lmt$
		and the average cap length $\Ncap$ of the two catastrophes and
		the rescue shown in the insets of Fig.\ 10 in the main text
		as a function of simulation time $\tsim$.
		The same color coding as in Fig.\ 10 in the main text is
		used and gray is used to show additional data before and after
		the highlighted parts in the insets of
		Fig.\ 10.
	}
	\label{fig:full_simulations_analysis}
\end{figure}

 \autoref{fig:full_simulations_analysis} shows additional data
 on the catastrophe and rescue events from Fig.\ 10
 in the main text (catastrophes from insets (A.C) and (C.C) and
 rescue from (C.R)). We show the mean protofilament length $\Lmt$
 together with additional information on the length
 fluctuations $\sigma_\ell$ of the individual
protofilament lengths and  the  average  GTP-cap
length $\Ncap$. When catastrophes become unavoidable, the cap length
has shrunken to  around $\Ncap \sim 2$, for rescues a cap length
around $\Ncap \sim 4$  seems necessary. Length fluctuations are
also  increased  if catastrophe or rescues are triggered.

\subsection*{Determination of catastrophe and rescue rates}

For the determination of catastrophe and rescue rates
in Fig.\ 8 in the main text we employed the
following algorithm.

\begin{figure}[ht!]
        \centering
	\includegraphics[width=0.45\linewidth]{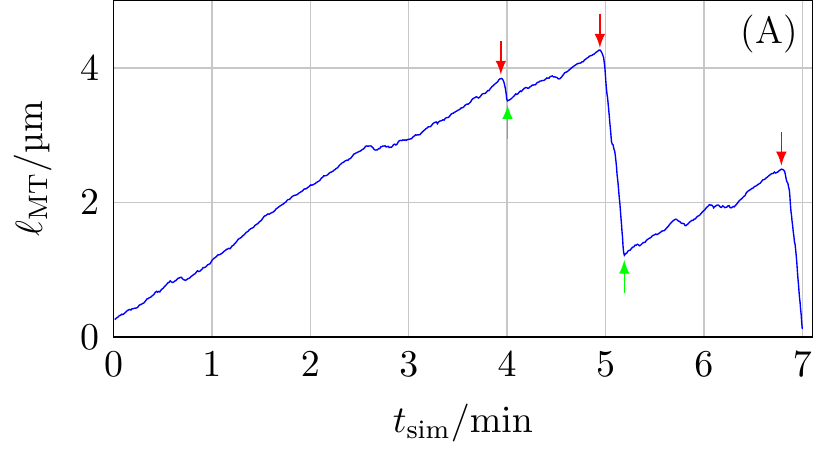}
	\hspace{0.01\linewidth}
	\includegraphics[width=0.45\linewidth]{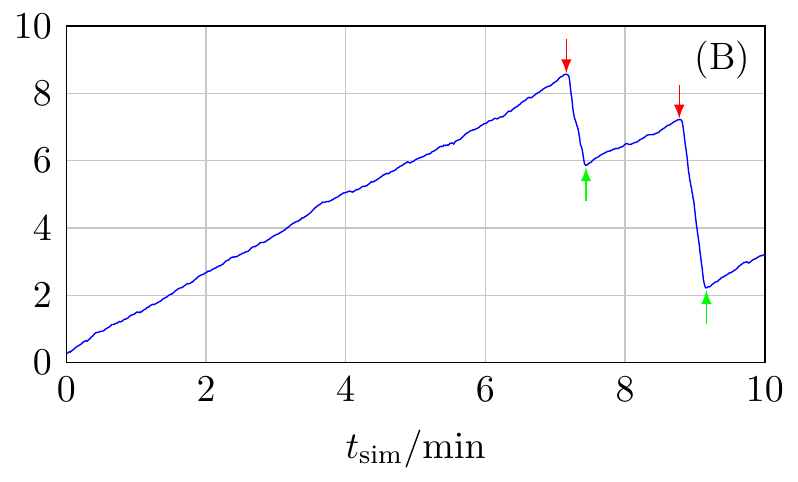}
        \caption{
          Two exemplary microtubule trajectories in which
          catastrophes and rescues determined
          by the algorithm are marked by
          red arrows pointing down and green arrows pointing up, respectively.
          (A) $\ctubVal{10}$, $\khydrVal{0.3}$
          (B) $\ctubVal{11}$, $\khydrVal{0.4}$.
        }
        \label{fig:marked_catastrophes_rescues}
\end{figure}

First a MT trajectory $\Lmt=\Lmt(\tsim)$  is classified into growth
and shrinkage intervals using a greedy threshold value $\Delta \ell_1=
\SI{50}{\nano\meter}$. We start at the beginning of an interval at
$\tsim =t_0$ and increase 
$\tsim$ searching for a suitable end $t_1$ of an interval. 
If $\Lmt(t_1) - \Lmt(t_0) \ge \Delta \ell_1$, i.e.,
if a MT has grown more than $\Delta \ell_1$, the interval
$[t_0,t_1]$
is classified as growth interval. Likewise, if
$\Lmt(t_1) - \Lmt(t_0)
\le -\Delta \ell_1$, i.e.,
if a MT has shrunken by more than $\Delta \ell_1$,
the interval $[t_0,t_1]$
is classified as growth interval. All plateaus, where the length changes by 
less than $\Delta \ell_1$ are ``absorbed'' into surrounding growth or
shrinkage  intervals. This part of the procedure gives a complete 
classification into a (not necessarily alternating) succession 
 of growth and shrinkage intervals. 

If the (n-1)-th interval is a growth (shrinkage) 
interval and the n-th interval a 
shrinkage (growth) interval the n-th interval is marked as possibly 
containing a catastrophe (rescue). 

Then we continue with the second part of the algorithm, where we 
employ a less greedy  threshold value $\Delta \ell_2=
\SI{300}{\nano\meter}$.
We search for a catastrophe time  $t_c$ and an enclosing interval 
$[t_{c-},t_{c+}]$ in a potential catastrophe containing 
 interval according to the following steps:
\begin{enumerate}
\item
We find a $t_{c-}< t_c$ with $t_c-t_{c-} < \SI{50}{\second}$ and
$\Lmt(t_c) - \Lmt(t_{c-}) \ge \Delta \ell_2$, i.e., 
the MT grows  by  $\Delta \ell_2$ within the previous 
$\SI{50}{\second}$ or less. We select the largest $t_{c-}$ fulfilling 
these criteria. 
\item 
We find a $t_{c+} > t_c$ with with $t_{c+}-t_{c} < \SI{50}{\second}$
and $\Lmt(t_{c}) - \Lmt(t_{c+}) \ge \Delta \ell_2$, i.e., 
the MT shrinks   by  $\Delta \ell_2$ within the next 
$\SI{50}{\second}$ or less. We select the smallest  $t_{c+}$ fulfilling 
these criteria. 
\item 
Among all possible $t_c$ in the potential catastrophe containing 
 interval, we choose the value producing the smallest enclosing 
interval $[t_{c-},t_{c+}]$ according to steps 1 and 2.
\end{enumerate}
For rescues, we proceed analogously.

Two exemplary simulation trajectories that we analyzed with the
algorithm are shown in \autoref{fig:marked_catastrophes_rescues}.

The algorithm identifies the points in time where catastrophes and
rescues happen and, thus, also gives access to the
times $\Delta t_\text{gr,k}$ that a MT grows before the $k$-th catastrophe
and times  $\Delta t_\text{gr,k}$ that a MT shrinks
before  the $k$-th rescue.
Catastrophe and recue rates are obtained as inverse of the averaged
average growth and shrinking times,
\begin{align}
  \omega_\text{cat} &= \left(\frac{1}{N_\text{cat} }\sum_{k=1}^{N_\text{cat}}
                      \Delta t_\text{gr,k}  \right)^{-1}
     \label{eq:wcat}\\
   \omega_\text{res} &= \left(\frac{1}{N_\text{res}} \sum_{k=1}^{N_\text{res}}
                       \Delta t_\text{sh,k}  \right)^{-1}.
       \label{eq:wres}
\end{align}

\subsection*{Theoretical spatial GTP distribution}

We employ a similar approach as Ref.\ \cite{Padinhateeri2012} and consider a
one-dimensional
MT (or single protofilament approximation) with polymerization rate $\kon$,
effective
depolymerization rate $\koffOneD$, and hydrolysis rate $\khydr$ (for all tubulin
dimers, including the terminal one).
In the steady state, 
the probability of the $i$-th tubulin dimer
counted from the plus end ($i = d(p) - d + 1$) to be a GTP tubulin dimer is
given by
\begin{equation}
	p_i
	= q^i
	= \left[ \frac{\kon + \koffOneD + \khydr}{2\koffOneD}
		\left( 1 - \sqrt{ 1 - \frac{4 \kon \koffOneD}{\left( \kon + \koffOneD + \khydr \right)^2}} \right)
	\right]^i
	\label{eq:pi}
\end{equation}
with $i = 1$ referring to the tubulin dimer directly at the plus end.
As we are measuring the normalized probability $\tilde{p}_i$ with
$\sum_{i=1}^\infty \tilde{p}_i= 1$,
we have to compare our simulation results with
\begin{equation}
	\tilde{p}_i
	= (1 - q) q^{i-1} .
		\label{eq:spatial_gtp_distribution_simulation}
\end{equation}
While we have a constant polymerization rate $\kon$ and hydrolysis rate $\khydr$
(with hydrolysis is not coupled to mechanics), there is no clear mapping of the
effective
one-dimensional depolymerization rate $\koff$ to our three-dimensional modelling
because lateral bond formation and rupture results in an
effective depolymerization rate.
When comparing our simulation results with the theoretical prediction
\eqref{eq:spatial_gtp_distribution_simulation}, we are using $\koff$ as a fitting
parameter and using $\kon$ and $\khydr$ from the simulation.

\begin{figure}[!ht]
	\centering
	\includegraphics[width=0.99\linewidth]{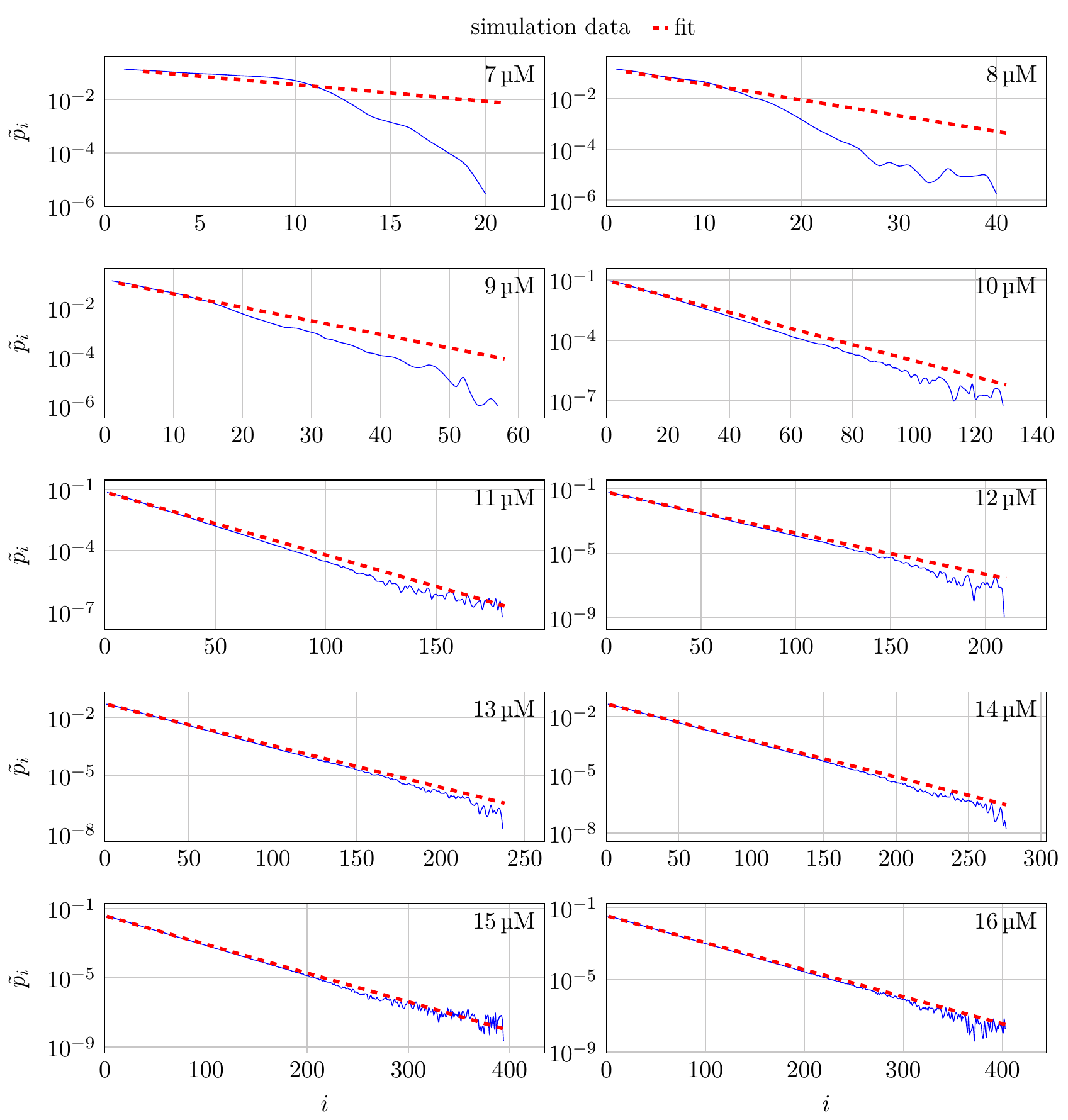}	\caption{
		Relative probability $\tilde{p}_i$ of the $i$-th tubulin dimer
		counted from the plus end being a GTP-tubulin dimer with
		$\koncVal{4}$, $\GlongVal{-9.3}$,
		$\klat = \SI{100}{\kBT \per\nano\meter\squared}$, and
		$\khydrVal{0.25}$.
	}
	\label{fig:spatial_gtp_distribution}
\end{figure}

\autoref{fig:spatial_gtp_distribution} compares the simulation results for
$\tilde{p}_i$ and the theoretical prediction for different free tubulin dimer
concentrations $c$ (and thus different $\kon$ values).
Once $\ctub$ is sufficiently large so that the MTs can reach a steady
state of growth, the prediction by the one-dimensional theory matches the
simulation data.

In the main text, we give
\begin{equation}
	0
	= -(\kon - \koffOneD) \frac{\text{d} \probGTP}{\text{d} x}
		- \langle \khydr \rangle(x) \,\probGTP(x)
	\label{eq:pGTPx}
\end{equation}
for the probability $\probGTP(x)$ to find a GTP-dimer at distance
$x = d(p) - d$ from the tip.
\eqref{eq:pGTPx} is a continuous version (with a
continuous $x\approx i - 1$) of the discrete master equation for $p_i$
that leads to the above result (\ref{eq:pi}).

Again, a direct comparison with our data is not possible because of the unknown
effective one-dimensional depolymerization rate $\koffOneD$.
However, \eqref{eq:pGTPx} can be rearranged for an explicit expression for
$\koffOneD$ so that we can calculate $\koffOneD(x)$:
\begin{equation}
	\koffOneD(x)
	= \kon + \left( \frac{\text{d} \probGTP}{\text{d} x} \right)^{-1}
		\langle \khydr \rangle(x) \,\probGTP(x) .
	\label{eq:koffOneD}
\end{equation}
It should be noted that $\koffOneD(x)$ is not the depolymerization rate of layer
$x = d(p) - d$ but the depolymerization rate of the last layer calculated using
the data from layer $x$.
If our data can be described by \eqref{eq:pGTPx}, we expect $\koffOneD(x)$ to be
independent of $x$.

\begin{figure}[!ht]
	\centering
	\includegraphics[width=0.99\linewidth]{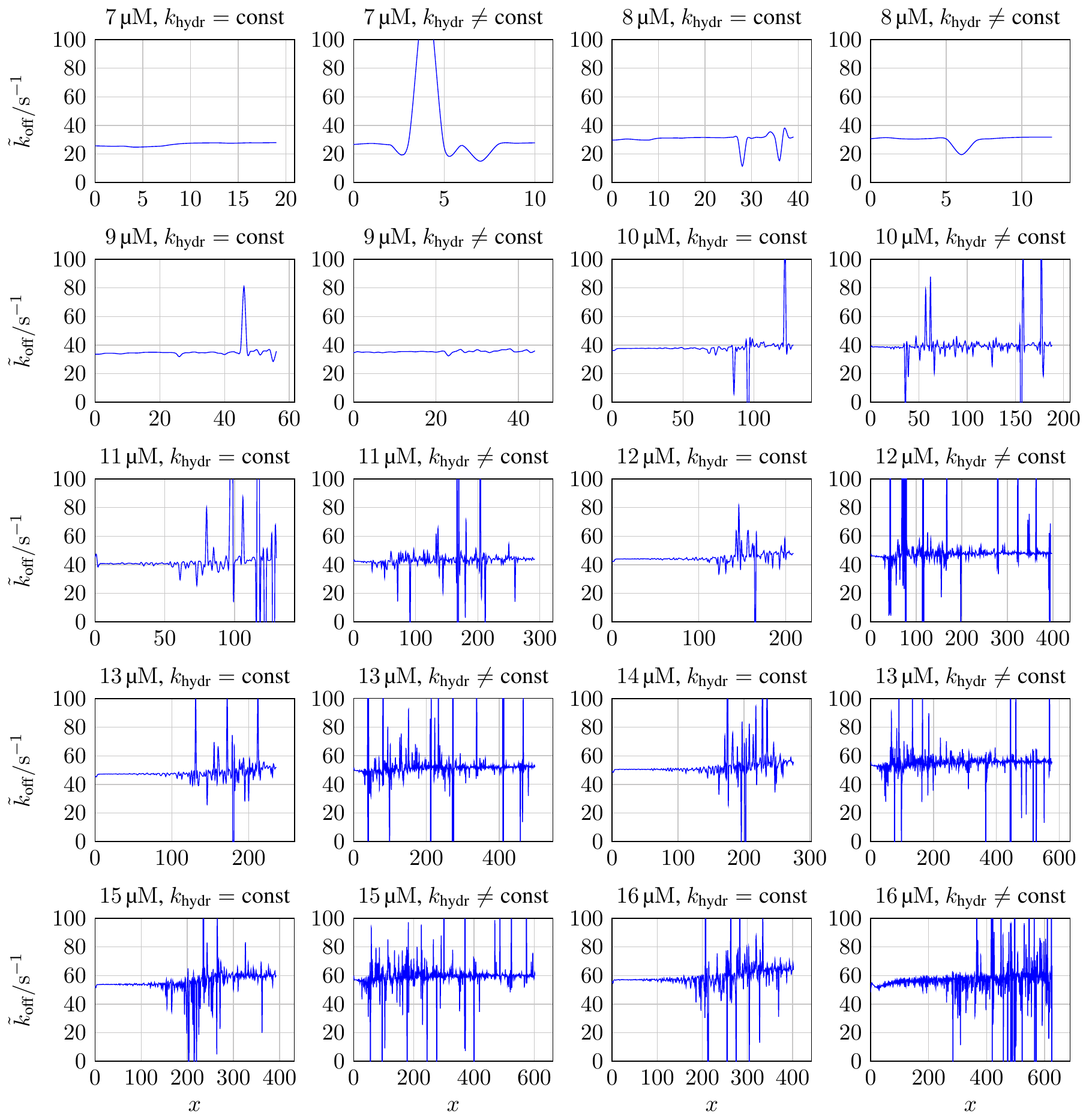}
	\caption{
		Depolymerization rate $\koffOneD$ calculated from the GTP-tubulin
		dimer probability distribution $\probGTP$ according to \eqref{eq:koffOneD}
		for a constant hydrolysis rate of $\khydrVal{0.25}$ and hydrolysis
		coupled to mechanics with $\khydrNVal{1.5}$.
		Values with
		$\koffOneD < \SI[per-mode=reciprocal]{0}{\per\second}$ and
		$\koffOneD > \SI[per-mode=reciprocal]{100}{\per\second}$ are cut
		off here as they are due to insufficient data statistics.
	}
	\label{fig:koffOneD_from_spatial_gtp_distribution}
\end{figure}

\autoref{fig:koffOneD_from_spatial_gtp_distribution} shows $\koffOneD(x)$ for
both a constant hydrolysis rate and hydrolysis coupled to mechanics.
The derivative was calculated using the symmetric derivative, except for the
first and last values, were a forward or backward derivative was used.
Ignoring numerical issues due to the discrete derivative and insufficient data
statistics for larger $x$ values, $\koffOneD(x)$ is sufficiently independent of
$x$ showing that \eqref{eq:pGTPx} can also be used to describe
the results of the
simulations in which hydrolysis is coupled to mechanics.

\clearpage

\subsection*{Minimization time comparison between constant hydrolysis rates and
hydrolysis rates coupled to mechanics}

\begin{figure}[!ht]
	\centering
	\includegraphics[width=0.99\linewidth]{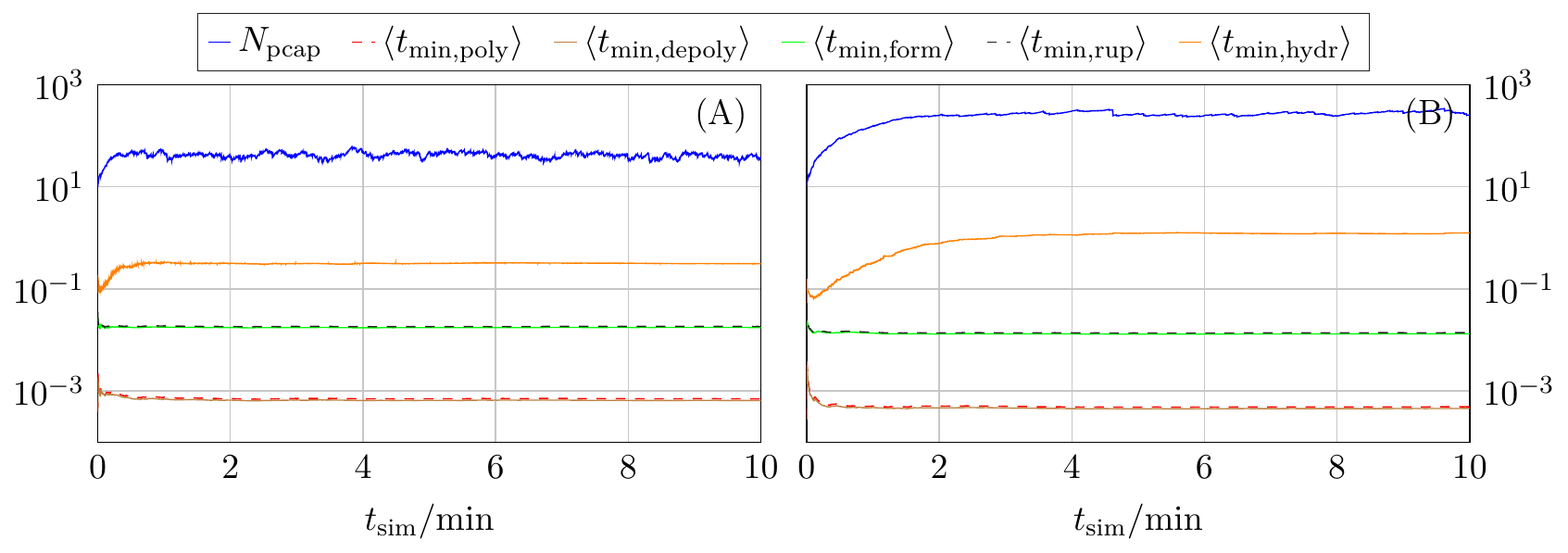}	\caption{
		Comparison of the porous cap length $\porousCapLength$ and the
		cumulative running averages $\langle t_{\text{min},i} \rangle$
		(in seconds) for minimization times after event $i$ for
		(A) a constant hydrolysis rate
		$\khydrVal{0.25}$ (run on a single core of an Intel Xeon CPU
		E5-2650 (Sandy Bridge) processor) and
		(B) hydrolysis coupled to mechanics with $\khydrNVal{1.5}$ (run
		on an single core of a Intel Xeon CPU E5-2630 v3 (Haswell)
		processor).
		Both plots show the results of one exemplary simulation that use
		$\koncVal{4}$, $\GlongVal{-9.3}$,
		$\klat = \SI{100}{\kBT \per\nano\meter\squared}$, and
		$\ctubVal{11}$.
	}
	\label{fig:minimization_time_comparison}
\end{figure}

To gain further insight into the additional amount of execution time
required by
simulations in which hydrolysis is coupled to mechanics,
\autoref{fig:minimization_time_comparison}
shows a comparison of average times for
minimization after each of the five possible chemical events polymerization,
depolymerization, bond formation, bond rupture, and hydrolysis.

While the average minimization time after polymerization, depolymerization, bond
formation, and bond rupture is not affected by mechanical feedback, the average
minimization time after hydrolysis increases in the presence of mechanical
feedback. 
In both cases, at a certain point that roughly matches the point in time when
the porous cap length $\porousCapLength$ reaches it steady state, the average
minimization time after hydrolysis does not change significantly anymore either.
For the two simulations shown in \autoref{fig:minimization_time_comparison},
minimizations after hydrolysis take four times longer for hydrolysis coupled to
mechanics than for a constant hydrolysis rate; the reason for this increase
is that the porous cap length is on
average more than six times longer such that minimization has to
be executed for up to six times more layers if a dimer deep in GDP-body
is hydrolyzed.
In both cases, the simulation spends around \SI{98}{\percent} of its time during
minimization.
Of that total minimization time, the simulation shown in
\autoref{fig:minimization_time_comparison}(A) uses roughly \SI{30}{\percent} for
minimizations after hydrolysis events, while simulation
\autoref{fig:minimization_time_comparison}(B) uses about \SI{66}{\percent} for
minimizations after hydrolysis events even though the difference between the
percentage of hydrolysis events from all events only increased by little more
than \num{0.1} percentage points.

\subsection*{Analysis of dilution simulations}

For the determination of the delay time $\Dtdelay$ after GTP-tubulin
dilution at time $\tdil$,
we employed the following algorithm.

\begin{figure}[ht!]
  \centering
      \includegraphics[width=0.45\linewidth]{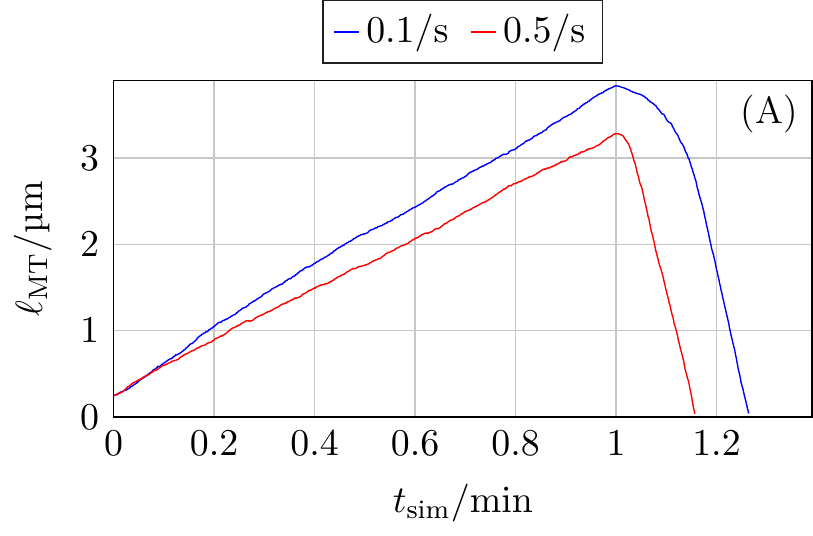}
	\hspace{0.01\linewidth}
       \includegraphics[width=0.45\linewidth]{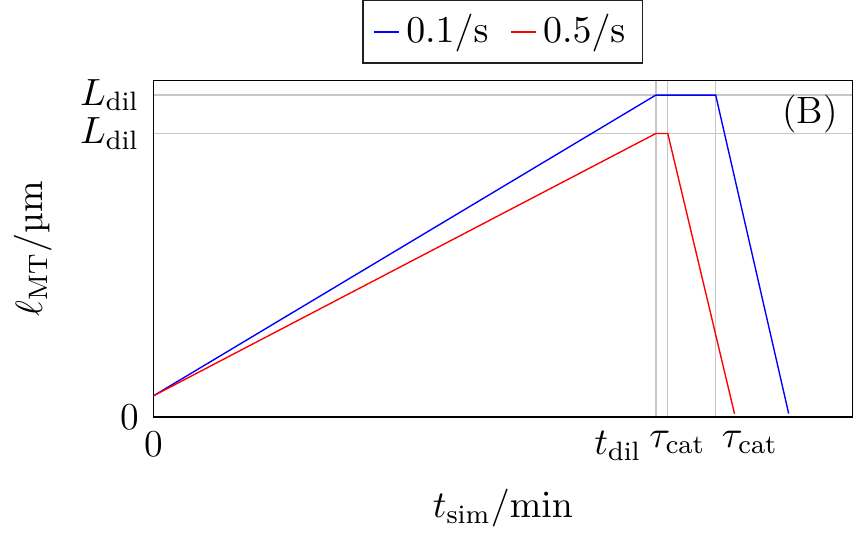}
       \caption{
          Example of how  (A) actual microtubule growth trajectories of dilution
           simulations with
           $\ctub=  \SI{16}{\micro\molar}$ and $\cdil =  \SI{0}{\micro\molar}$,
           and two different values of $\khydr$
            are simplified into (B) a growth, delay, and
         shrinkage phase.
        }
        \label{fig:dilution_trajectories}
\end{figure}

First,  we  fit a linear  growth law $\Lmt(\tsim) = \vgro \tsim + \Lmt(0)$
to the MT length data for simulation times $\tsim\le \tdil$ up to the dilution. 
This determines $\Lmt(\tdil)$.
Then we fit a linear shrinking law
$\Lmt(\tsim) = \vshr (\tsim-\tau_{\rm cat})  + \Lmt(\tdil)$ ($\vshr<0$) to
the shrinking part of the trajectory after dilution and delay. 
This determines the catastrophe time $\tau_{\rm cat}> \tdil$
as intersection point
with the dilation plateau $\Lmt(\tsim) = \Lmt(\tdil)$, which we fit
for $\tdil < \tsim <\tau_{\rm cat}$. 
The delay time is given by $\Dtdelay = \tau_{\rm cat}  - \tdil$. 

Two exemplary simulation trajectories that we analyzed with the
algorithm are shown in \autoref{fig:dilution_trajectories}.

\clearpage

\subsection*{Supplementary Figures}

\begin{figure}[!ht]
	\centering
	\nocite{Walker1988}
       \includegraphics[width=0.99\linewidth]{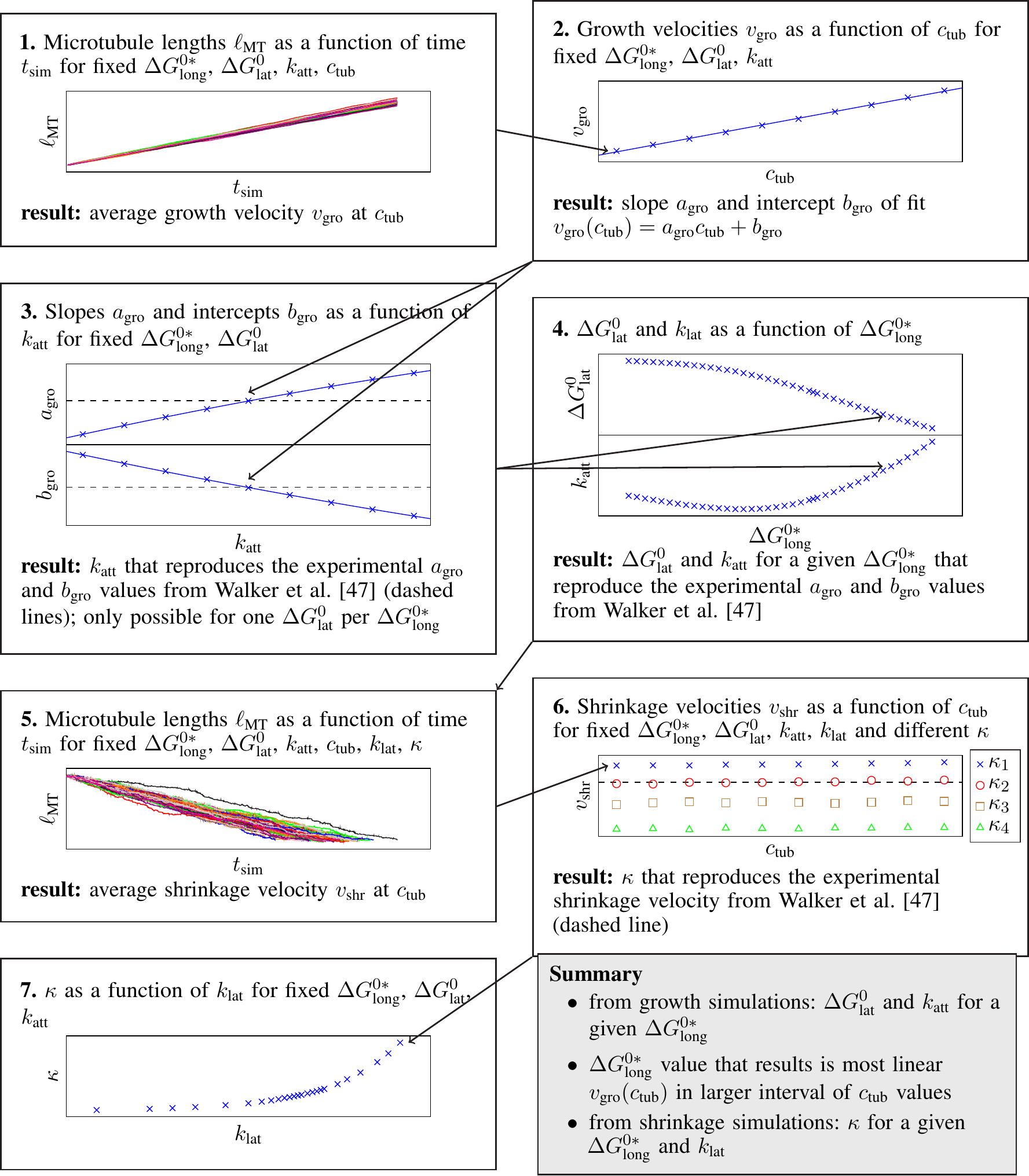}	\caption{
		Schematic illustration of how the model parameters $\katt$ and $\Glat$
		as a function of $\Glong$ (for a given $\konc$) from a large number of
		individual length trajectories of growing MTs (step 1) and $\klat$
		as a function of $\kcurl$ from trajectories of shrinking MTs (step 5).
	}
	\label{fig:parameter_determination}
\end{figure}

\begin{figure}[!ht]
	\centering
       \includegraphics[width=0.99\linewidth]{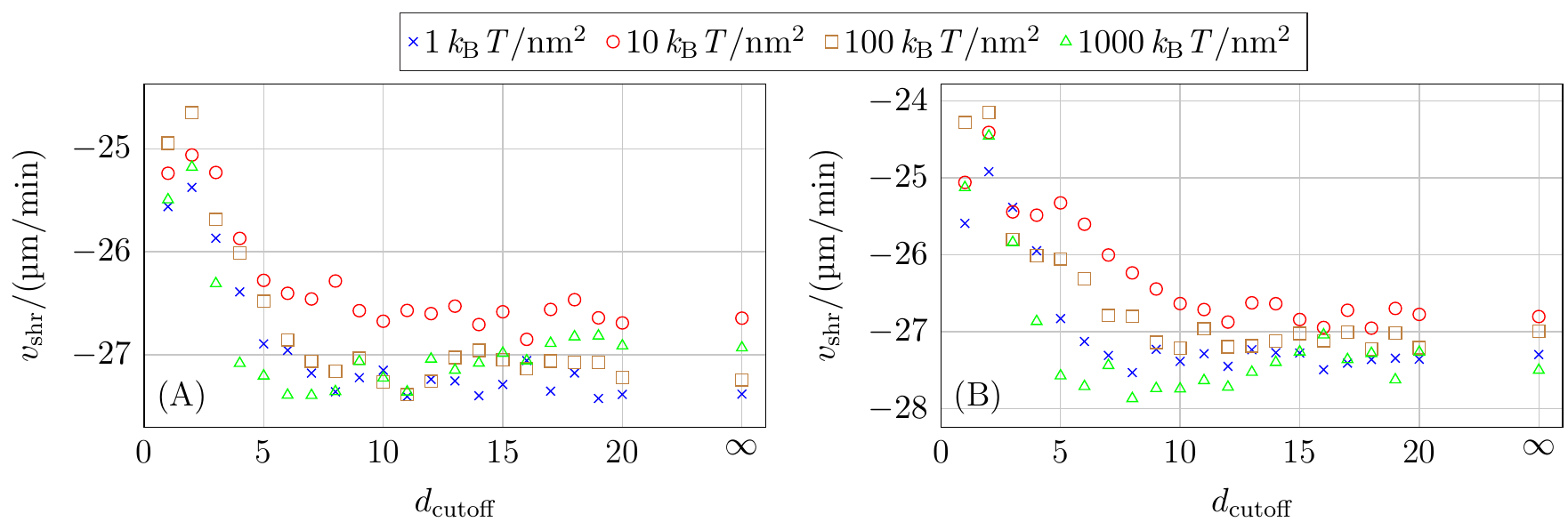}		\caption{
		Influence of the minimization cutoff $\dcutoff$ on the shrinkage velocity for
		different values of $\klat$ and
		(A) $\koncVal{2}$, $\GlongVal{-9.7}$, and initial MTs consisting of
		$\NGDP = 20$ and $\NGTP = 0$ per protofilament,
		(B) $\koncVal{4}$, $\GlongVal{-9.3}$, and initial MTs consisting of
		$\NGDP = 50$ and $\NGTP = 0$ per protofilament.
              }
        \label{fig:minimization_cutoff}
\end{figure}

\begin{figure}[!ht]
	\centering
        \includegraphics[width=0.99\linewidth]{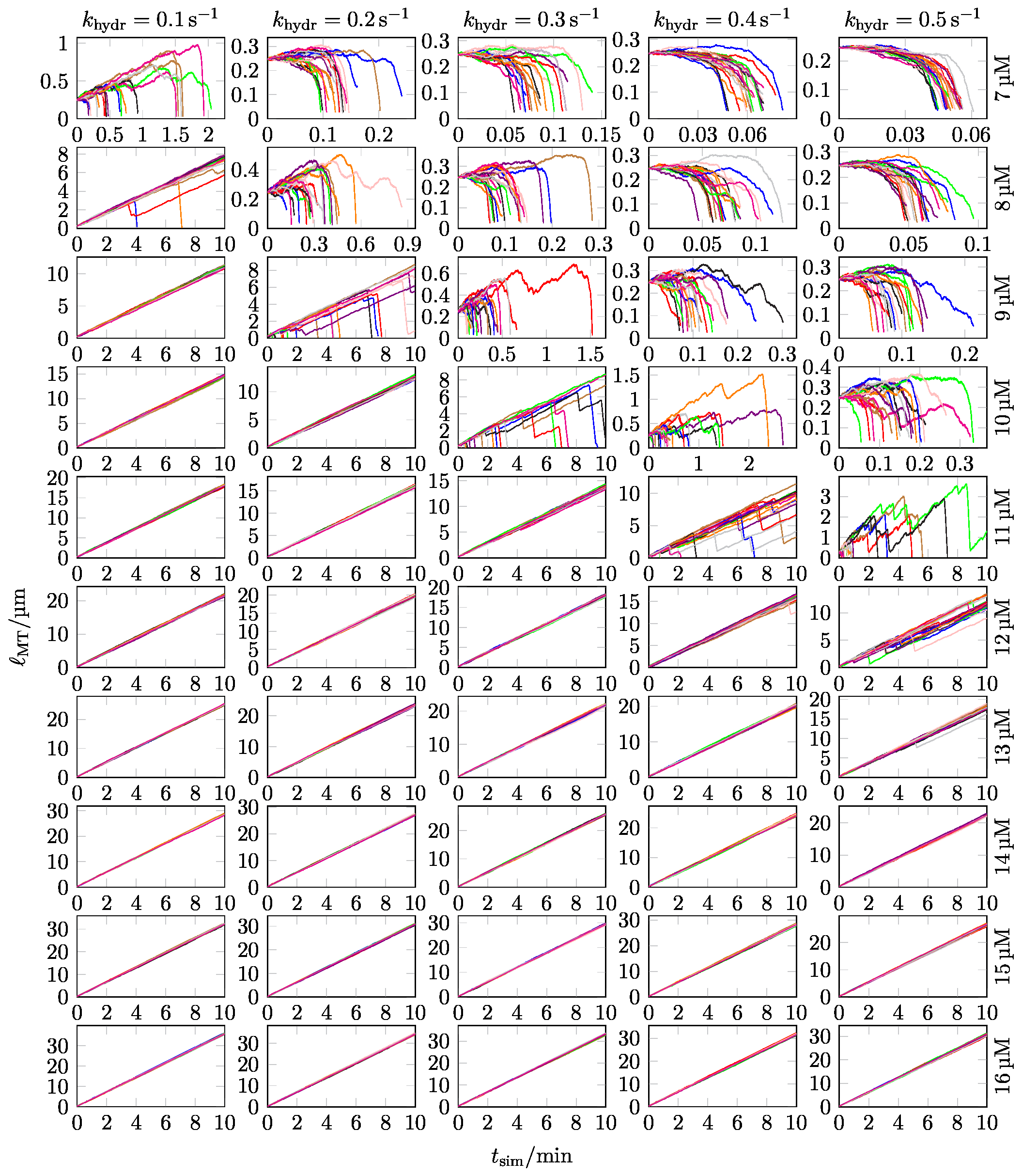}
	\caption{
		MT length $\Lmt$ as a function of time $\tsim$ for 20 different simulations with
		$\koncVal{4}$, $\GlongVal{-9.3}$,
		$\klat = \SI{100}{\kBT \per\nano\meter\squared}$, ten different values of
		$\ctub$, and five different values of $\khydr$.
		The initial MTs consist of $\NGDP = 20$ and $\NGTP = 10$ per protofilament.
		(extended version of Fig.\ 7 in the main text)
	}
	\label{fig:full_simulation_klat100}
\end{figure}

\begin{figure}[!ht]
	\centering
        \includegraphics[width=0.99\linewidth]{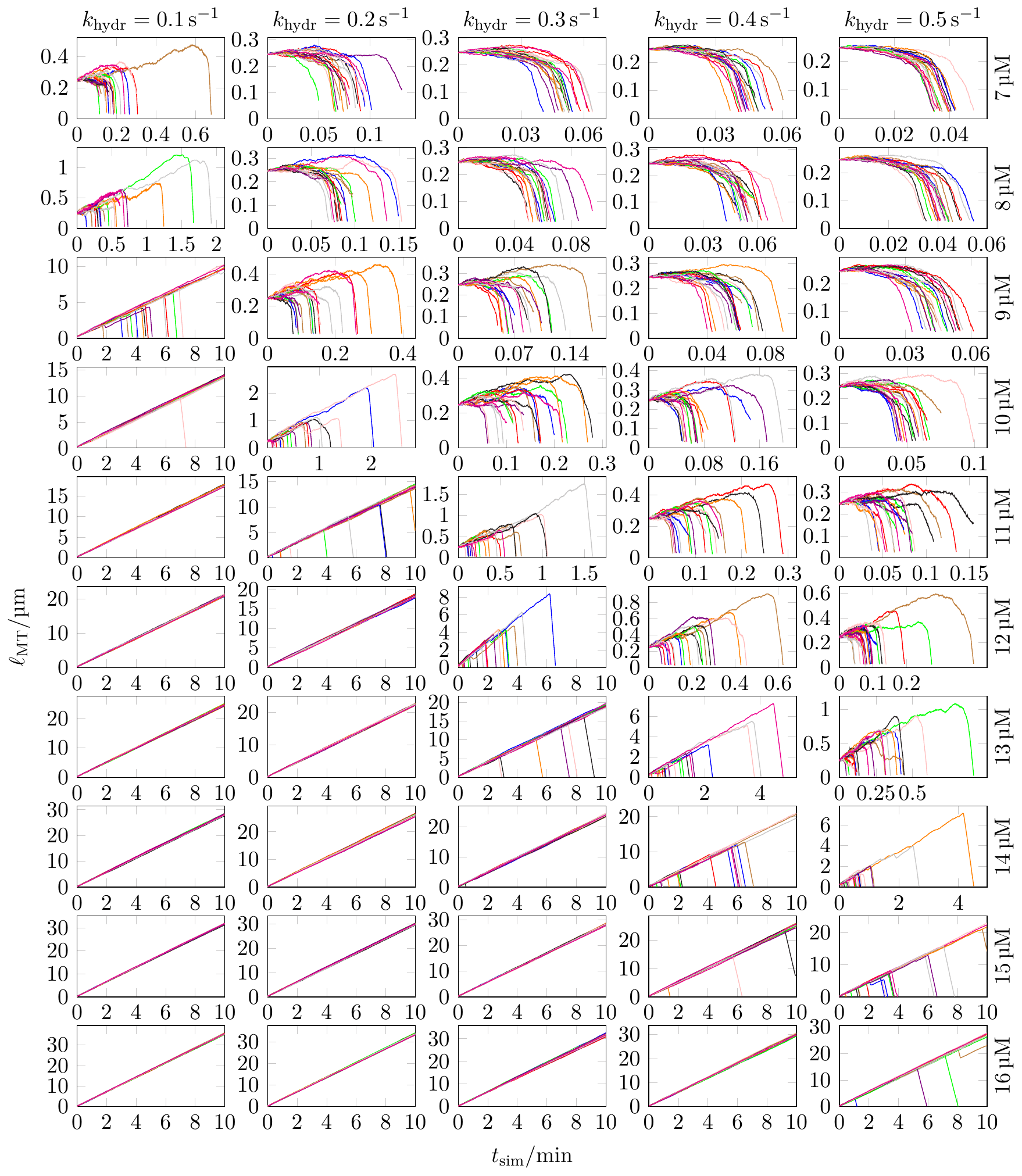}
	\caption{
		MT length $\Lmt$ as a function of time $\tsim$ for 20 different simulations with
		$\koncVal{4}$, $\GlongVal{-9.3}$,
		$\klat = \SI{1}{\kBT \per\nano\meter\squared}$, ten different values of $\ctub$,
		and five different values of $\khydr$.
		The initial MTs consist of $\NGDP = 20$ and $\NGTP = 10$ per protofilament.
	}
	\label{fig:full_simulation_klat1}
\end{figure}

\begin{figure}[!ht]
	\centering
        \includegraphics[width=0.99\linewidth]{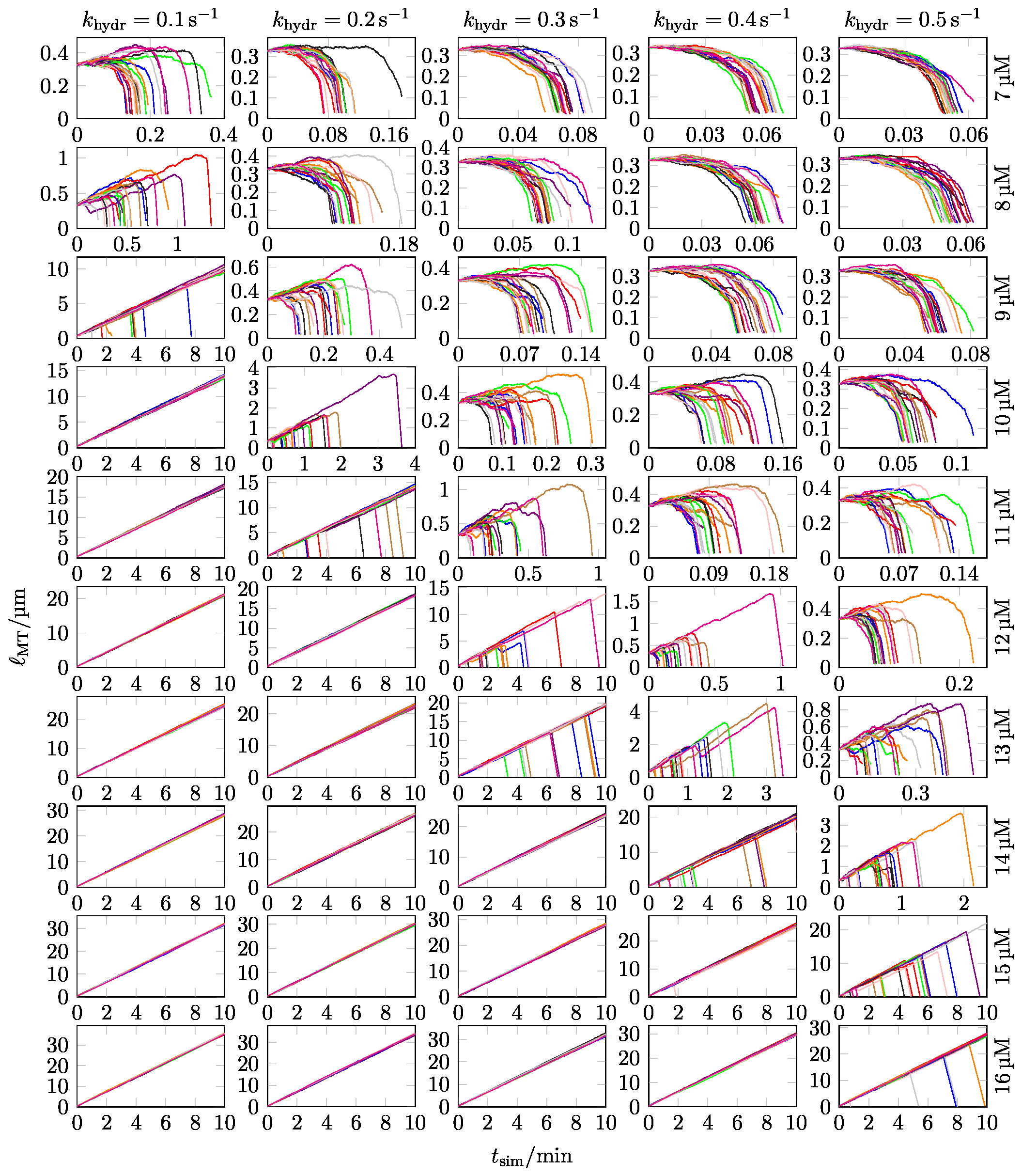}
	\caption{
		MT length $\Lmt$ as a function of time $\tsim$ for 20 different simulations with
		$\koncVal{4}$, $\GlongVal{-9.3}$,
		$\klat = \SI{1}{\kBT \per\nano\meter\squared}$, ten different values of $\ctub$,
		and five different values of $\khydr$.
		The initial MTs consist of $\NGDP = 20$ and $\NGTP = 20$ per protofilament.
	}
	\label{fig:full_simulation_klat1_cap20}
\end{figure}

\begin{figure}[!ht]
	\centering
        \includegraphics[width=0.99\linewidth]{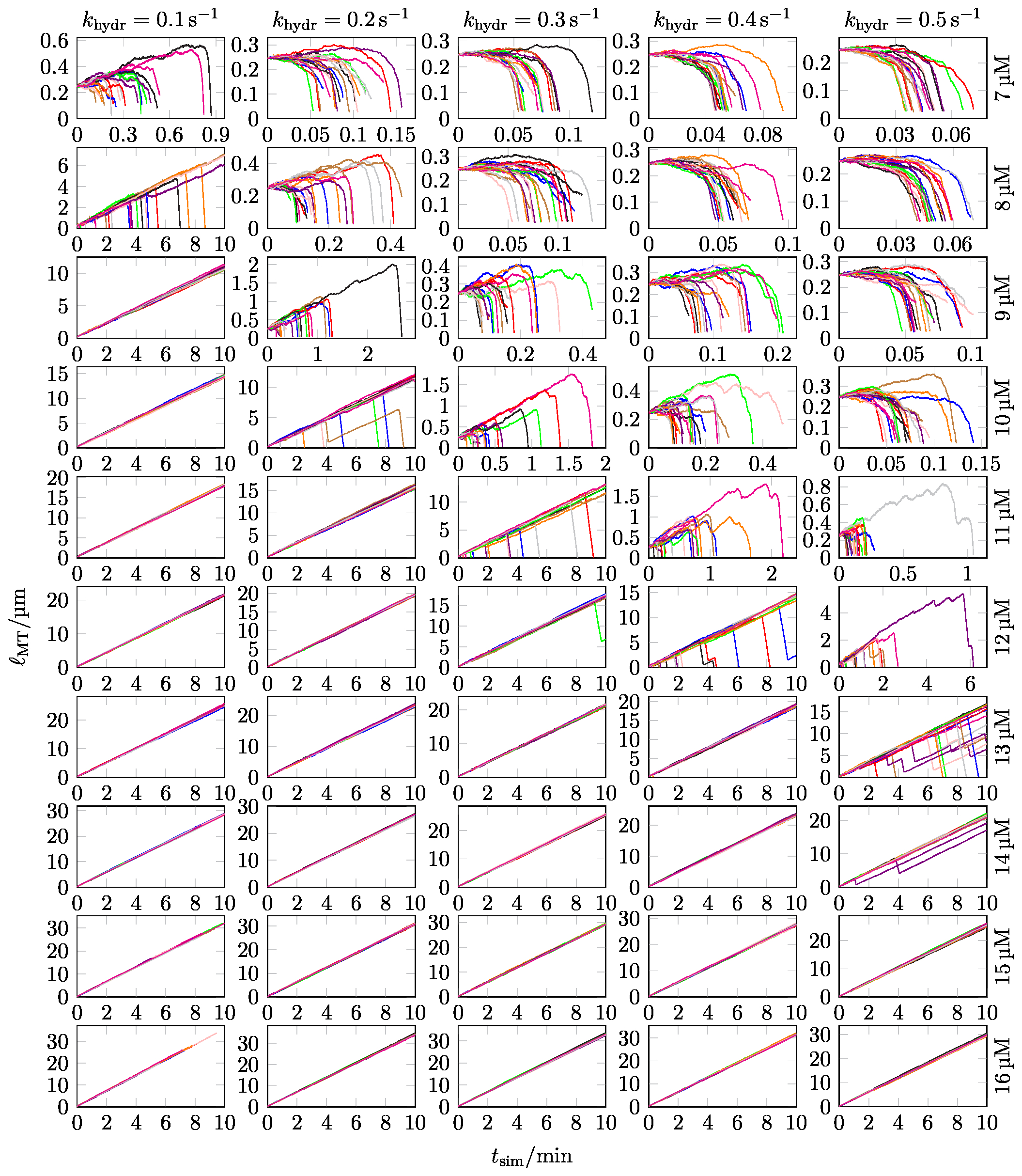}
	\caption{
		MT length $\Lmt$ as a function of time $\tsim$ for 20 different simulations with
		$\koncVal{4}$, $\GlongVal{-9.3}$,
		$\klat = \SI{10}{\kBT \per\nano\meter\squared}$, ten different values of
		$\ctub$, and five different values of $\khydr$.
		The initial MTs consist of $\NGDP = 20$ and $\NGTP = 10$ per protofilament.
		Due to runtime constraints, some of the simulations were not able to reach
		$\tsim = \SI{10}{\minute}$.
	}
	\label{fig:full_simulation_klat10}
\end{figure}

\begin{figure}[!ht]
	\centering
        \includegraphics[width=0.99\linewidth]{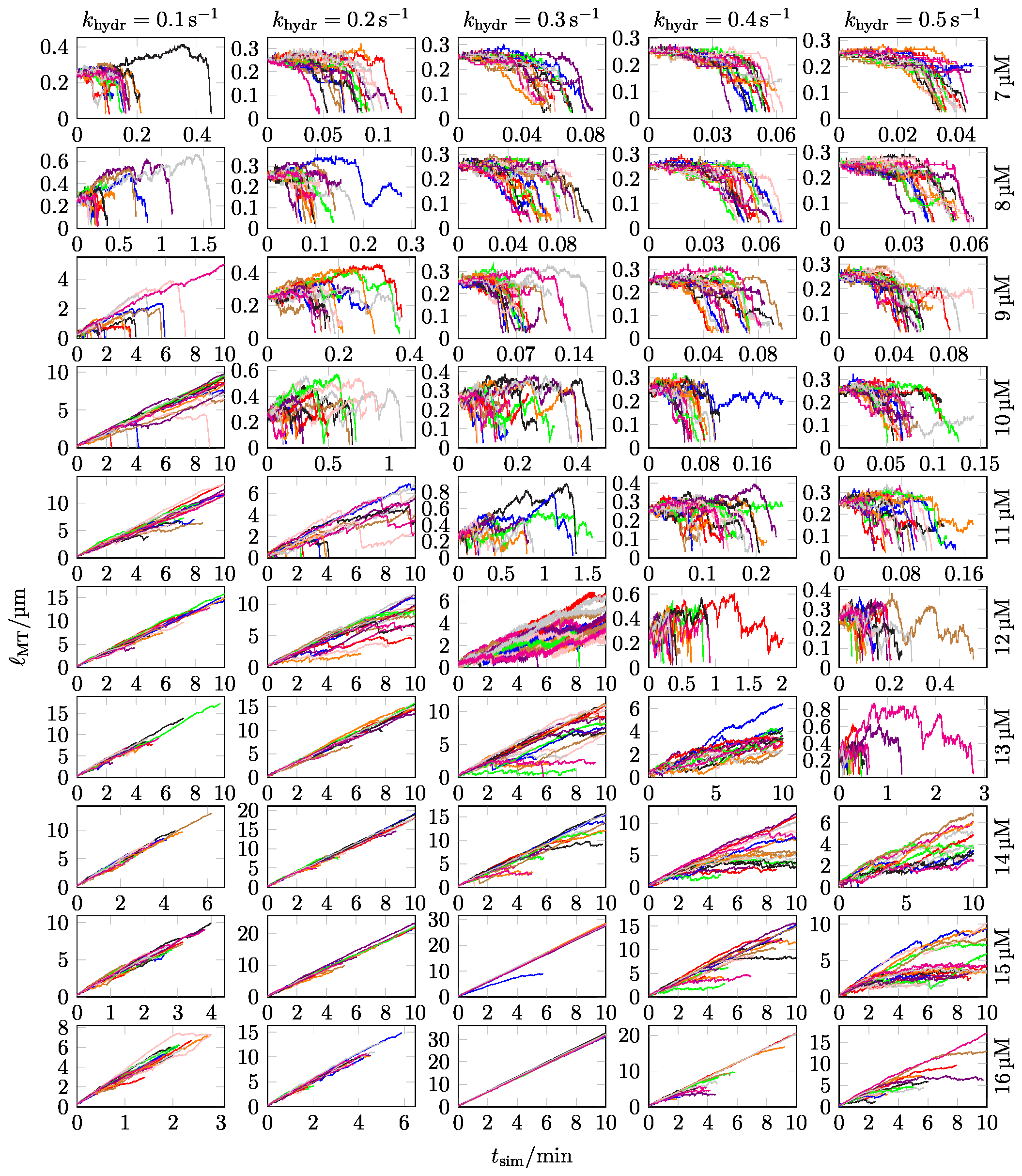}
	\caption{
		MT length $\Lmt$ as a function of time $\tsim$ for 20 different simulations with
		$\koncVal{4}$, $\GlongVal{-9.3}$,
		$\klat = \SI{20000}{\kBT \per\nano\meter\squared}$, ten different values of
		$\ctub$, and five different values of $\khydr$.
		The initial MTs consist of $\NGDP = 20$ and $\NGTP = 10$ per protofilament.
		Due to runtime constraints, several simulations were not able to reach
		$\tsim = \SI{10}{\minute}$.
	}
	\label{fig:full_simulation_klat20000}
\end{figure}

\begin{figure}[!ht]
	\centering
        \includegraphics[width=0.99\linewidth]{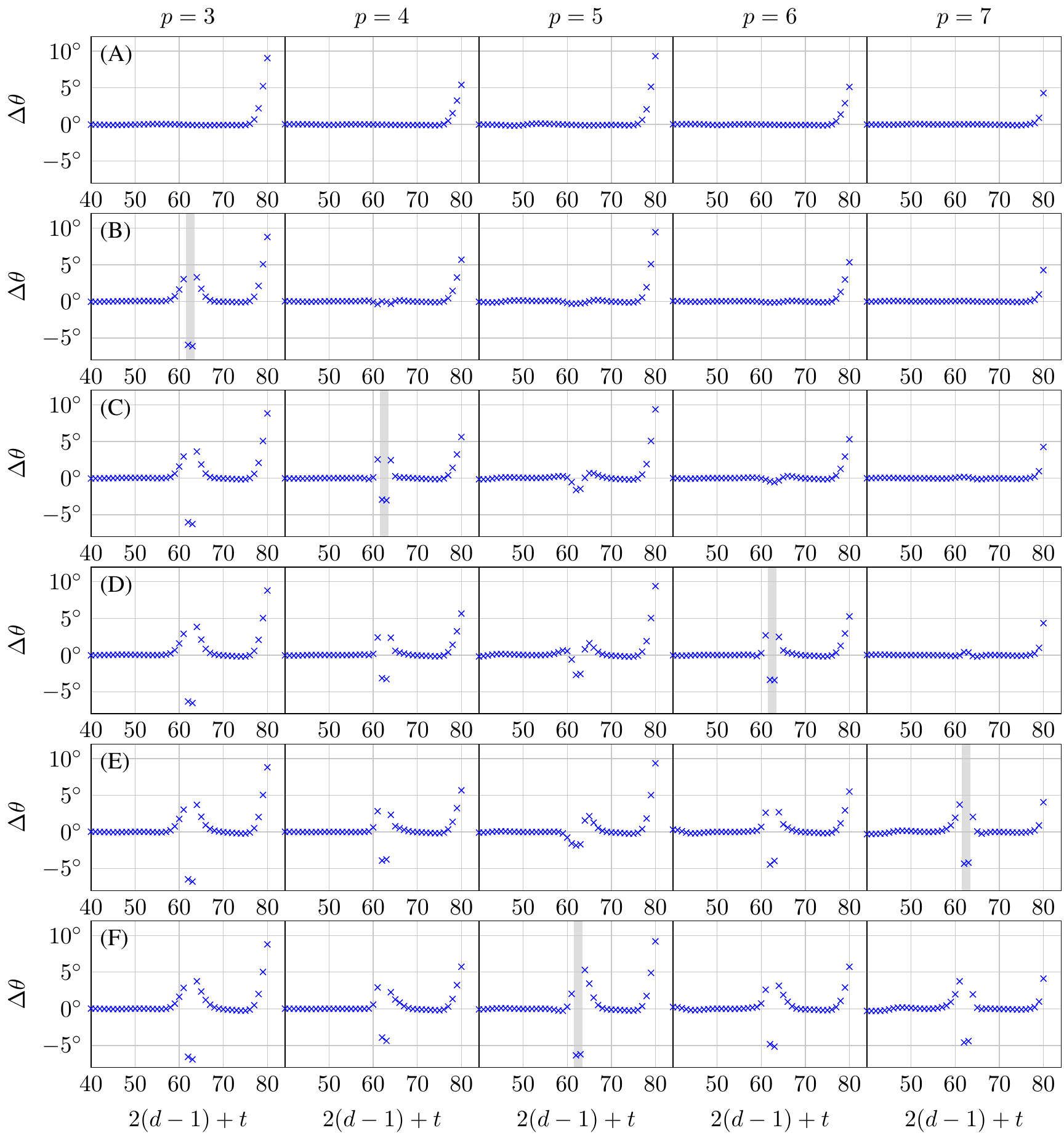}
	\caption{
		Influence of neighboring GTP-dimers in GDP-body on bending angles.
		(A) Starting from an initial MT with $\NGDP = 40$ and $\NGTP = 0$ using our
		standard parameter set from Table 2 in the main text,
		we have created a sequence in
		which the following GDP-dimers in layer $d = 31$ were exchanged with GTP-dimers
		to measure the bending angles of the GTP dimers:
		(B) $(3,31)$, (C) $(4,31)$, (D) $(6,31)$, (E) $(7,31)$, and (F) $(5,31)$.
		The highlighted intervals show the tubulin monomers that bend inward due to all
		previous changes from GDP to GTP.
              }
              \label{fig:S14_Fig}
\end{figure}

\begin{figure}[!ht]
	\centering
        \includegraphics[width=0.99\linewidth]{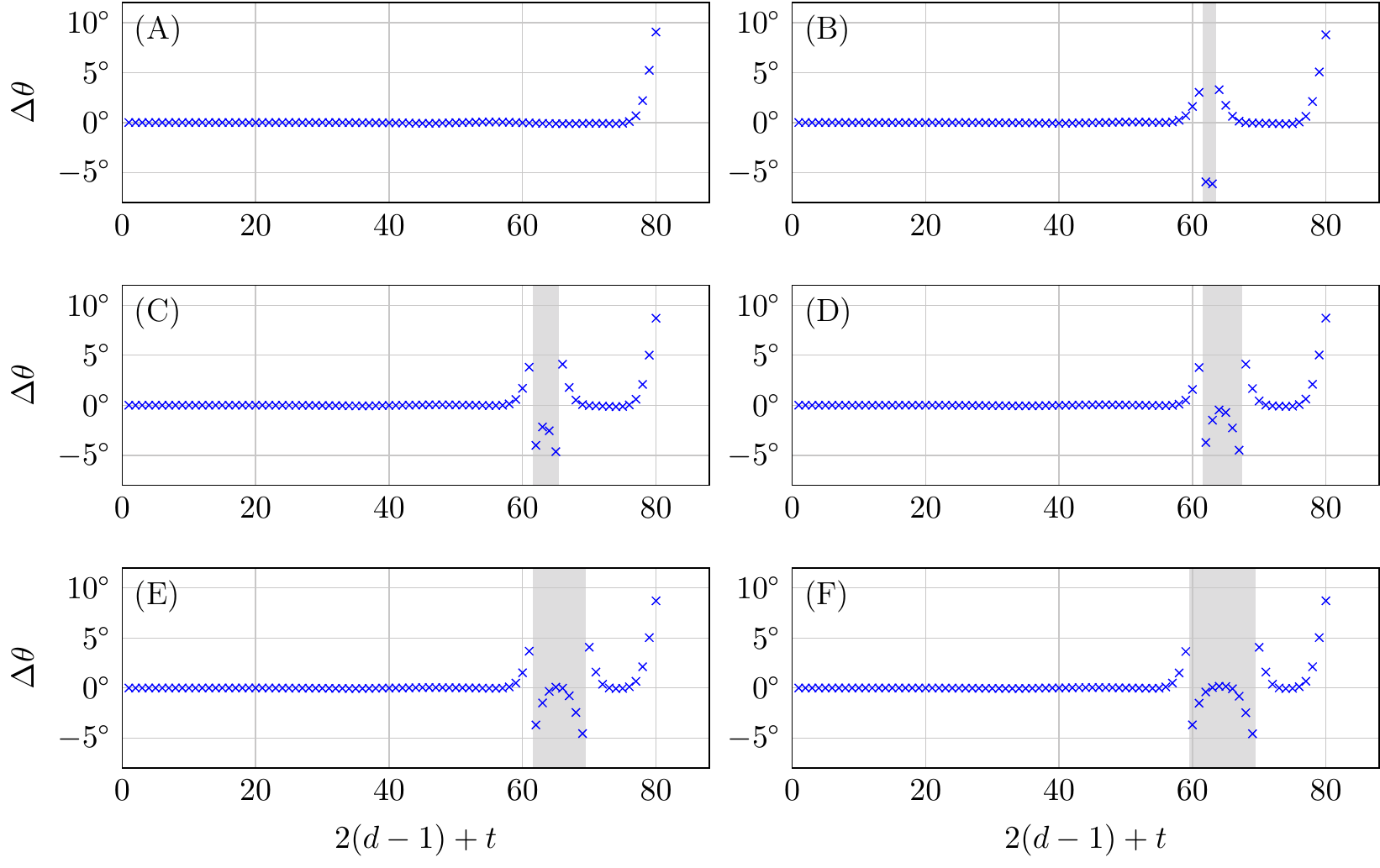}
	\caption{
		Influence of GTP-dimers in same protofilament in GDP-body on bending angles.
		(A) Starting from an initial MT with $\NGDP = 40$ and $\NGTP = 0$ using our
		standard parameter set from Table 2 in the main text,
		we have created a sequence in
		which the following GDP-dimers in protofilament $p = 3$ were exchanged with
		GTP-dimers to measure the bending angles of the GTP-dimers:
		(B) $(3,31)$, (C) $(3,32)$, (D) $(3,33)$, (E) $(3,34)$, and (F) $(3,30)$.
		The highlighted intervals show all tubulin monomers that bend inward
		due to all previous changes from GDP to GTP.
              }
              \label{fig:S15_Fig}
\end{figure}

\clearpage

\subsection*{Videos}

We supply videos of the microtubule (MT) growth simulations
from Figs.\ 10 and 13 in the main text in a two- and three-dimensional
representation.
The videos use the following color coding:
\begin{itemize}
	\item	alpha-tubulin monomers are bright green,
	\item	GTP-beta tubulin monomers are dark green,
	\item	GDP-beta tubulin monomers are orange.
\end{itemize}
\texttt{MT1\_2d.mp4} and \texttt{MT1\_3d.mp4} show the growth of the
MT from  Fig.\ 10(A) in the main text. 
\texttt{MT2\_2d.mp4} and \texttt{MT2\_3d.mp4} show the growth of the
MT from  Fig.\ 10(C) in the main text.
\texttt{MT3\_2d.mp4} and \texttt{MT3\_3d.mp4} show the growth of the
MT from Fig.\ 13(A) in the main text.


\end{document}